%% file: HIG-17-012_temp.tex
\pdfoutput=1

\documentclass[11pt,twoside,a4paper,cmspaper,final,collab]{cms-tdr}
\def\svnVersion{467387P}\def\cmsCernNoTag{CERN-EP-2018-009}\def\cmsCernDate{\today}\def\cmsMessage{Published in the Journal of High Energy Physics as \href{http://dx.doi.org/10.1007/JHEP06(2018)127}{\doi{10.1007/JHEP06(2018)127.}}}

\begin{document}\cmsNoteHeader{HIG-17-012}

\hyphenation{had-ron-i-za-tion}
\hyphenation{cal-or-i-me-ter}
\hyphenation{de-vices}
\RCS$HeadURL: svn+ssh://svn.cern.ch/reps/tdr2/papers/HIG-17-012/trunk/HIG-17-012.tex $
\RCS$Id: HIG-17-012.tex 467343 2018-07-04 09:09:29Z covarell $

\newlength\cmsFigWidth
\newlength\cmsTabSkip\setlength{\cmsTabSkip}{1ex}

\newcommand{\rowgroup}[1]{\hspace{-1em}#1}

\newcommand{\MT}{\ensuremath{m_{\mathrm{T}}\xspace}}
\newcommand\sss{}
\newcommand{\usedLumi}{35.9\fbinv}
\newcommand{\ff}{\ensuremath{4 \textrm{f}}\xspace}
\newcommand{\ggF}{\ensuremath{\Pg\Pg\mathrm{F}}\xspace}
\newcommand{\qqZZ}{\ensuremath{\cPq\cPaq\to\cPZ\cPZ}}
\newcommand{\ggZZ}{\ensuremath{\Pg\Pg\to\cPZ\cPZ}}
\newcommand{\GX}{\ensuremath{\Gamma_{\PX}}}
\newcommand{\mX}{\ensuremath{m_{\PX}}}
\newcommand{\mZZ}{\ensuremath{m_{\cPZ\cPZ}}}
\newcommand{\mZZtilde}{\ensuremath{\tilde{m}_{\cPZ\cPZ}}}
\newcommand{\tltq}{\ensuremath{2 \ell 2 \cPq}}
\newcommand{\PX}{\ensuremath{\mathrm{X}}}
\newcommand{\PV}{\ensuremath{\mathrm{V}}}
\renewcommand{\LL}{\ensuremath{\ell^+\ell^-}}
\newcommand{\XZZ}{\ensuremath{\mathrm{X}\to\cPZ\cPZ}}
\newcommand{\Ztoqq}{\ensuremath{\cPZ\to\qqbar}\xspace}
\newcommand{\mlljj}{\ensuremath{m_{\ell\ell jj}}}
\newcommand{\mllJ}{\ensuremath{m_{\ell\ell J}}}
\newcommand{\EMJWG}{\ensuremath{\Pepm\PGmmp}\xspace}
\newcommand{\nsubj}{\ensuremath{\tau_{21}}}
\newcommand{\Zhad}{\ensuremath{\cPZ_\textrm{had}}}
\newcommand{\MZhad}{\ensuremath{m(\Zhad)}}
\newcommand{\Dbkg}{\ensuremath{D_{\mathrm{bkg}}}}
\newcommand{\ZJJMELA}{\ensuremath{\mathcal{D}_\textrm{bkg}^\textrm{Zjj}}}
\newcommand{\VBFMELA}{\ensuremath{\mathcal{D}_\textrm{2jet}^\textrm{VBF}}}
\newcommand{\ZLEPLOW}{60}
\newcommand{\ZLEPHIGH}{120}
\newcommand{\PTLEPHIGH}{40}
\newcommand{\LSBLOW}{40}
\newcommand{\LSBHIGH}{70}
\newcommand{\SRLOW}{70}
\newcommand{\SRHIGH}{105}
\newcommand{\USBLOW}{135}
\newcommand{\USBHIGH}{180}
\newcommand{\VHADRESOLVEDPTCUT}{100}
\newcommand{\VHADMERGEDPTCUT}{170}
\newcommand{\DILEPTONPTCUT}{100}
\newcommand{\VHADPTSWITCH}{300}
\newcommand{\DILEPTONPTSWITCH}{200}
\newcommand{\MZZCUT}{500}
\newcommand{\TAUCUT}{0.6}

\cmsNoteHeader{HIG-17-012}
\title{Search for a new scalar resonance decaying to a pair of \cPZ\ bosons in proton-proton collisions at $\sqrt{s} = 13\TeV$}

\date{\today}

\abstract{
A search for a new scalar resonance decaying to a pair of $\cPZ$ bosons is performed in the mass range from 130\GeV to 3\TeV, and for various width scenarios. The analysis is based on proton-proton collisions recorded by the CMS experiment at the LHC in 2016, corresponding to an integrated luminosity of \usedLumi at a center-of-mass energy of 13\TeV. The
$\cPZ$ boson pair decays are reconstructed using the $4\ell$, $2 \ell 2 \cPq$, and $2\ell2\nu$ final states, where $\ell = \Pe$ or \PGm.
Both gluon fusion and electroweak production of the scalar resonance are considered, with a free parameter
describing their relative cross sections. A dedicated categorization of events, based on the kinematic properties of
associated jets, and matrix element techniques are employed for an optimal signal and background separation. A description of the interference between signal and background amplitudes for a resonance of an arbitrary width is included.
No significant excess of events with respect to the standard model expectation is observed and limits are set on the product of the cross section for a new scalar boson and the branching fraction for its decay to $\cPZ\cPZ$ for a large range of masses and widths.
}

\hypersetup{%
pdfauthor={CMS Collaboration},%
pdftitle={Search for a new scalar resonance decaying to a pair of \cPZ\ bosons in proton-proton collisions at sqrt(s) = 13 TeV},%
pdfsubject={CMS},%
pdfkeywords={Heavy scalars, matrix-element technique, ZZ final states}}

\maketitle
\ifthenelse{\boolean{cms@external}}{}{
\clearpage
}

\section{Introduction}
\label{sec:Introduction}

The standard model (SM) of particle physics postulates the existence of a single Higgs boson as the manifestation
of a scalar field responsible for electroweak (EW) symmetry
breaking~\cite{StandardModel67_1, Englert:1964et,Higgs:1964ia,Higgs:1964pj,Guralnik:1964eu,StandardModel67_2,StandardModel67_3}.
The ATLAS and CMS Collaborations have discovered a boson with a mass close to 125\GeV~\cite{Aad:2012tfa,Chatrchyan:2012xdj,Chatrchyan:2013lba}
with properties consistent with those expected for the SM Higgs boson~\cite{Chatrchyan:2012jja,Chatrchyan:2013mxa,Khachatryan:2014kca,Aad:2013xqa,Aad:2015mxa}, and no other fundamental particle that would require explanation beyond the SM (BSM) has been discovered to date.
Nonetheless, searches for BSM physics are motivated by a number of phenomena such as the presence of dark matter or baryon asymmetry in the universe that are not explained
by the SM. Extensions of the SM that attempt
to address these questions include two-Higgs-doublet models (2HDM)~\cite{Branco:2011iw}, of which supersymmetry is an example,
or other models predicting an extended Higgs-like EW singlet~\cite{Caillol:2013gqa}.
In the following, we denote the recently discovered scalar boson as $\PH(125)$.
The search for a heavy scalar partner of the $\PH(125)$, which we will generically
denote as $\PX$, is the subject of this paper.

The $\cPZ\cPZ$ decay has a sizable branching fraction
for a SM-like Higgs boson for masses larger than the $\cPZ$ boson pair production threshold, $2m_\cPZ$, and is one of the main discovery channels
for masses less than $2m_\cPZ$~\cite{Aad:2012tfa,Chatrchyan:2012xdj,Chatrchyan:2013lba}. Since the mass of
a new state $\PX$ is unknown, the search is performed over a wide range of masses from 130\GeV up to 3\TeV.
Three final states
are considered: $4\ell$, \tltq, and $2\ell2\nu$, with $\ell= \Pe$ or \PGm.
Previous searches for a new boson decaying to $\cPZ\cPZ$ or $\PW\PW$ pairs
have been reported
by the CMS~\cite{Khachatryan:2015cwa} and ATLAS~\cite{Aad:2015kna,Aad:2015agg}
Collaborations at the CERN LHC, using proton-proton collisions recorded at center-of-mass energies of 7 and 8\TeV, where no significant excess was observed.
A data set of proton-proton collisions recorded at a center-of-mass energy of 13\TeV by the CMS experiment in 2016 is used in this analysis,
corresponding to an integrated luminosity of $\usedLumi$.

The approach adopted in this analysis treats a new $\PX$ boson in a model-independent way.
For any given mass $\mX$ of the $\PX$ boson, both its width $\GX$ and production mechanism are assumed
to be unknown. In this analysis, $\mX$ and $\GX$ refer to the mass and width of the scalar boson that enter the propagator.
No modification from the complex-pole scheme~\cite{Kauer:2012hd,Goria:2011wa} is considered.
The two dominant production mechanisms of a scalar boson are gluon fusion (\ggF) and
EW production, the latter dominated by vector boson fusion (VBF) with a small contribution of
production in association with an EW boson $\cPZ\PH$ or $\PW\PH$ ($\PV\PH$).
We define the parameter $f_{\mathrm{VBF}}$
as the fraction of the EW production cross section with respect to the total cross section.
The three parameters $\mX$, $\GX$, and $f_{\mathrm{VBF}}$ are scanned over
a wide range of allowed phase space, and limits are set on the $\Pp\Pp\to\PX\to\cPZ\cPZ$ cross section.

The new state $\PX$ can potentially have a large value $\GX$:
in this case, there is sizable interference between the $\PX\to \cPZ\cPZ\to \ff$ amplitude and that of the SM background process $\cPZ\cPZ / \cPZ\gamma^*\to \ff$, where f denotes any fermion.
The interference distorts both the kinematic distributions and overall yield of the BSM contribution.
The SM background includes the contribution from the $\PH(125)\to\cPZ\cPZ\to \ff$ decays, which yields a nonnegligible off-shell contribution above the $2m_\cPZ$ threshold~\cite{Kauer:2012hd}.
The above interference effect is present in both \ggF and EW processes and is taken into account in this analysis.
The reported cross-section limits correspond to the signal-only contribution as it would be in the absence of interference.
A novel feature in this analysis is the inclusion of all of the above effects in a parametric way
in a likelihood fit to the data. The matrix element (ME) formalism is used both for the parameterization
of the likelihood and for the construction of the observables optimal for event categorization.

The paper is organized as follows.
In Section~\ref{sec:CMS}, the CMS detector and event reconstruction techniques are presented.
Monte Carlo (MC) simulation of the signal and background processes is described in Section~\ref{sec:MC}.
Matrix element methods are discussed in Section~\ref{sec:mela}.
Event selection and categorization in each channel are presented in Section \ref{Section_selection}.
Modeling of the signal distributions and background estimation techniques are described in Section~\ref{Section_parameterization}.
Systematic uncertainties are summarized in Section~\ref{sec:Systematics}.
In Section~\ref{sec:Results} results are presented, and we conclude in Section~\ref{sec:Summary}.

\section{The CMS detector and event reconstruction}
\label{sec:CMS}

The CMS detector comprises a silicon pixel and strip tracker, a lead tungstate crystal electromagnetic
calorimeter (ECAL), and a brass and scintillator hadron calorimeter (HCAL), each composed of a barrel and two endcap sections,
all within a superconducting solenoid of 6\unit{m} internal diameter and providing a magnetic field of 3.8\unit{T}.
Outside of the solenoid are the gas-ionization detectors for muon measurements, which are embedded
in the steel flux-return yoke outside the solenoid. The detection layers are made using three technologies: drift tubes,
cathode strip chambers, and resistive-plate chambers. Extensive forward calorimetry
complements the coverage provided by the barrel and endcap detectors.  A more detailed
description of the CMS detector, together with a definition of the coordinate system
and the relevant kinematic variables used, can be found in Ref.~\cite{Chatrchyan:2008zzk}.

The particle-flow (PF) event algorithm~\cite{Sirunyan:2017ulk} reconstructs and identifies each individual particle with an optimized combination of information from the various elements of the CMS detector.
The reconstructed vertex with the largest value of summed physics-object $\pt^2$ is taken to be the primary $\Pp\Pp$ interaction vertex. The physics objects are the jets, clustered using the jet finding algorithm~\cite{Cacciari:2008gp,Cacciari:2011ma} with the tracks assigned to the vertex as inputs, and the associated missing transverse momentum, taken as the negative vector sum of the \pt of those jets.
The energy of photons is obtained from the ECAL
measurement, corrected for zero-suppression effects. The energy of electrons is determined from a combination of the
electron momentum at the primary interaction vertex as determined by the tracker, the energy of the corresponding ECAL cluster,
and the energy sum of all bremsstrahlung photons spatially compatible with originating from the electron track.
The momentum of muons is obtained from the curvature of the corresponding tracks in the tracker and the muon systems~\cite{Chatrchyan:2012xi}.
The energy of charged hadrons is determined from a combination of their momentum measured in the tracker and
the matching ECAL and HCAL energy deposits, corrected for zero-suppression effects and for the response function
of the calorimeters to hadronic showers. Finally, the energy of neutral hadrons is obtained from the corresponding
corrected ECAL and HCAL energy.
The missing transverse momentum vector \ptvecmiss is defined as the projection onto the plane perpendicular to the beam axis of the negative vector sum of the momenta of all reconstructed particle-flow objects in an event. Its magnitude is referred to as \ptmiss.
The correction mentioned above also applies to the determination of \ptmiss.

Collision events are selected by high-level trigger algorithms~\cite{Khachatryan:2016bia} that require the presence of leptons passing loose identification and isolation
requirements. The main triggers for this analysis select a pair of electrons or muons. Triggers selecting an \Pe\PGm pair
are also used for the $4\ell$ channel and in control samples for $2\ell2\Pq$ and $2\ell2\nu$. The minimal \pt of the leading electron (muon) is 23 (17)\GeV, while that of the subleading lepton is 12 (8)\GeV. Isolated single-electron (muon) triggers with minimal \pt of 27 (22)\GeV are also employed to complement the double-lepton triggers.

Electrons are measured in the ECAL in the pseudorapidity range $\abs{\eta}< 2.4$.
The momentum resolution
for electrons with $\pt\approx45\GeV$ from $\cPZ \to \Pe \Pe$ decays ranges from 1.7\% for nonshowering electrons
in the barrel region to 4.5\% for showering electrons in the endcaps~\cite{Khachatryan:2015hwa}.
Muons are measured in the range $\abs{\eta}< 2.4$.
Muons are reconstructed by combining information from the silicon tracker and the muon system~\cite{Chatrchyan:2012xi}. The matching between the inner and outer tracks proceeds either outside-in, starting from a track in the muon system, or inside-out, starting from a track in the silicon tracker. In the latter case, tracks that match track segments in one or two (out of four) layers of the muon system are also considered in the analysis to collect very low \pt muons that may not have sufficient energy to penetrate the entire muon system.
Matching muons to tracks measured in the silicon tracker results in a relative
\pt resolution for muons with $20 <\pt < 100\GeV$ of 1.3--2.0\% in the barrel and better than 6\% in the endcaps.
The \pt resolution in the barrel is better than 10\% for muons with \pt up to 1\TeV~\cite{Chatrchyan:2012xi}.

Hadronic jets are clustered from the four-momenta of the particles in a jet reconstructed by the PF algorithm,
using the \FASTJET software package~\cite{Cacciari:2011ma}. Jets are clustered using the anti-\kt algorithm~\cite{Cacciari:2008gp}
with a distance parameter equal either to 0.4 (``AK4 jets'') or 0.8 (``AK8 jets'').
Charged PF constituents not associated with the primary vertex
are not used in the jet clustering procedure.

Jet energy momentum is determined as the vectorial sum of all particle four-momenta in the jet. Jets are reconstructed
in the range $\abs{\eta}< 4.7$.
An offset correction is applied to jet energy momenta to account for the contribution from additional proton proton interactions
in the same or neighboring bunch crossings (pileup). These corrections are derived from simulation, and are confirmed with in situ measurements
of the energy momentum balance in dijet, multijet, $\gamma + \text{jet}$ and leptonically decaying $\cPZ + \text{jets}$ events~\cite{Khachatryan:2016kdb}. Additional selection criteria are applied to each event to remove spurious jet like features originating from isolated noise patterns in certain HCAL regions.

\section{Monte Carlo simulation}
\label{sec:MC}

Signal events with SM like couplings are generated at next to leading order (NLO) in quantum chromodynamics (QCD) with
\POWHEG\ 2.0~\cite{Frixione:2007vw,Bagnaschi:2011tu,Nason:2009ai,Nason:2004rx,Alioli:2010xd}
for the \ggF and VBF production modes. The
decays $\PX \to \cPZ\cPZ \to 4\ell$, $2\ell2\cPq$, and $2\ell2\nu$ are
modeled with \textsc{JHUGen}\ 7.0.2~\cite{Gao:2010qx,Bolognesi:2012mm,Anderson:2013afp,Gritsan:2016hjl},
including corrections for the $\cPZ\cPZ$ branching fraction, and correct modeling of the angular correlation among the fermions.
A wide range of masses $m_{\PX}$ from 100\GeV to 3\TeV is generated with the width $\Gamma_{\PX}$
set according to the SM Higgs boson expectation for $m_{\PX}$ up to 1\TeV. For higher masses, we choose the width $\Gamma_{\PX} = 0.5m_{\PX}$, which approximately corresponds to the SM Higgs boson prediction for $m_{\PX} =  1\TeV$. The samples are used to derive a generic signal parameterization.

While NLO accuracy in QCD is used in production, no modeling of the interference with background
is included at this stage of the simulation.
The MELA matrix element package~\cite{Gao:2010qx,Bolognesi:2012mm,Anderson:2013afp,Gritsan:2016hjl},
based on \textsc{JHUGen} for both $\PH(125)$ and $\PX$ signal,
and on \MCFM\ 7.0~\cite{MCFM,Campbell:2011bn,Campbell:2013una} for the continuum background, allows modeling of interference of a broad $\PX$ resonance
with SM background in either \ggF or EW production, the latter including VBF
and $\PV\PH$ processes.

The loop induced production of two $\cPZ$ bosons, gg $\to \cPZ\cPZ / \cPZ\gamma^*\to \ff$ background, including the
off shell tail of the $\PH(125)$, is modeled at leading order (LO) in QCD with \MCFM.
The corresponding background from EW production, $\Pq \Pq' \cPZ\cPZ / \cPZ\gamma^*\to \ff \Pq \Pq'$
is modeled at LO in QCD with \textsc{Phantom}\ 1.2.8~\cite{Ballestrero:2007}.
For both \ggF and VBF simulation, the factorization and renormalization
scales are chosen as $m_{\cPZ\cPZ}/2$, and NNPDF3.0 parton distribution functions (PDFs)~\cite{Ball:2014uwa} are adopted.
In order to include higher order QCD corrections to gluon fusion production, LO, NLO, and next to next to leading order (NNLO) signal cross section calculations
are performed using the \MCFM\ and \textsc{hnnlo} v2 programs~\cite{Catani:2007vq,Grazzini:2008tf,Grazzini:2013mca}
for a wide range of masses using the narrow width approximation.
The ratio between the NNLO and LO, or between the NLO and LO, is used as a
weight depending on the $\ff$ invariant mass (K factor).
While this procedure is directly applicable for the signal, it is approximate for the background. However, an
NLO calculation is available~\cite{Caola:2015psa,Melnikov:2015laa} for the background in the mass range
$2m_Z<m_{4\ell}<2m_{t}$. There is a good agreement between the NLO K factors calculated for signal
and background and any differences set the scale of systematic uncertainties in this procedure,
for which we assign a 10\% uncertainty.
Event yields for the $\PH(125)$ boson production are normalized to the cross section at NNLO in QCD and NLO in EW for \ggF~\cite{Anastasiou:2016cez} and others taken from Ref.~\cite{deFlorian:2016spz}.

The MELA package is also used to reweight the \POWHEG{}/\textsc{JHUGen}, \MCFM, or \textsc{Phantom} signal
samples to model various values of $\mX$ and $\GX$, as well as the interference with the background component.

The background from the production of two $\cPZ$ bosons from quark antiquark annihilation, $\cPq \overline{\cPq} \to \cPZ\cPZ/ \cPZ\gamma^*\to \ff$, is evaluated at NLO with \POWHEG~\cite{Nason:2013ydw}
and \MGvATNLO 2.3.2~\cite{MadGraph}.
The $\PW\cPZ$ production is generated at LO with \PYTHIA\ 8.212~\cite{Sjostrand:2008za}, normalized to NNLO in QCD accuracy~\cite{Grazzini:2016swo}.
The $\cPZ + \text{jets}$ (\cPZ\ $\to \ell^+ \ell^-$) simulation is made of a composite sample comprising
a set of exclusive LO samples with various associated parton multiplicities,
including a dedicated sample with associated \cPqb\ quark production. These samples are produced at LO with
\MGvATNLO and corrected
to NLO QCD accuracy with a K factor depending on the \pt of the dilepton pair,
derived from \MGvATNLO simulation at NLO with FxFx merging scheme~\cite{Frederix:2012ps}.
The simulation of top quark antiquark pair production, \ttbar{}, is performed with \POWHEG{} at NLO in QCD~\cite{Frixione:2007nw}.

All generated samples are interfaced with \PYTHIA,
configured with the CUETP8M1 tune~\cite{CUETP8M1}
for simulation of parton showers, hadronization, and underlying event effects.
All simulated events are further processed with a \GEANTfour{} based description~\cite{Agostinelli2003250}
of the CMS detector and reconstructed with the same algorithms as used for data.
Supplementary minimum bias (pileup) interactions are added to the simulated events with a multiplicity determined such as to match that observed in data.

\section{Matrix element techniques}
\label{sec:mela}

The ME method in this study is utilized in three ways. First, it is used to apply weights
to generated events from various models to avoid having to fully simulate the samples, as
discussed in Section~\ref{sec:MC}. Second, the ME method is used to create a model
of a broad high mass resonance $\PX$, including its interference with the SM background, to be used in the
likelihood fit. Finally, this method is used to create optimal discriminants for either categorization
of events according to likely production mechanism, or to separate signal from the dominant background.

The ME calculations are performed using the {\sc MELA}
package,
which provides the full set of processes studied in this paper and uses \textsc{JHUGen} matrix elements
for the signal and \MCFM matrix elements for the background.
The signal includes both the four fermion kinematic properties for the decay
$\PX \to \cPZ\cPZ \to \ff$, and the kinematical properties of associated particles in the $\PX$ + 2jets, VBF, $\cPZ\PH$, $\PW\PH$ production.
The background includes $\Pg\Pg$ or $\qqbar\to\cPZ\cPZ$ / $\cPZ\gamma^*$ / $\gamma^*\gamma^*$ / $\cPZ\to \ff$ processes,
VBF production of a $\cPZ$ boson pair, the associated production of a \cPZ\ pair with a third vector boson,
and the production of a single $\cPZ$ boson in association with jets.

\begin{figure}[!htb]
\centering
\includegraphics[width=0.45\textwidth]{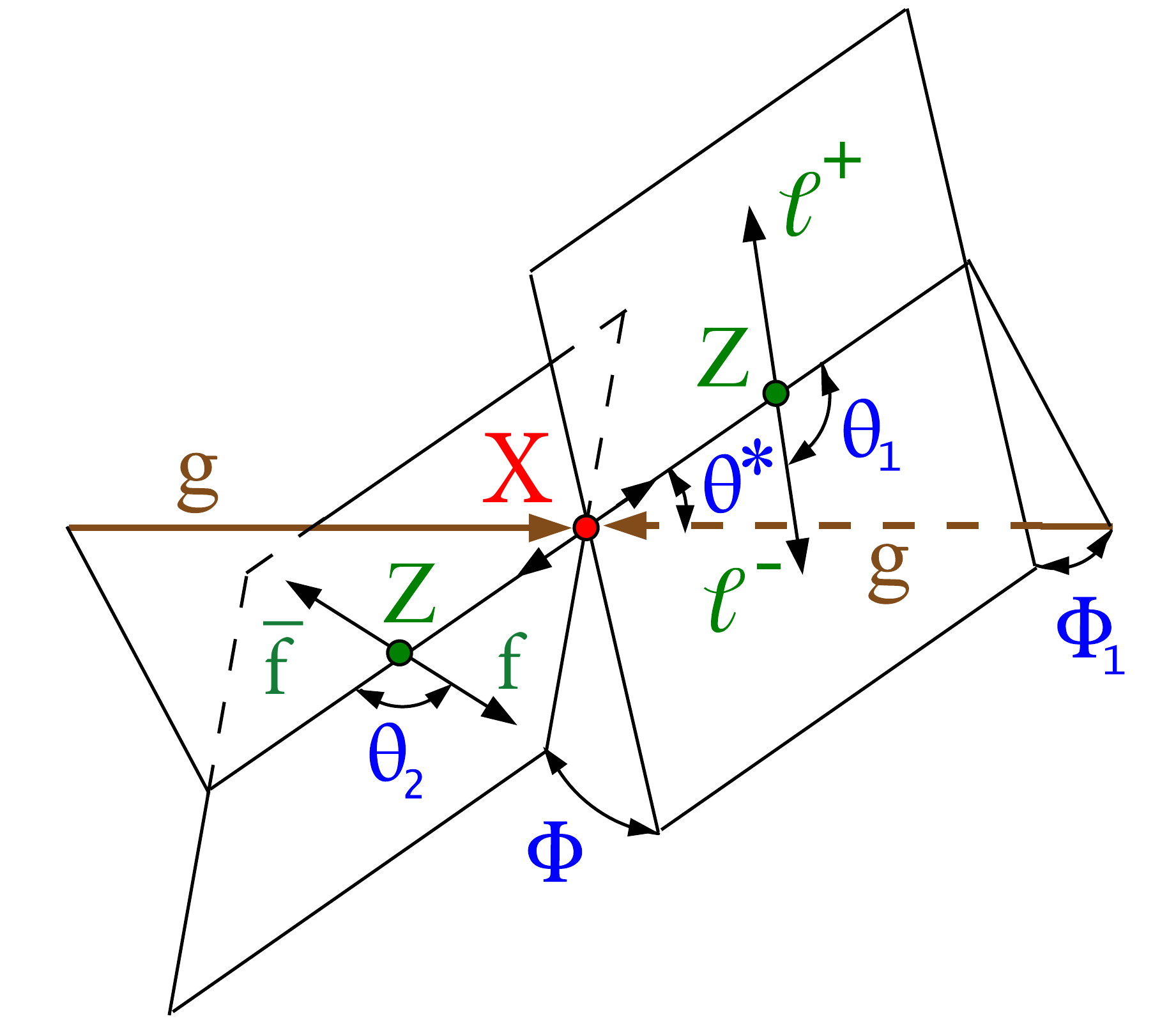}
\includegraphics[width=0.45\textwidth]{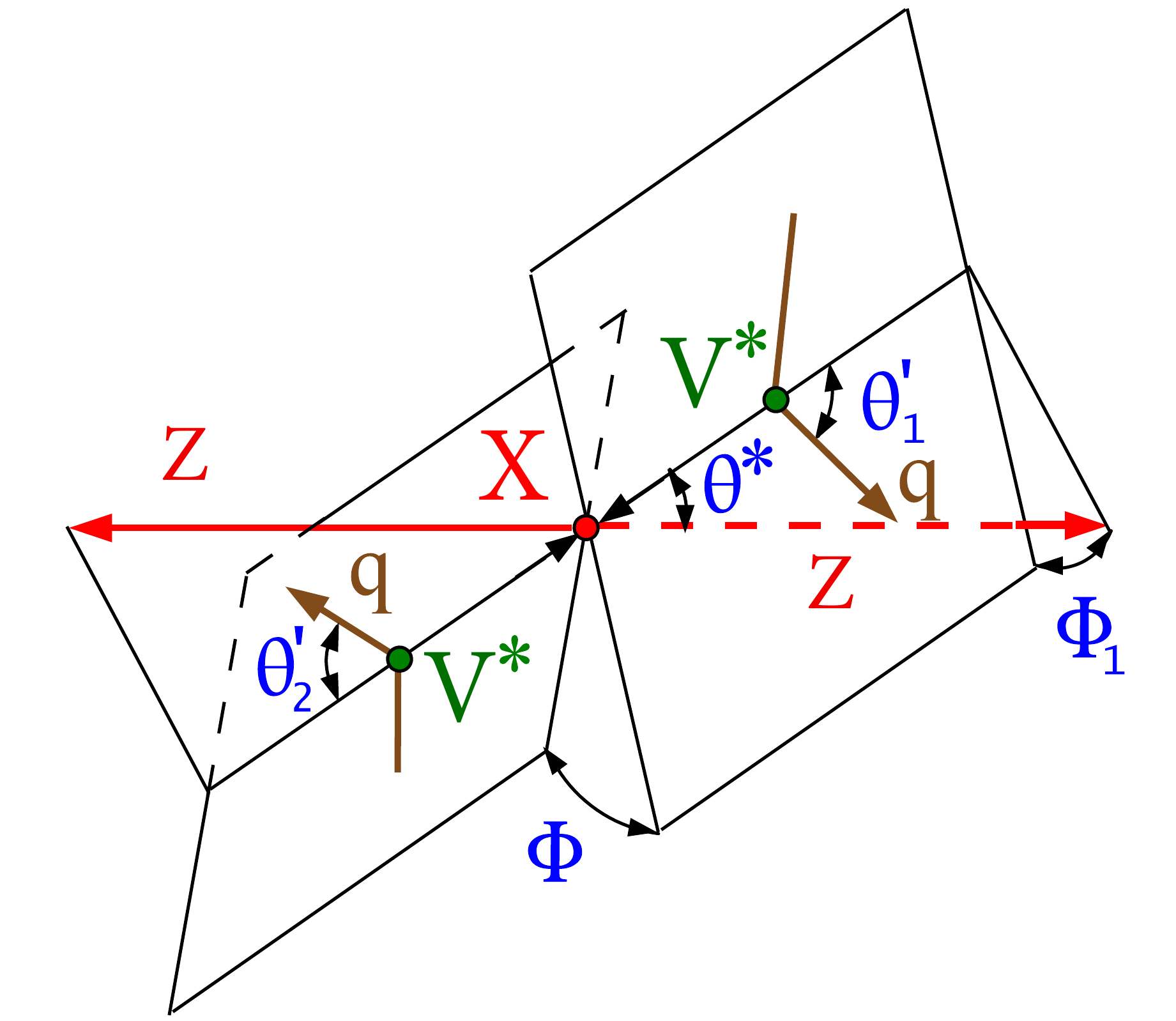}
\caption
{
Illustration of an $\PX$ boson production from \ggF, $\Pg\Pg\to \PX\to \PZ\PZ\to (\ell^+\ell^-)(f\overline{f})$ (left), and VBF, $\Pq{\Pq^\prime}\to \Pq{\Pq^\prime} \PX \to \Pq{\Pq^\prime}\PZ\PZ$ (right). The five angles shown in blue and the invariant masses of the two vector bosons shown in green fully characterize either the production or the decay chain. The angles are defined in either the \PX\ or \PV\ boson rest frames~\cite{Gao:2010qx,Anderson:2013afp}.
\label{fig:decay}
}
\end{figure}

Two of the final states studied in this analysis, $\PX \to \cPZ\cPZ \to 4\ell$ and $2\ell2\cPq$
provide full information about the kinematic properties of the process in both production and decay. This is illustrated in
Fig.~\ref{fig:decay}, where a complete set of angles and invariant masses, denoted as $\vec\Omega$, fully defines the four vectors
of all involved particles in the center of mass frame~\cite{Gao:2010qx,Anderson:2013afp}.
The overall boost of the system depends on QCD effects beyond LO
(in the transverse plane) or PDFs (in the longitudinal direction). Therefore, in these two channels, matrix element calculations
are used to create discriminants optimal either for categorization of the production mechanism or to separate signal
from background using production and decay information.

The discriminant sensitive to the VBF signal topology with two energetic and forward associated jets
is calculated as~\cite{Khachatryan:2015cwa, Khachatryan:2015mma}
\begin{eqnarray}
\label{eq:vbfmela}
\VBFMELA =
\left[1+
\frac{ \mathcal{P}_{\mathrm{\PX JJ}} (\vec\Omega^{\mathrm{\PX+JJ}} | \mZZ) }
{\mathcal{P}_{\mathrm{VBF}}  (\vec\Omega^{\mathrm{\PX+JJ}} | \mZZ)  }
\right]^{-1}
,
\end{eqnarray}
where $\mathcal{P}_{\mathrm{VBF}}$ and $\mathcal{P}_{\mathrm{\PX JJ}}$ are probabilities obtained from the
\textsc{JHUGen} matrix elements for the VBF and \ggF production processes
in association with two jets ($\PX+2$\,jets).
This discriminant is equally efficient in separating VBF from either $\Pg\Pg\to\PX+2$\,jets signal or $\Pg\Pg$
or $\qqbar\to 2\ell2\Pq+2$\,jets background because jet correlations in these processes are distinct from the VBF process. Being independent of the
type of fermions produced in the $\cPZ$ boson decay, it is used in both the
$\PX \to \cPZ\cPZ \to 4\ell$ and $\PX \to \cPZ\cPZ \to 2\ell2\cPq$ analyses.

In addition, in the $\PX \to \cPZ\cPZ \to 4\ell$ analysis, the dominant background originates from the
 $\qqbar\to\cPZ\cPZ$ / $\cPZ\gamma^*$ / $\gamma^*\gamma^* \to 4\ell$ process.
 Therefore, the discriminant sensitive to the $\PX \to \cPZ\cPZ \to 4\ell$ kinematic properties and
 optimal for suppression of the dominant background is defined as
\begin{eqnarray}
\label{eq:Dbkgmela}
{\cal D}_{\mathrm{bkg}}^{\mathrm{kin}} =
\left[1+  \frac{{\cal P}_{\qqbar\to 4\ell} (\vec\Omega^{\PX\to4\ell} | \mZZ)}
{{\cal P}_{\PX\to4\ell} (\vec\Omega^{\PX\to4\ell} | \mZZ)} \right]^{-1}\,.
\end{eqnarray}

In the $\PX \to \cPZ\cPZ \to 2\ell2\cPq$ analysis, the dominant background originates from the
 \cPZ\ + 2 jets process. Therefore, the discriminant sensitive to the $\PX\to \cPZ\cPZ\to 2\ell2\Pq$ kinematic properties is calculated as
\begin{eqnarray}
\label{eq:Zjjmela}
\ZJJMELA =
\left[1+  \frac{{\cal P}_\text{Zjj} (\vec\Omega^{\PX\to2\ell2q} | \mZZ)}
{{\cal P}_{\PX\to2\ell2\Pq} (\vec\Omega^{\PX\to2\ell2\Pq} | \mZZ)} \right]^{-1}\,.
\end{eqnarray}
In Eqs.~(\ref{eq:Dbkgmela}) and~(\ref{eq:Zjjmela}), ${\cal P}_{\PX\to4\ell}$ and ${\cal P}_{\PX\to2\ell2\Pq}$ are the probabilities for the signal, while ${\cal P}_{\qqbar\to 4\ell}$ and ${\cal P}_\text{Zjj}$ are the probabilities for the dominant background processes.

\section{Event selection and categorization}
\label{Section_selection}

The searches in the three final states cover different mass ranges. The $4\ell$ final state has the smallest backgrounds, so the search is performed over the full range from 130\GeV to 3\TeV. The $2\ell2\nu$ final state suffers from large $\cPZ + \text{jets}$ background in the low mass region, and the search range is thus restricted to be between 300\GeV and 3\TeV. For the same reason, the $2\ell2\Pq$ final state search is performed between 550\GeV and 3\TeV. Event selections are optimized for the search ranges in each final state.

Leptons are reconstructed as described in Section~\ref{sec:CMS}. Electrons are also required to pass
identification criteria based on
observables sensitive to the bremsstrahlung along the electron trajectory, the geometrical and
momentum energy matching between the
electron trajectory and the associated energy cluster in the ECAL, the shape of the electromagnetic shower in the ECAL,
and variables that discriminate against electrons originating from photon conversions.
Independent selection criteria on such observables are applied in the $2\ell2\nu$ channel,
while a multivariate discriminant based on them is adopted in the $4\ell$ and $2\ell2\Pq$
channel to retain high efficiency for low \pt leptons.
Muons are selected among the reconstructed muon track candidates by applying minimal requirements on the track in both
the muon and inner tracker system, and requiring small associated energy deposits in the calorimeters. For muon \pt above 200\GeV, the additional lever arm provided by the outer muon detectors becomes a significant advantage; therefore the charge and momentum are extracted from the combined trajectory fit for the outside in muons, while otherwise tracks found in the silicon tracker are used.

Electrons and muons with high \pt are required in the $2\ell2\Pq$ ($>$24\GeV) and $2\ell2\nu$ ($>$25\GeV) final states, while low \pt ($>$7\GeV for electrons and $>$5\GeV for muons) leptons are also retained in the $4\ell$ final state to ensure high efficiency for masses less than $2m_{Z}$. To suppress nonprompt leptons, the impact parameter in three dimensions of the lepton track, with respect to the primary vertex, is required to be less than 4 times its uncertainty ($\abs{\mathrm{SIP_{3D}}} < 4$).

In addition, an isolation requirement of ${\cal I}^{\ell}<0.35$ is imposed to select prompt leptons, where the isolation ${\cal I}^{\ell}$ is defined as
\ifthenelse{\boolean{cms@external}}{
\begin{equation}\begin{split}
\label{eqn:pfiso}
{\cal I}^{\ell} \equiv& \Big( \sum \PT^\text{charged} +
                                 \max\left[ 0, \sum \PT^\text{neutral}
                                 +\\
                                  &\sum \PT^{\Pgg}
                                 - \PT^\mathrm{PU}(\ell) \right] \Big)
                                 / \PT^{\ell}.
\end{split}
\end{equation}
}{
\begin{equation}
\label{eqn:pfiso}
{\cal I}^{\ell} \equiv \Big( \sum \PT^\text{charged} +
                                 \max\left[ 0, \sum \PT^\text{neutral}
                                 +
                                  \sum \PT^{\Pgg}
                                 - \PT^\mathrm{PU}(\ell) \right] \Big)
                                 / \PT^{\ell}.
\end{equation}
}
The three involved sums run over the \pt of charged hadrons originating
from the primary vertex, of neutral hadrons and of photons in a cone of angular
radius $\Delta R=0.3$ around the lepton direction.

Since the isolation variable is particularly sensitive to energy deposits from pileup interactions, a $\PT^\text{PU}(\ell)$ contribution is subtracted, using two different techniques.
For muons, we define $\PT^\mathrm{PU}(\Pgm) \equiv 0.5 \sum_i \PT^{\mathrm{PU}, i}$, where $i$ runs over the momenta of the charged hadron PF candidates not originating from the primary vertex, and the factor of 0.5 accounts for the fraction of neutral particles. For electrons, an area based subtraction technique~\cite{Cacciari:2007fd,Cacciari:2008gn,Cacciari:2011ma}, as implemented in \FASTJET, is used, in which $\PT^\mathrm{PU}(\Pe) \equiv \rho A_\text{eff}$, where the effective area $ A_\text{eff}$ is the geometric area of the isolation cone scaled by a factor that accounts for the residual dependence of the average pileup as a function of $\eta$, and $\rho$ is the median of the energy density distribution of neutral particles within the area of any jet in the event.

In the $4\ell$ and $2\ell 2\Pq$ final states, an algorithm is used to recover the final state radiation (FSR) from leptons.
Photons reconstructed by the PF algorithm within $\abs{\eta_\Pgg}< 2.4$ are considered as FSR candidates if they satisfy
$\PT^{\Pgg} > 2\GeV$ and ${\cal I}^{\ell}< 1.8$~\cite{Sirunyan:2017exp}.
Associating every such photon to the closest selected lepton in the event, photons that do not satisfy
$\Delta R(\cPgg,\ell)/(\PT^{\Pgg})^{2}<0.012$ and $\Delta R(\cPgg,\ell)<0.5$ are discarded.
The lowest $\Delta R(\cPgg,\ell)/(\PT^{\Pgg})^{2}$ photon candidate for every lepton, if any, is retained.
The photons identified as FSR are excluded from any isolation computations.

The momentum scale and resolution for electrons and muons are calibrated in bins of $\pt^\ell$ and $\eta^\ell$
using the decay products of known dilepton resonances.
The electron momentum scale in data is corrected with a $\cPZ \to \Pe\Pe$ sample, by adjusting the peak of the reconstructed
dielectron mass spectrum to that expected from simulation.
A Gaussian smearing is applied
to electron energies in simulation such that the $\cPZ \to \Pe\Pe$ mass resolution agrees with the one observed in data.
Muon momenta are calibrated based on a Kalman filter approach~\cite{Fruhwirth:1987fm}, using $\PJGy$ meson and $\cPZ$ boson decays.

A ``tag-and-probe'' technique~\cite{CMS:2011aa} based on inclusive samples of $\cPZ$ boson events in data
is used to correct the efficiency of the reconstruction and selection
for prompt electrons and muons in several bins of $\PT^\ell$ and $\eta^\ell$.
The difference in the efficiencies measured in simulation and data is used to correct the selection
efficiency in the simulated samples.

The jets in the three analyses must satisfy $\pt^{\text{jet}}>30\GeV$
and $\abs{\eta^{\text{jet}}}<4.7$ and be separated from all selected leptons by $\Delta R(\ell/\cPgg,\text{jet})>0.4$. The analyses
use \cPqb\ tagged jets of $\abs{\eta^{\text{jet}}}<2.5$ for event categorization and selection, where a \cPqb\ jet is tagged using the combined secondary vertex algorithm ~\cite{Chatrchyan:2012jua,CMS-PAS-BTV-15-001}
based on the impact parameter significance of the tracks associated with the jet, with respect to the primary vertex. The loose working point is used, corresponding to an efficiency of 80\% and a mistag rate of 10\% for light quark jets.

The main feature distinguishing the two dominant $\PX$ boson production mechanisms (\ggF and VBF) is the presence of associated jets and the kinematic
correlation between such jets and the $\PX$ boson. In order to gain sensitivity to the production process of the $\PX$ boson,
events are split into categories based on such kinematic correlations. In the case of fully reconstructed final states,
$\PX\to 4\ell$ and $2\ell2$q, a ME technique is used to categorize events based on the correlation between the two forward jets and
the $\PX$ boson candidate, while in the $2\ell2\nu$ final state a simpler correlation between the two jets is used.

Subsequent event selections differ depending on the considered final state and are described for each final state in the following.

\subsection{\texorpdfstring{$\PX\to\cPZ\cPZ\to4\ell$}{to4l}}
\label{sec:XZZ4l}

The $\PX\to \cPZ\cPZ\to 4\ell$ analysis uses the same selection as in the measurements
of the properties of the $\PH (125)$ boson in the $\PH\to\cPZ\cPZ\to4\ell$ decay channel~\cite{Sirunyan:2017exp}.
The $\cPZ$ candidates are formed from pairs of leptons of the same flavor and opposite charge
($\Pe^{+}\Pe^{-}$, $\Pgm^{+}\Pgm^{-}$) and are required to pass the invariant mass selection $12 < m_{\ell^+\ell^-} < 120\GeV$. The flavors of involved leptons define three mutually exclusive channels: $4\Pe$, $4\Pgm$, and $2\Pe 2\Pgm$.
\cPZ\ candidates are combined into $\cPZ\cPZ$ candidates, wherein we denote as $\cPZ_1$ the $\cPZ$ candidate
with an invariant mass closest to the nominal $\cPZ$ boson mass~\cite{Olive:2016xmw}, and the other \cPZ\ candidate $\cPZ_2$.
To be considered for the analysis, $\cPZ\cPZ$ candidates have to pass a set of kinematic requirements.
The $\cPZ_1$ invariant mass is required to be larger than $40\GeV$.
All leptons are separated in angular space by at least $\Delta R(\ell_i, \ell_j) > 0.02$.
At least two leptons are required to have $\pt > 10\GeV$ and at least one is required to have $\pt > 20\GeV$.
In the $4\Pgm$ and $4\Pe$ channels, where an alternative $\cPZ_a \cPZ_b$ candidate can be built out of the
same four leptons, candidates with $m_{\cPZ_b}<12\GeV$ are removed if $\cPZ_a$ is closer to the nominal
$\cPZ$ boson mass than $\cPZ_1$ is.

In Ref.~\cite{Sirunyan:2017exp}, six categories are defined based on the number and types of particles
associated with the $\PH(125)$ boson. Here we follow the same approach with some optimization specific for a high mass search.
Two categories dedicated to the production mechanisms are used: VBF jets and inclusive; to further improve the efficiency
in the electron channels at high \pt, a relaxed selection electron (RSE) category is added.
The $\abs{\mathrm{SIP}_{\mathrm 3D}} < 4$ requirement in the standard electron selection removes fake electrons from photon conversions, which are not dominant at high masses. The requirement becomes the main cause of efficiency losses at high \pt.
The second cause of the efficiency loss, particularly at high masses, is the opposite sign lepton charge requirement,
as the charge misidentification rate increases with lepton \pt. Thus, a relaxed selection
removing both requirements on at most one pair of electrons is applied for $m_{4\ell} > 300$\GeV. The detailed categorization is structured as follows:
\begin{itemize}
	\item  \textbf{VBF-tagged} requires exactly four leptons selected with regular criteria. In addition, there must be either two or three jets among which at most one is \cPqb\ tagged, or at least four jets and no \cPqb\ tagged jets, and ${\cal D}^\mathrm{VBF}_{\mathrm{2jet}}$ following Eq.~(\ref{eq:vbfmela}) is required to pass a mass dependent selection;
\item  \textbf{Untagged} consists of the remaining events with regularly selected leptons;
\item  \textbf{RSE} contains events from the relaxed electron selection that are not in the regular electron selection and for which $m_{4\ell} > 300\GeV$.
\end{itemize}
When more than two jets pass the selection criteria, which happens in about half of the cases, the two \pt-leading jets are selected for matrix element calculations.

\begin{figure}[htbp]
\centering
\includegraphics[width=0.45\textwidth]{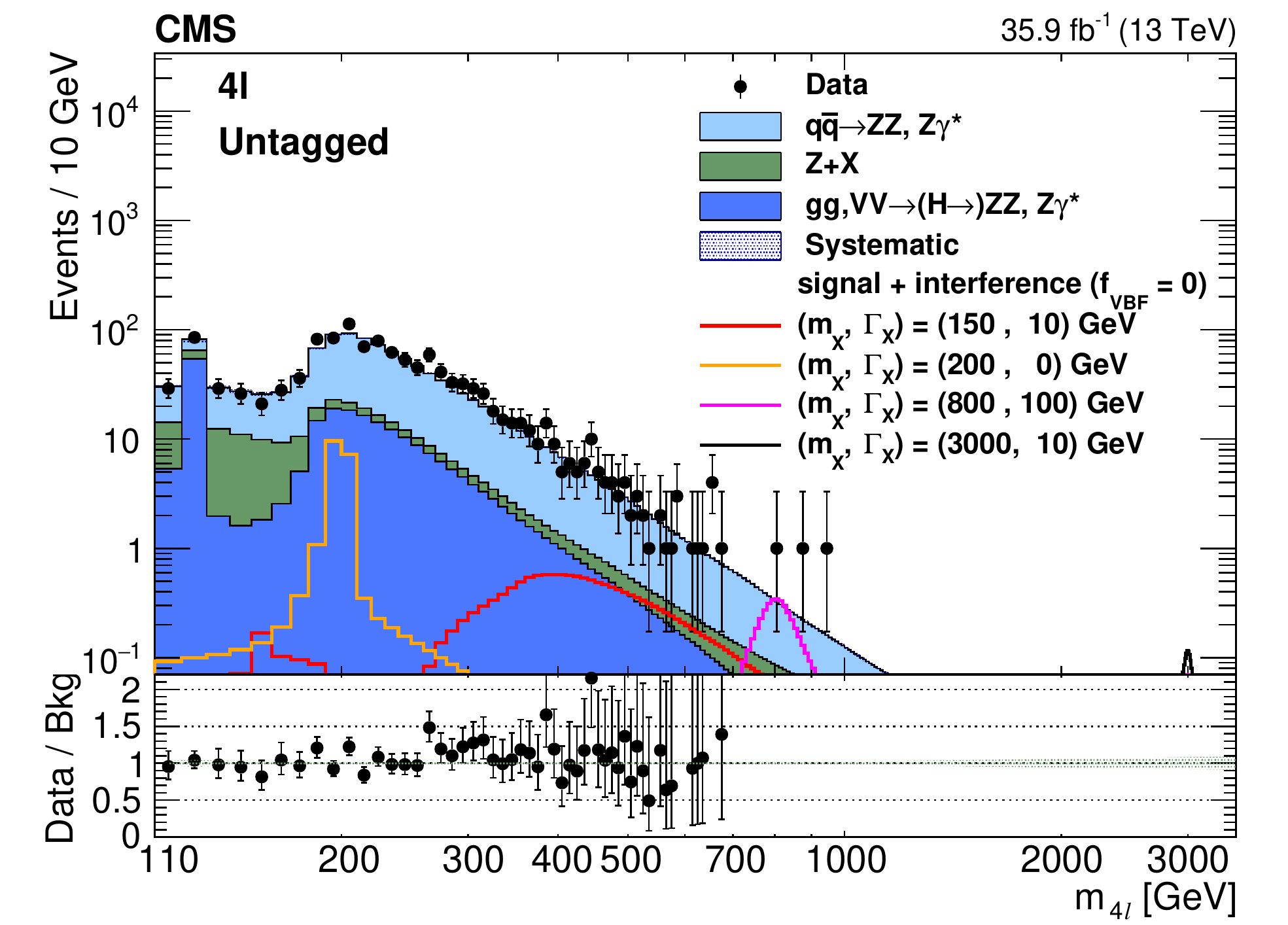}
\includegraphics[width=0.45\textwidth]{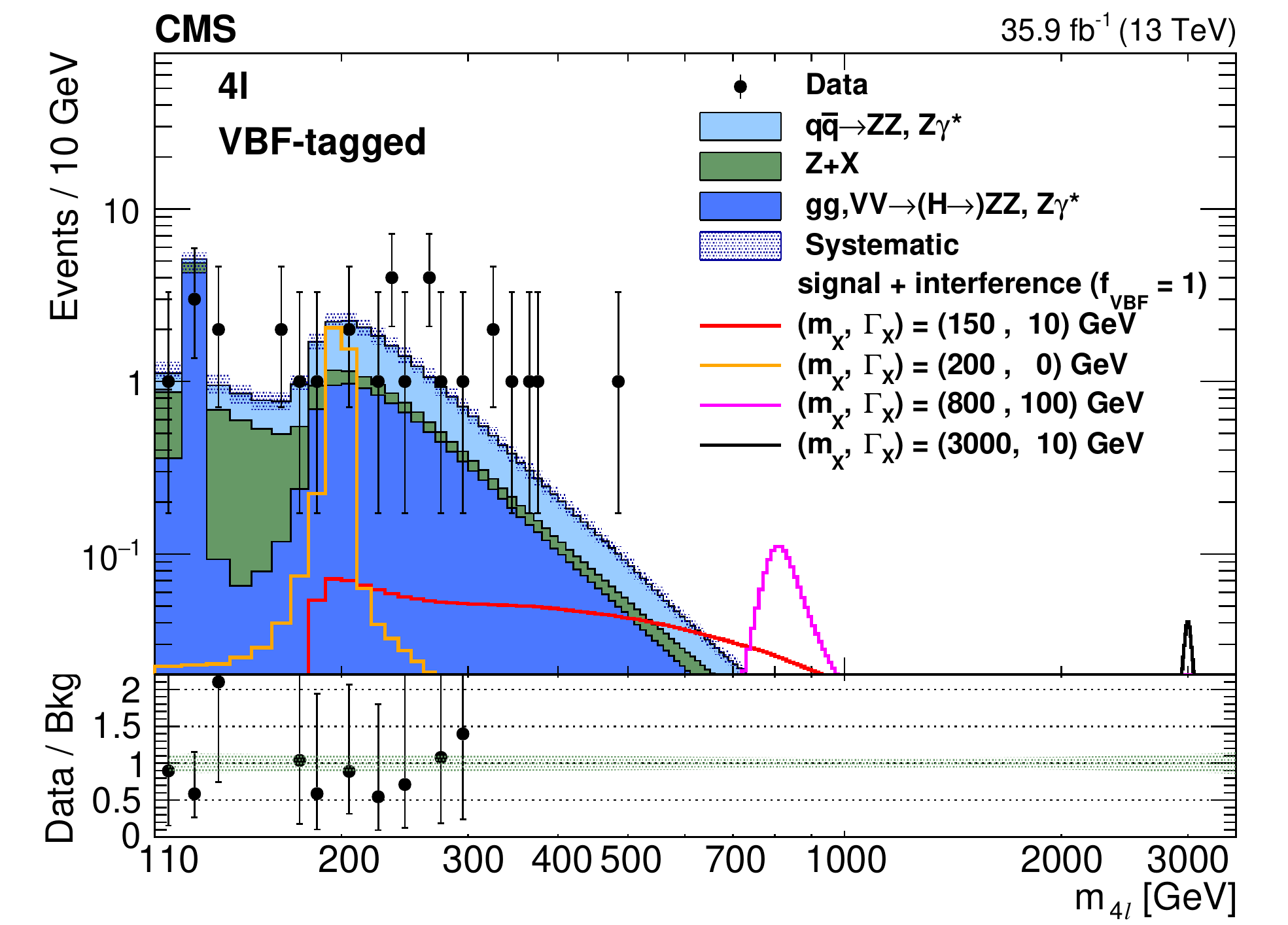}
\includegraphics[width=0.45\textwidth]{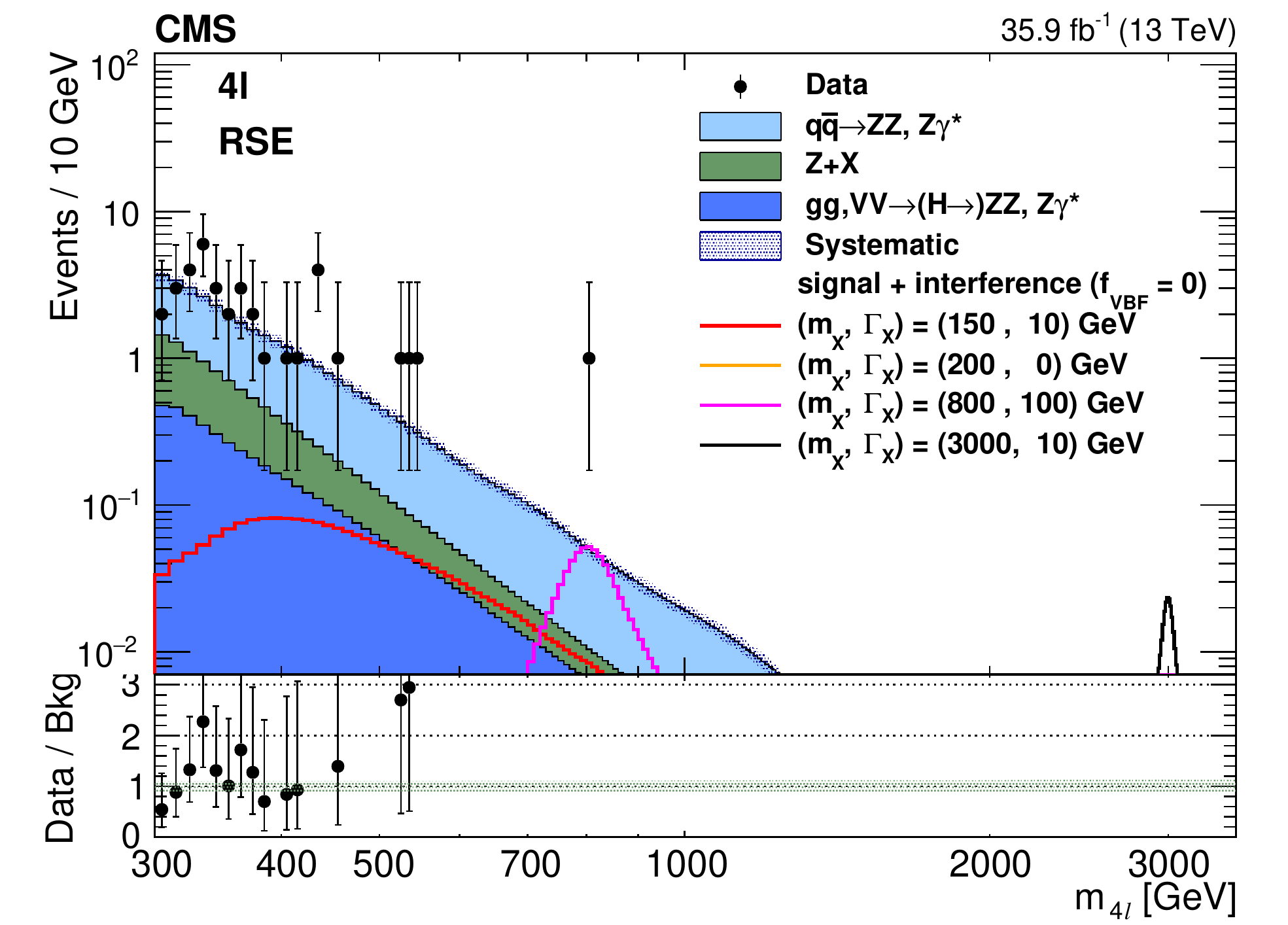}
\caption{
	Distributions of the four lepton invariant mass in the untagged (upper left plot), VBF-tagged (upper right plot) and RSE (lower plot) categories. Signal expectations including the interference effect for several mass and width hypotheses are shown. The signals are normalized to the expected upper limit of the cross section derived from this final state. Lower panels show the ratio between data and background estimation in each case.\label{fig:m4l}
}
\end{figure}

\begin{figure}[htbp]
\centering
\includegraphics[width=0.45\textwidth]{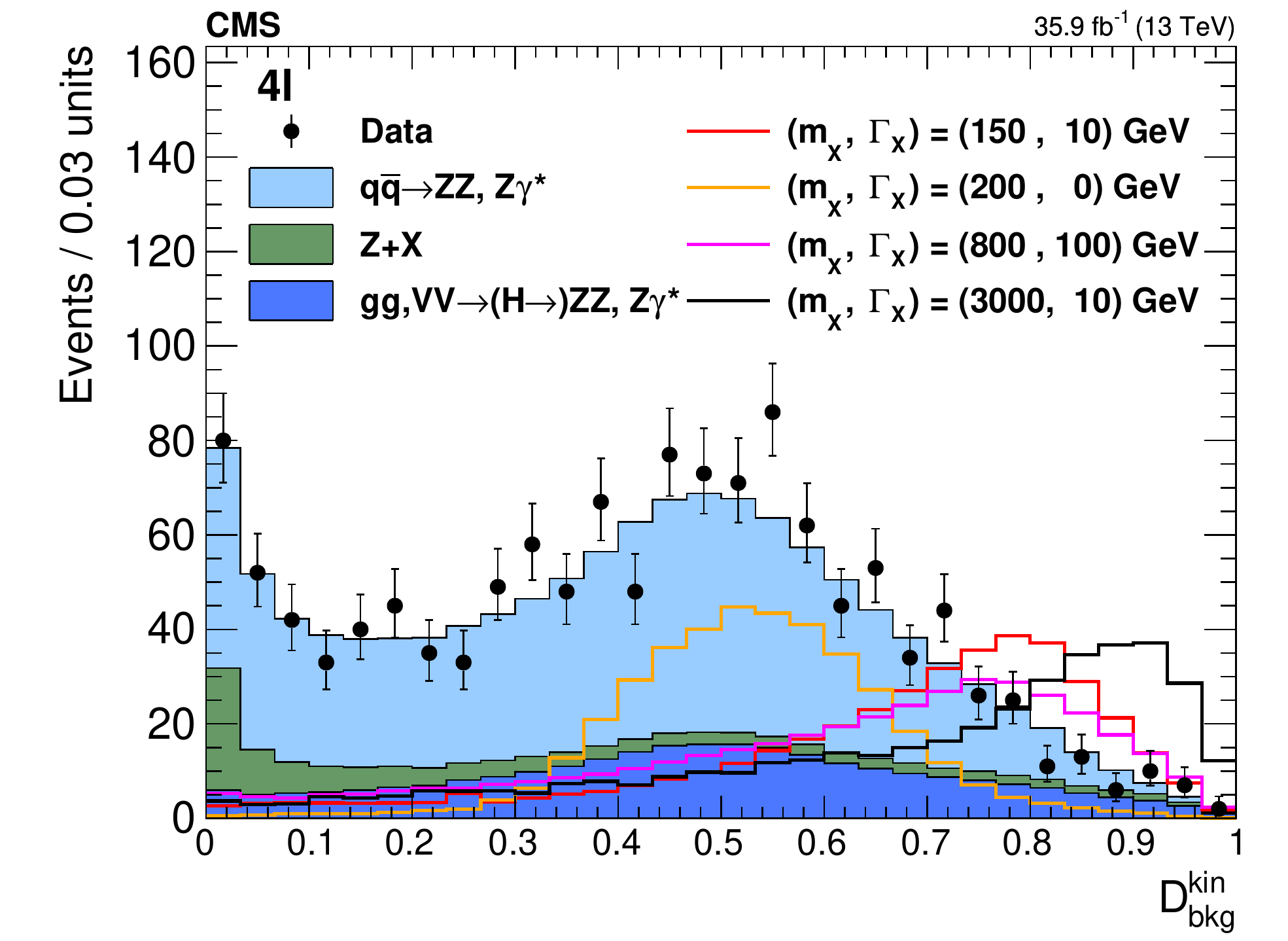}
\caption{
	Distributions of ${\cal D}_{\mathrm{bkg}}^{\mathrm{kin}}$ for all selected events. Signal expectations including the interference effect for several mass and width hypotheses are shown. The signals are normalized to a total of 400 events. \label{fig:kd_4l}
}
\end{figure}

As a result of the above categorization, events are split into eight categories:
{$4\Pe$},  {$4\mu$},  {$2\Pe2\mu$}, in either the {VBF-tagged} or the {untagged} category, or {$4\Pe$} and {$2\Pe2\mu$} in the RSE category. Each event is characterized by two observables $(m_{4\ell}$ and ${\cal D}_{\mathrm{bkg}}^{\mathrm{kin}})$ that are shown in Fig.~\ref{fig:m4l} and Fig.~\ref{fig:kd_4l}, together with several signal hypotheses.

\subsection{\texorpdfstring{$\PX\to\cPZ\cPZ\to2\ell2\Pq$}{to2l2q}}
\label{sec:XZZ2l2q}

In the $\PX\to \cPZ\cPZ\to 2\ell2\Pq$ analysis, events are selected by combining leptonically
and hadronically decaying \cPZ\ candidates. The lepton pair selection is similar to the four-lepton analysis:
pairs of opposite sign and same flavor electrons or muons
with invariant mass between \ZLEPLOW{} and \unit{\ZLEPHIGH}{\GeV} are constructed.
A $\PT > \unit{\PTLEPHIGH}{\GeV}$ requirement is applied on at least one
of the leptons in the pair, and a minimum dilepton \PT{} of \unit{\DILEPTONPTCUT}{\GeV} is imposed to reject Drell--Yan events with small hadronic recoil.

Hadronically decaying $\cPZ$ boson candidates (\Zhad) are reconstructed using two distinct
techniques, which are referred to as ``resolved'' and ``merged'' in the following.
In the resolved case, the two quarks from the $\cPZ$ boson decay form two distinguishable AK4 jets, while in the merged case a single AK8 jet with a large \pt is taken as a \Zhad.

In the merged jet case, a pruning algorithm is applied to the AK8 jet~\cite{prune,substructure}.
The goal of the algorithm is to recluster the jet constituents, while applying
additional requirements that eliminate soft, large angle QCD radiation that
artificially increases the jet mass relative to the nominal $\cPZ$ boson mass.
We adopt the unified nomenclature \MZhad\ to refer to the hadronically decaying \cPZ\ candidate mass, corresponding to the dijet invariant mass in the resolved case and the jet pruned mass in the merged case. The reconstructed \Zhad\ is required to have an invariant mass around the $\cPZ$ boson mass: $\LSBLOW < \MZhad < \unit{\USBHIGH}{\GeV}$
and $\PT > \unit{\VHADRESOLVEDPTCUT}\ (\VHADMERGEDPTCUT) {\GeV}$ in the resolved (merged) case. Merged jets must also be separated from all selected leptons by $\Delta R(\ell,\text{jet})>0.8$.
In addition, in the merged jet selection we exploit substructure techniques
commonly used in searches including Lorentz boosted bosons in the final state~\cite{subjettiness}.
The $N$-subjettiness $\tau_N$ is defined as
\begin{equation}
\tau_N = \frac{1}{d_0}
\sum_k
\PT{}_{,k}
\min(
\Delta R_{1,k},\Delta R_{2,k},\,\ldots,\Delta R_{N,k}
),
\end{equation}
where the index $k$ runs over the jet constituents and the distances $\Delta R_{N,k}$ are calculated with respect to the axis of the $n^\textrm{th}$ subjet.
The normalization factor $d_0$ is calculated as $d_0 = \sum_k \PT{}_{,k}R_0$, setting $R_0$ to the jet radius of the original jet.
Jets with smaller $\tau_N$ are more compatible with the $N$-subjets configuration.
We use the ratio of 2-subjettiness over 1-subjettiness, $\nsubj = \tau_2/\tau_1$, as the discriminating variable for the jet substructure
and impose a $\nsubj < \TAUCUT$ requirement on merged \Zhad\ candidates.

Events that pass the above selection and additionally have \MZhad\ in the range
[\SRLOW, \SRHIGH]\GeV
form the signal region, covering 1--2 standard deviations dijet mass resolution.
On the other hand, events that have \MZhad\ in the range
[\LSBLOW, \LSBHIGH]\GeV or [\USBLOW, \USBHIGH]\GeV
form the sideband regions and are retained for background estimation.

An arbitration procedure is used to rank multiple \Zhad\ candidates reconstructed in a single event:
merged candidates have precedence over resolved candidates if
they have $\PT > \unit{\VHADPTSWITCH}{\GeV}$ and the accompanying leptonically decaying \cPZ\ candidate has $\PT(\LL) > \unit{\DILEPTONPTSWITCH}{\GeV}$;
resolved candidates have precedence otherwise. Within each selection category the candidate with the largest \PT{} has priority over the others.

The hadronically and leptonically decaying $\cPZ$ boson candidates are combined to form a resonance candidate. In order to improve the $\cPZ\cPZ$ invariant mass resolution in the resolved jet case, a kinematic fit is performed using a mass constraint on the
intermediate decay \Ztoqq{}. The constraint improves the signal resolution by 7--10\%.
When a candidate belongs to the signal region, we reevaluate the kinematical distributions of final state particles (here the \PT{} of the two jets
forming the $\cPZ$ boson of the resonance candidate) with a constraint on the reconstructed $\cPZ$ boson mass to follow the $\cPZ$ boson line shape.
For each event, the likelihood is maximized and the \PT{} of the jets is updated.
After refit, the mass of the $\cPZ$ boson candidate and \mZZ{} are recalculated. This procedure is not applied to events in the sidebands, where \MZhad\ is
very different from the nominal $\cPZ$ boson mass.

The reconstructed $\cPZ\cPZ$ candidate mass \mZZ{} denotes the dilepton + dijet mass \mlljj{} in the resolved case and the dilepton + merged jet invariant mass \mllJ{} in the merged case.
A requirement of $\mZZ{}> \unit{\MZZCUT}{\GeV}$ is imposed to reduce the $\cPZ + \text{jets}$ background.

To increase the sensitivity to the different production modes, events are categorized into VBF and inclusive types. Furthermore, since a large fraction of signal events is enriched with \cPqb\ quark jets due to the presence of $\cPZ\to\bbbar$ decays, a dedicated category is defined. The definitions are as follows:
\begin{itemize}
	\item  \textbf{VBF-tagged} requires two additional and forward jets besides those constituting the hadronic $\cPZ$ boson candidate; a mass dependent selection criterion on ${\cal D}^\mathrm{VBF}_{\mathrm{2jet}}$ is applied;
	\item  \textbf{\cPqb\ tagged} consists of the remaining events with two \cPqb\ tagged jets (in the resolved case) or two \cPqb\ tagged subjets from the hadronic $\cPZ$ boson candidate;
	\item  \textbf{Untagged} consists of the remaining events.
\end{itemize}

\begin{figure}[htbp]
   \centering
   \includegraphics[width=0.45\textwidth]{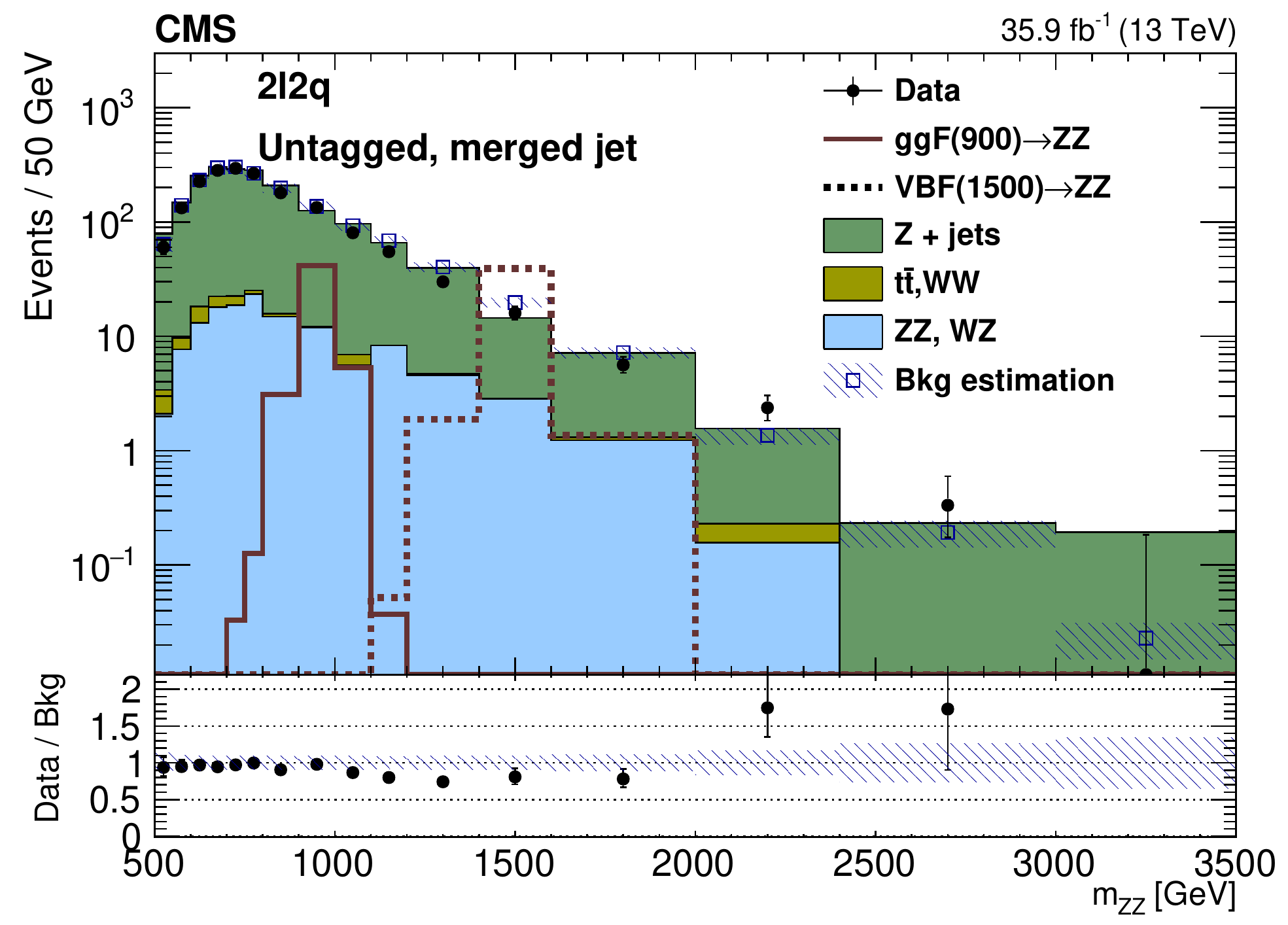}
   \includegraphics[width=0.45\textwidth]{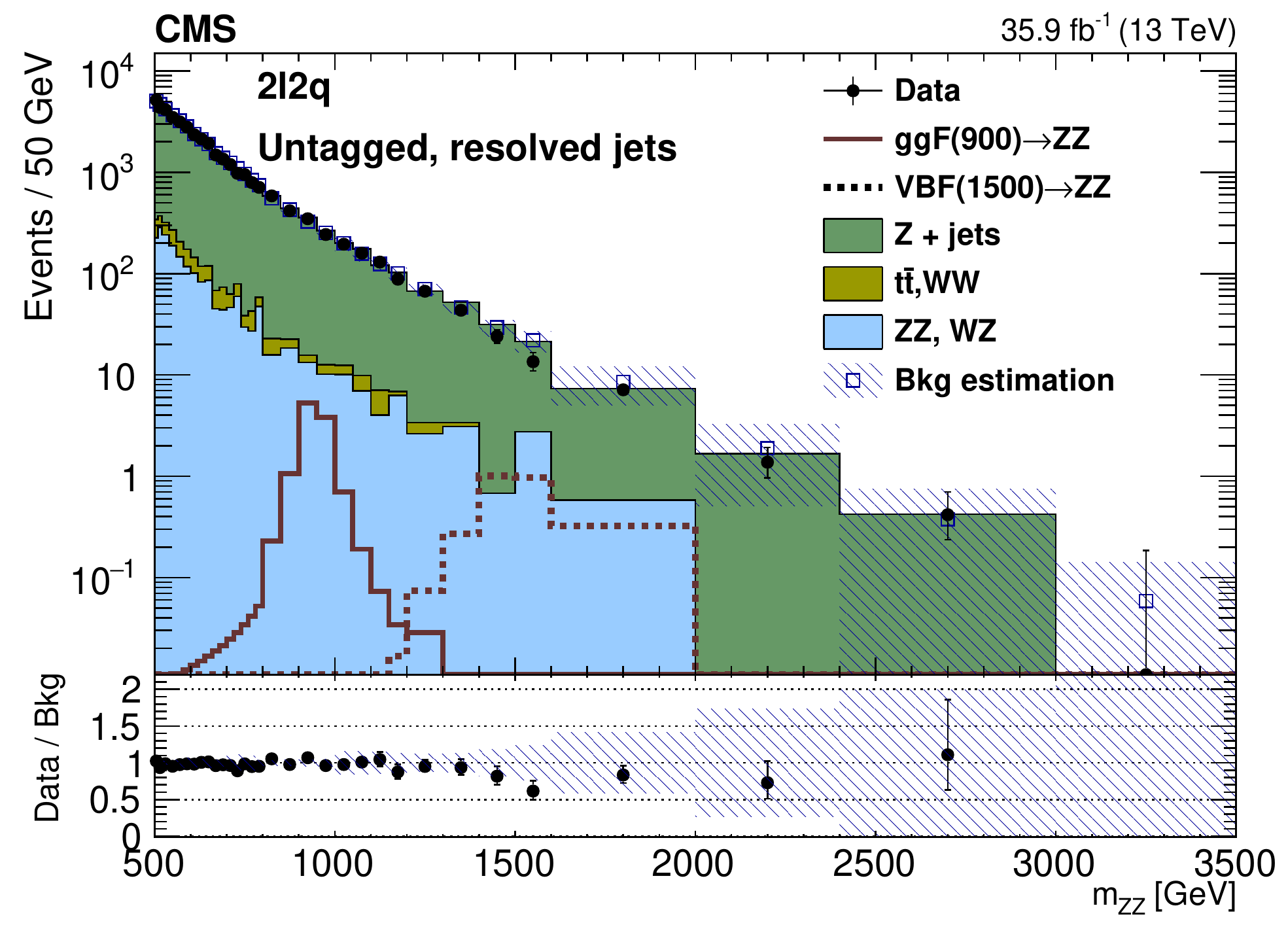}\\
   \includegraphics[width=0.45\textwidth]{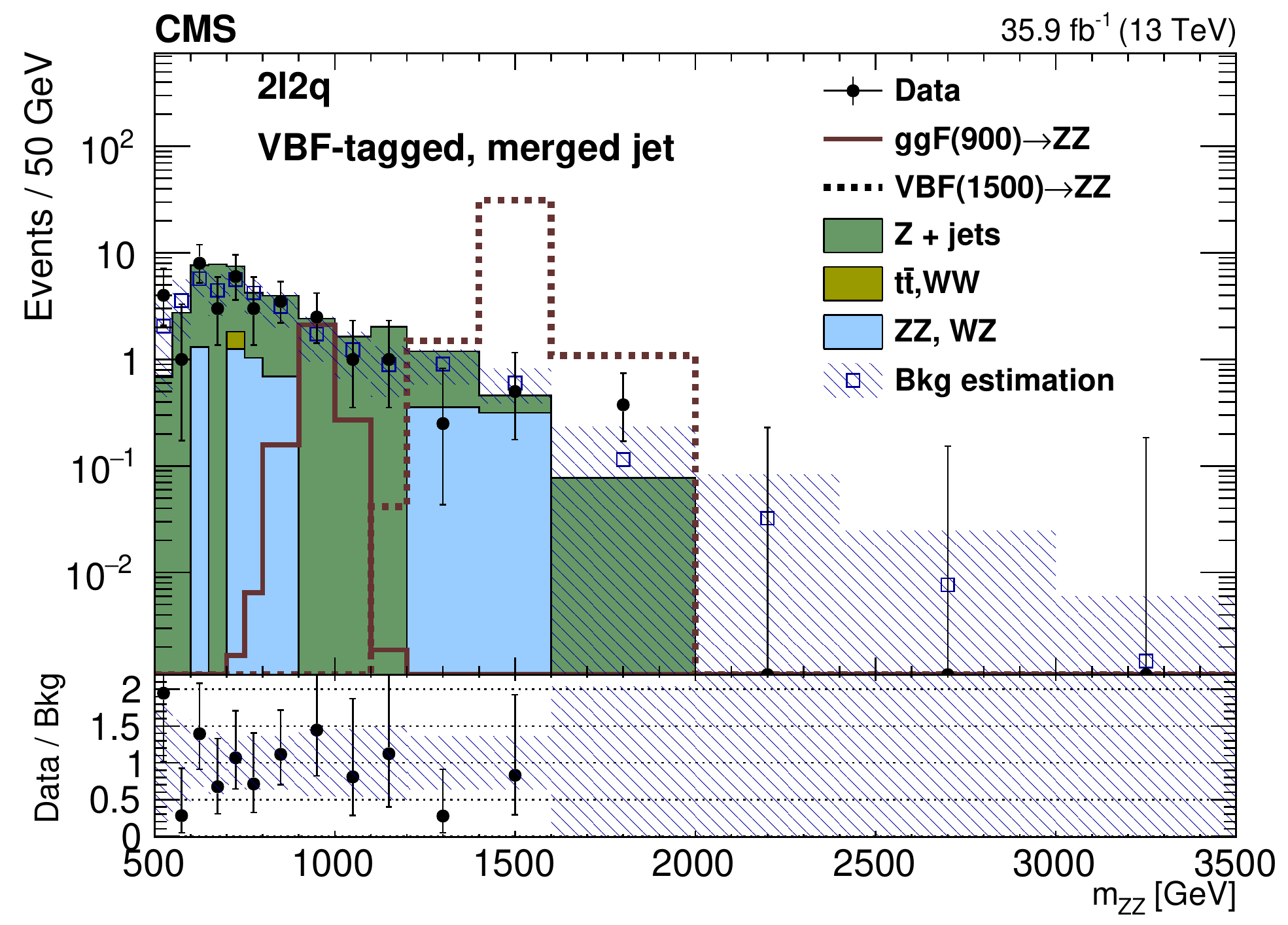}
   \includegraphics[width=0.45\textwidth]{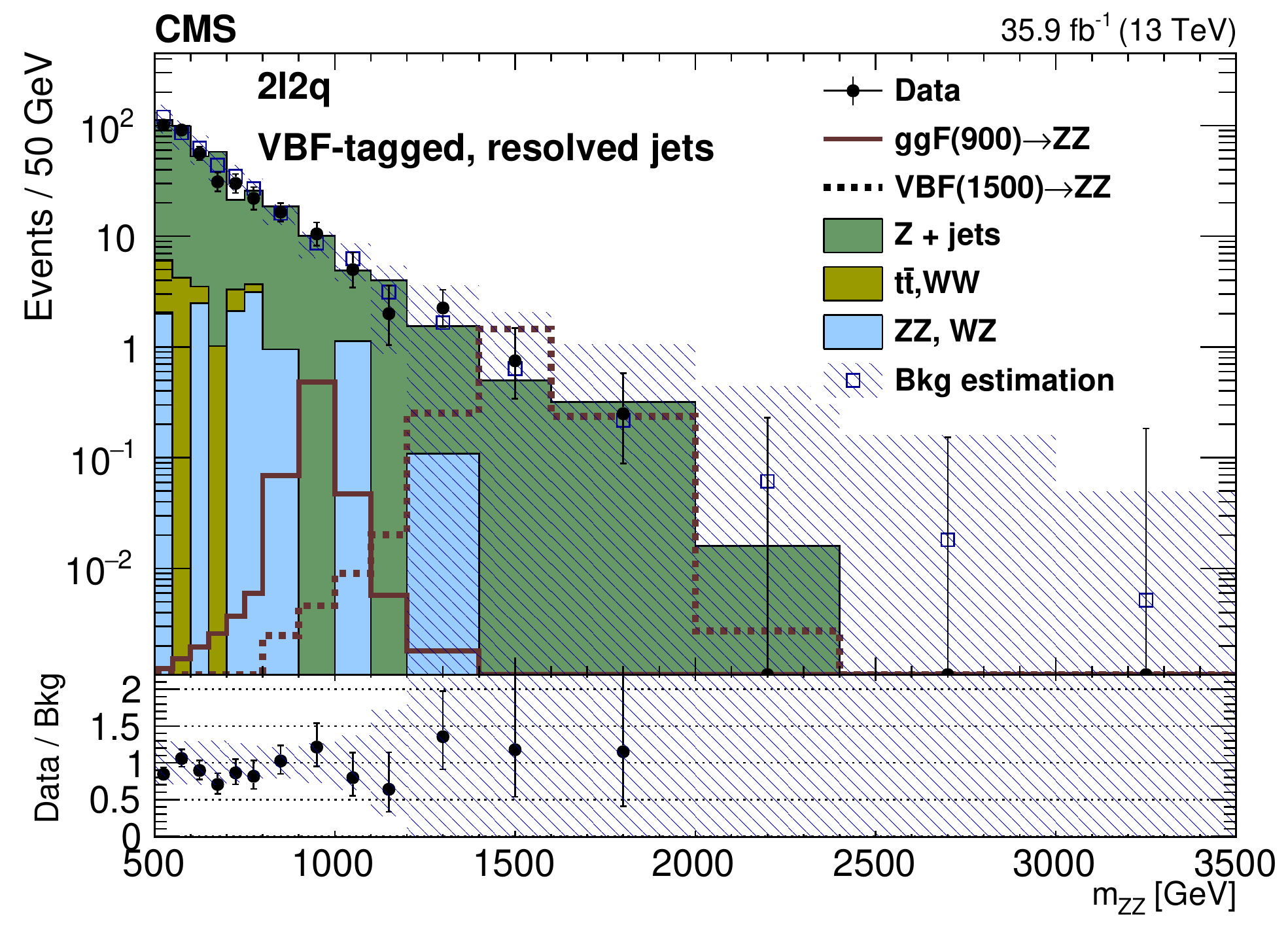}
   \includegraphics[width=0.45\textwidth]{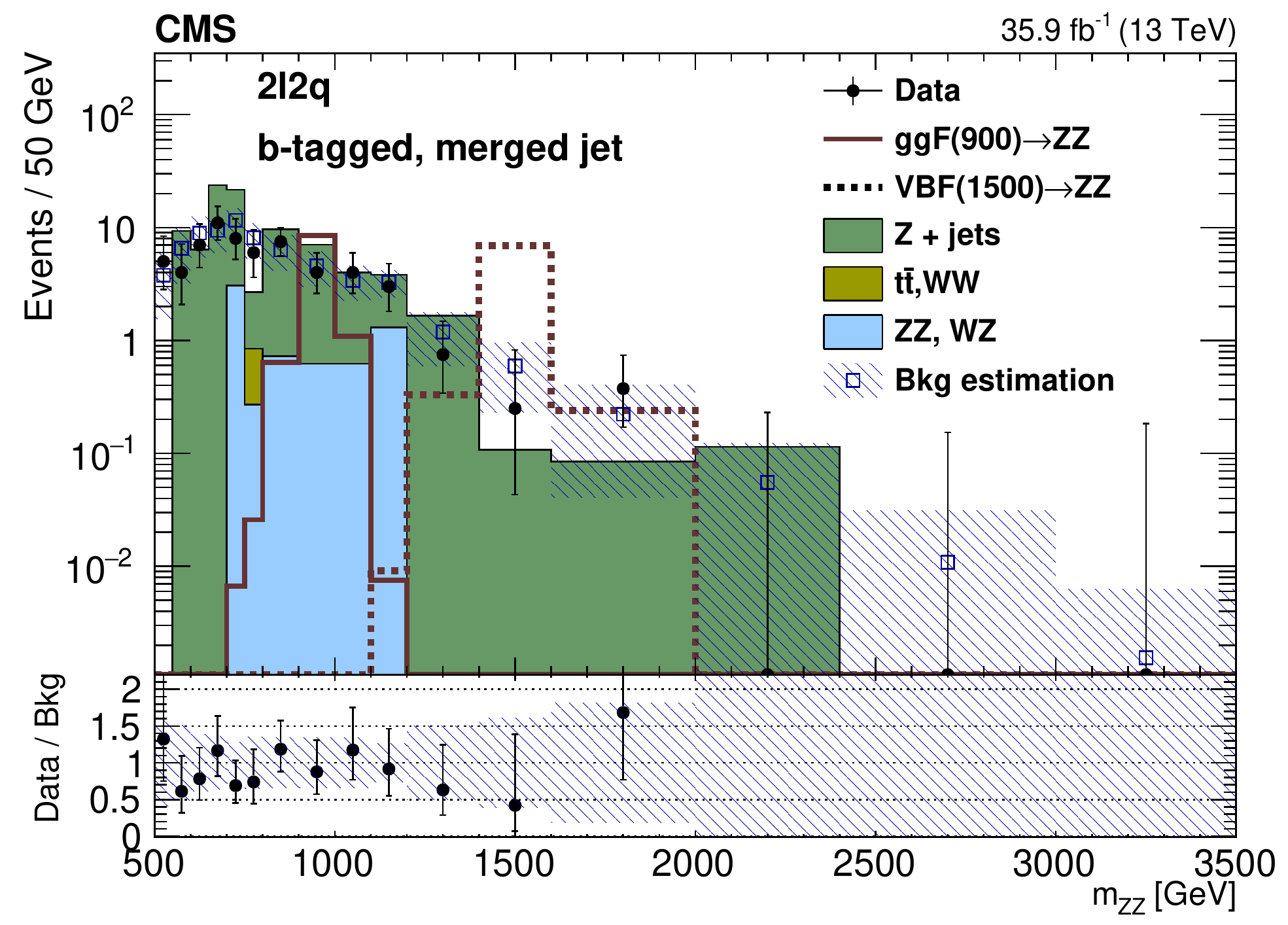}
   \includegraphics[width=0.45\textwidth]{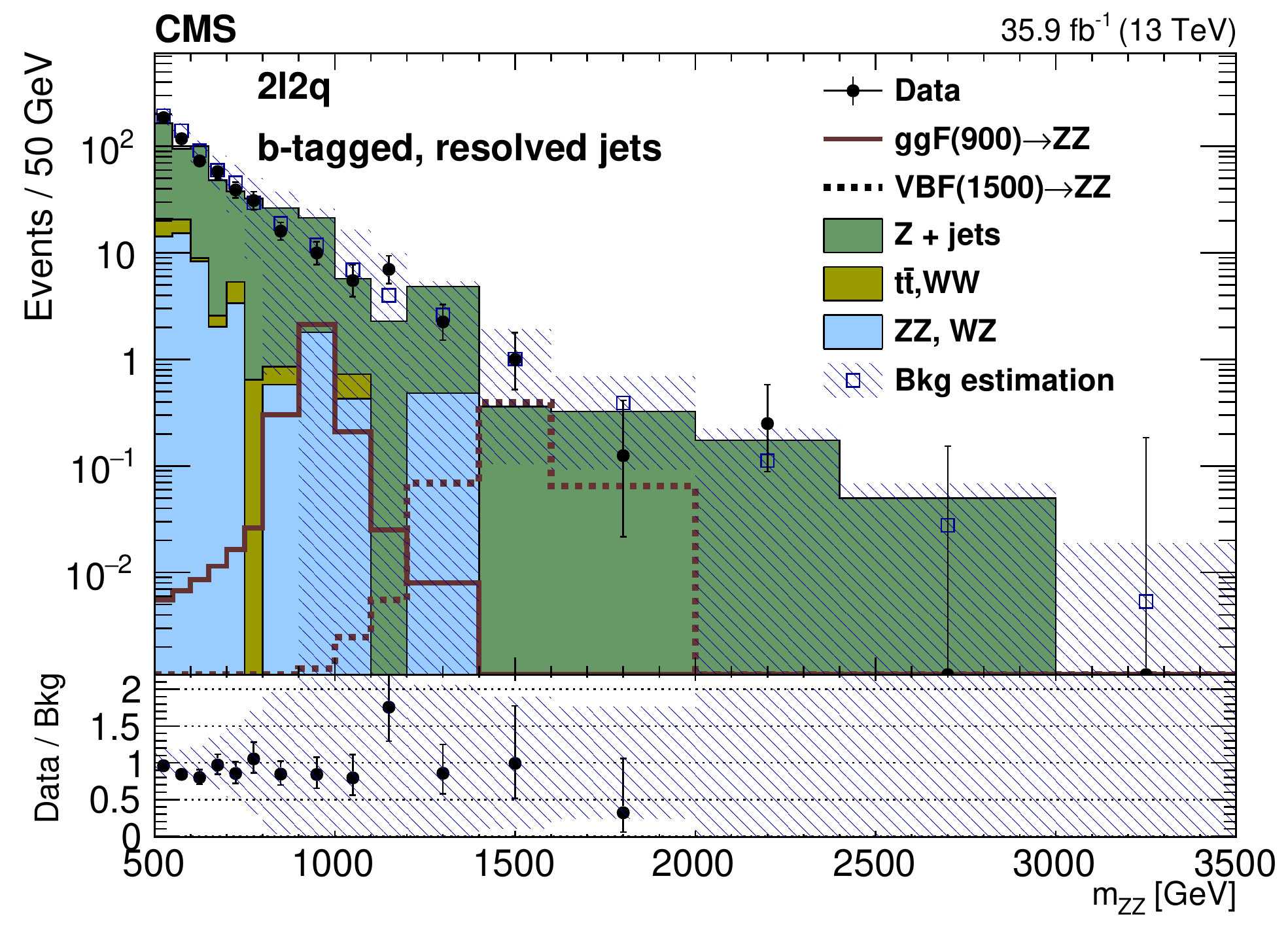}\\
   \caption{Distributions of the invariant mass \mZZ{} in the signal region for the merged (left) and resolved (right) case for the different categories in the $2{\ell}2\Pq$ channel. The points represent the data, the stacked histograms the expected backgrounds from simulation, and the open histograms the expected signal. The blue hatched bands refer to the sum of background estimates derived from either simulation or control samples in data, as described in the text. Lower panels show the ratio between data and background estimation in each case.}
   \label{fig:ZZmass_untag}
\end{figure}

\begin{figure}[htbp]
   \centering
   \includegraphics[width=0.45\textwidth]{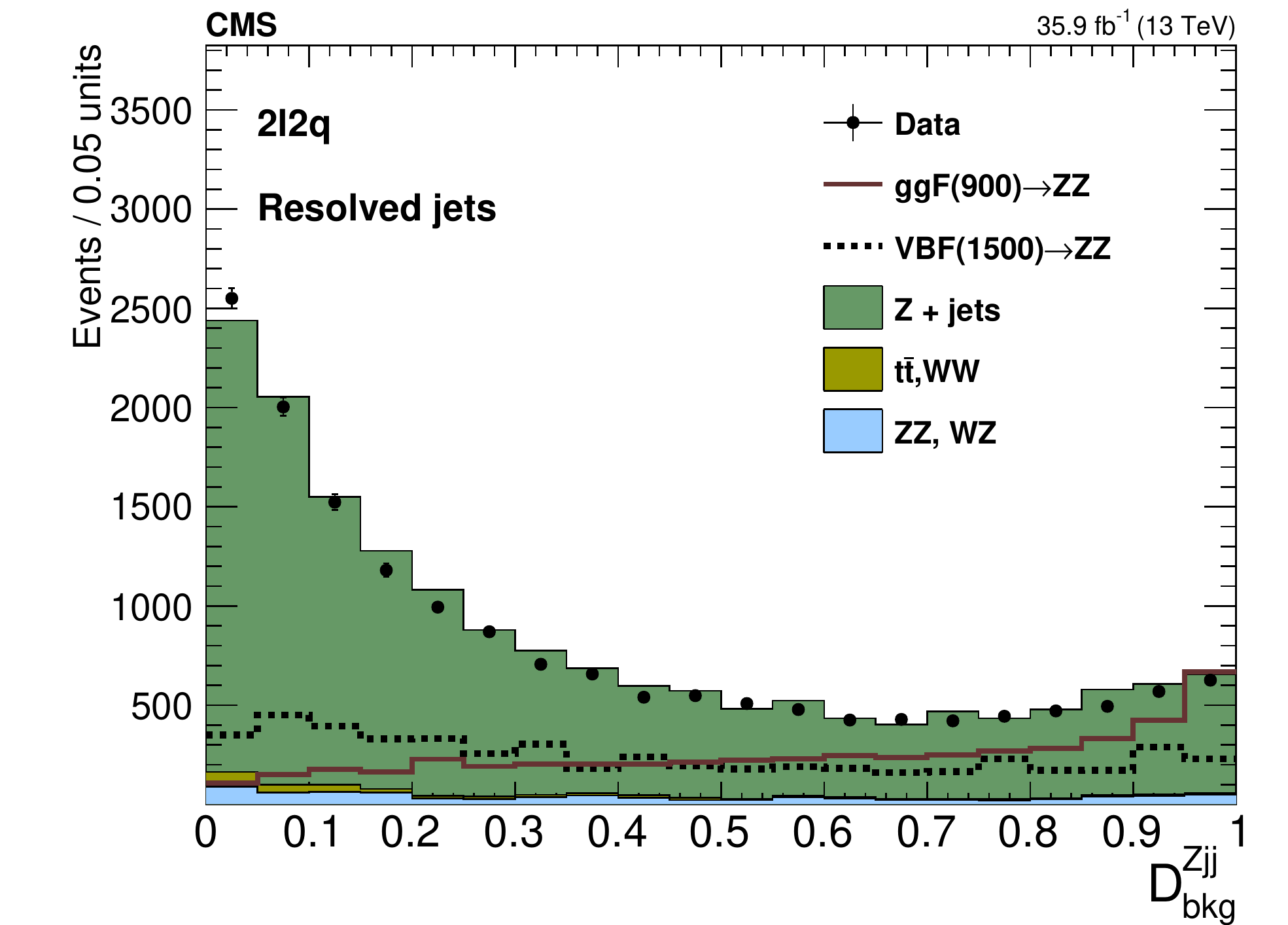}
   \includegraphics[width=0.45\textwidth]{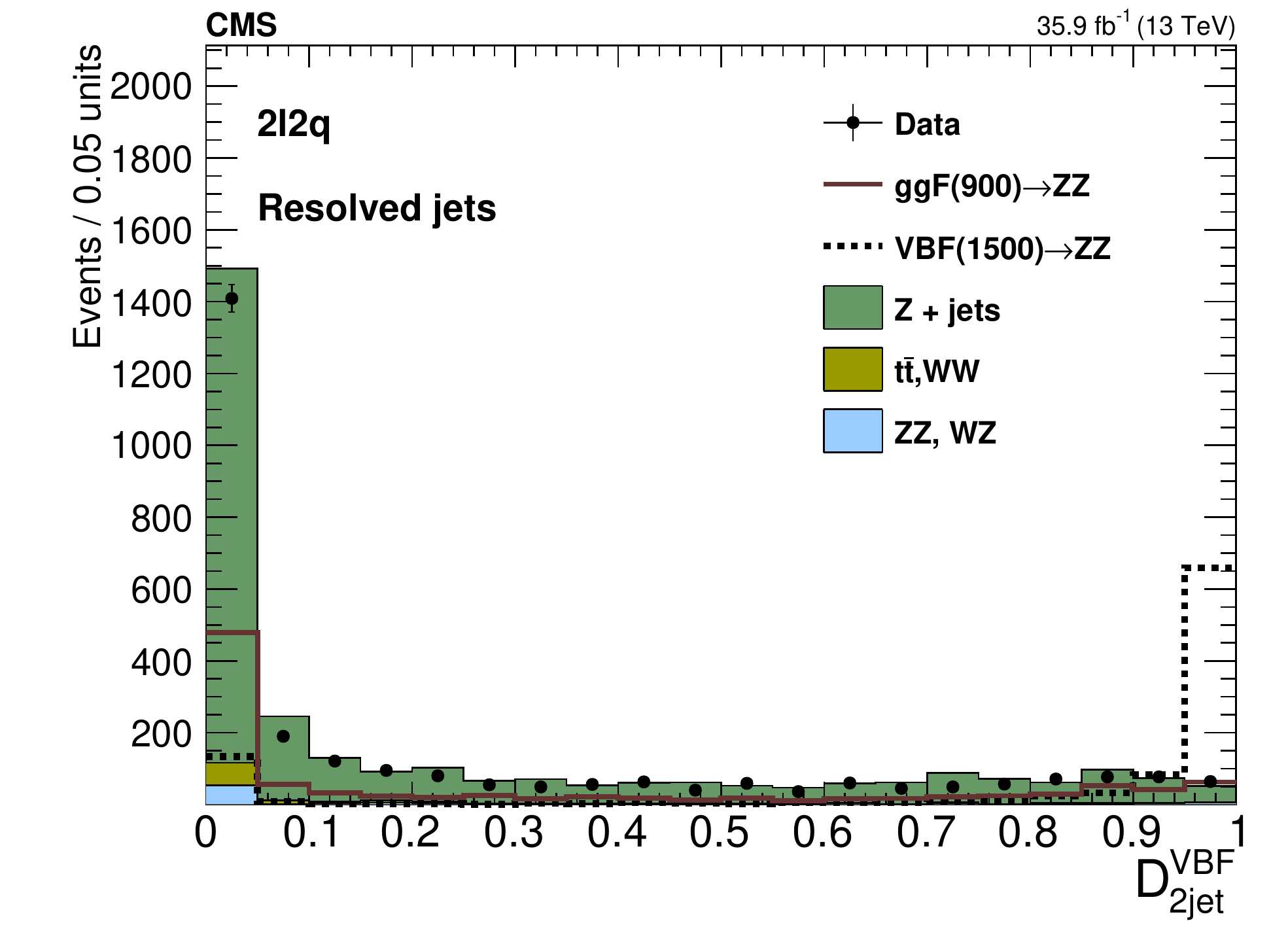}\\
   \caption{Distributions of the \ZJJMELA (left) and \VBFMELA (right) discriminants in the signal region for the resolved selection. The points represent the data, the stacked histograms the expected background from simulation, and the open histograms the expected signal.}
   \label{fig:ZZmela_untag}
\end{figure}

As a result of this categorization, events are split into twelve categories:
 {$2\Pe 2\Pq$} or  {$2\mu 2\Pq$}, either {VBF-tagged},  {{b}-tagged}, or {untagged}, and each with either {merged jets} or {resolved jets}.
Each event is characterized by the two observables $(\mZZ, \ZJJMELA)$. Figure~\ref{fig:ZZmass_untag} shows the invariant mass distribution for merged and resolved events in each category after the selection. Figure~\ref{fig:ZZmela_untag} shows the \ZJJMELA and \VBFMELA distributions for resolved events in each category together after the selection.

\subsection{\texorpdfstring{$\PX\to\cPZ\cPZ\to2\ell2\nu$}{to2l2nu}}
\label{sec:XZZ2l2nu}

In the $\PX\to \cPZ\cPZ\to 2\ell2\nu$ channel, events are selected by combining dilepton $\cPZ$ boson candidates with relatively large \ptmiss.
Events are selected requiring two leptons of the same flavor
that have an
invariant mass within a 30\GeV window centered on the nominal $\cPZ$
boson mass.
For X boson masses considered in this analysis ($>$300\GeV),
the $\cPZ$ bosons from the X boson decay are typically produced
with a large $\PT$.
To suppress the bulk of
the $\cPZ + \text{jets}$ background, the $\PT$ of the dilepton system is therefore required to be greater than
55\GeV, and a \ptmiss threshold of 125\GeV is imposed.
The region of large \ptmiss is contaminated by $\cPZ + \text{jets}$ events in which the \ptmiss
is largely due to mismeasurements of the jet energies.
To suppress this contribution, events are removed if the
azimuthal angle between the \ptmiss and the closest
jet with $\pt>30\GeV$ is smaller than 0.5 radians. An additional selection requirement $\abs{\Delta\phi(\cPZ,\ptvecmiss)} > 0.5$ is placed in order to remove events for which the instrumental \ptmiss is not well controlled.

Top quark decays are often associated with the production of leptons and missing transverse momentum in the final state but are also characterized by the presence of jets originating from \cPqb\ quarks (\cPqb\ jets).
The top quark background is suppressed by applying a veto on events having a \cPqb\ tagged jet with
$\pt > 30 \GeV$.
To reduce the $\PW\cPZ$ background in which both bosons decay leptonically,
any event with an additional $\Pe$ ($\Pgm$)  passing loose identification and isolation criteria with $\pt > 10\ (3) \GeV$ is rejected.

We select events with $\ptmiss\geq125\GeV$ and fit the transverse mass \MT\ distribution for the selected events.
The \ptmiss requirement rejects background processes that could lead to high \MT\ because of the kinematic properties of
the dilepton pair in the event.
The \ptmiss criterion is optimized based on expected signal significance.
The significance is found to be quite stable with the chosen \ptmiss requirement for masses above 400\GeV.

The transverse mass is reconstructed from the dilepton and \ptmiss system via the following definition :
\begin{equation}
	\MT^{2} = \left( \sqrt{{\PT(\ell\ell)}^{2} + {m(\ell\ell)}^{2}} + \sqrt{{\ptmiss}^{2} +  m_{\cPZ}^{2}} \right)^{2} - (\ptvec(\ell\ell) + \ptvecmiss)^{2},
\end{equation}
where $\ptvec(\ell\ell)$ and $m(\ell\ell)$ are the transverse momentum and invariant mass of the dilepton system, respectively.
In order to maximize the sensitivity, the search is carried out in different jet multiplicity categories defined as follows:

\begin{itemize}
	\item\textbf{VBF-tagged}: in this category we require two or more jets in the forward region with a pseudorapidity gap ($\abs{\Delta\eta}$) between the two leading jets greater than 4, and a minimal invariant mass of those two jets of 500\GeV. The two leptons forming the $\cPZ$ boson candidate are required to lie between these two jets in $\eta$, while no other jets ($\PT >  30\GeV$) are allowed in this central region;
    \item\textbf{$\mathbf{\geq 1}$-jet}: events with at least one reconstructed jet with $\PT >  30\GeV$, but failing the VBF selection;
    \item\textbf{0-jet}: events without any reconstructed jet with $\PT > 30\GeV$.
\end{itemize}

\begin{figure}[htbp]
   \centering
   \includegraphics[width=0.45\textwidth]{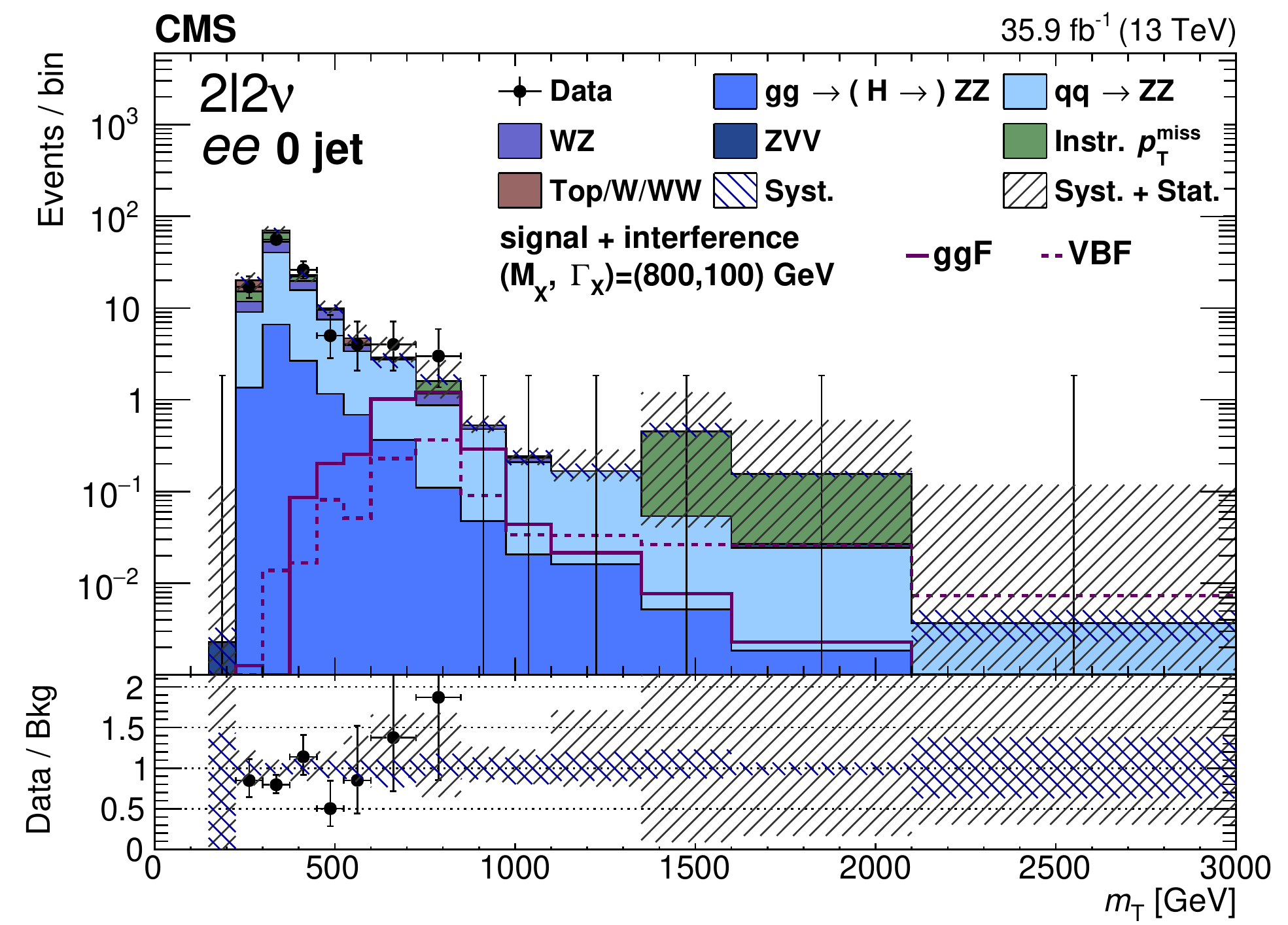}
   \includegraphics[width=0.45\textwidth]{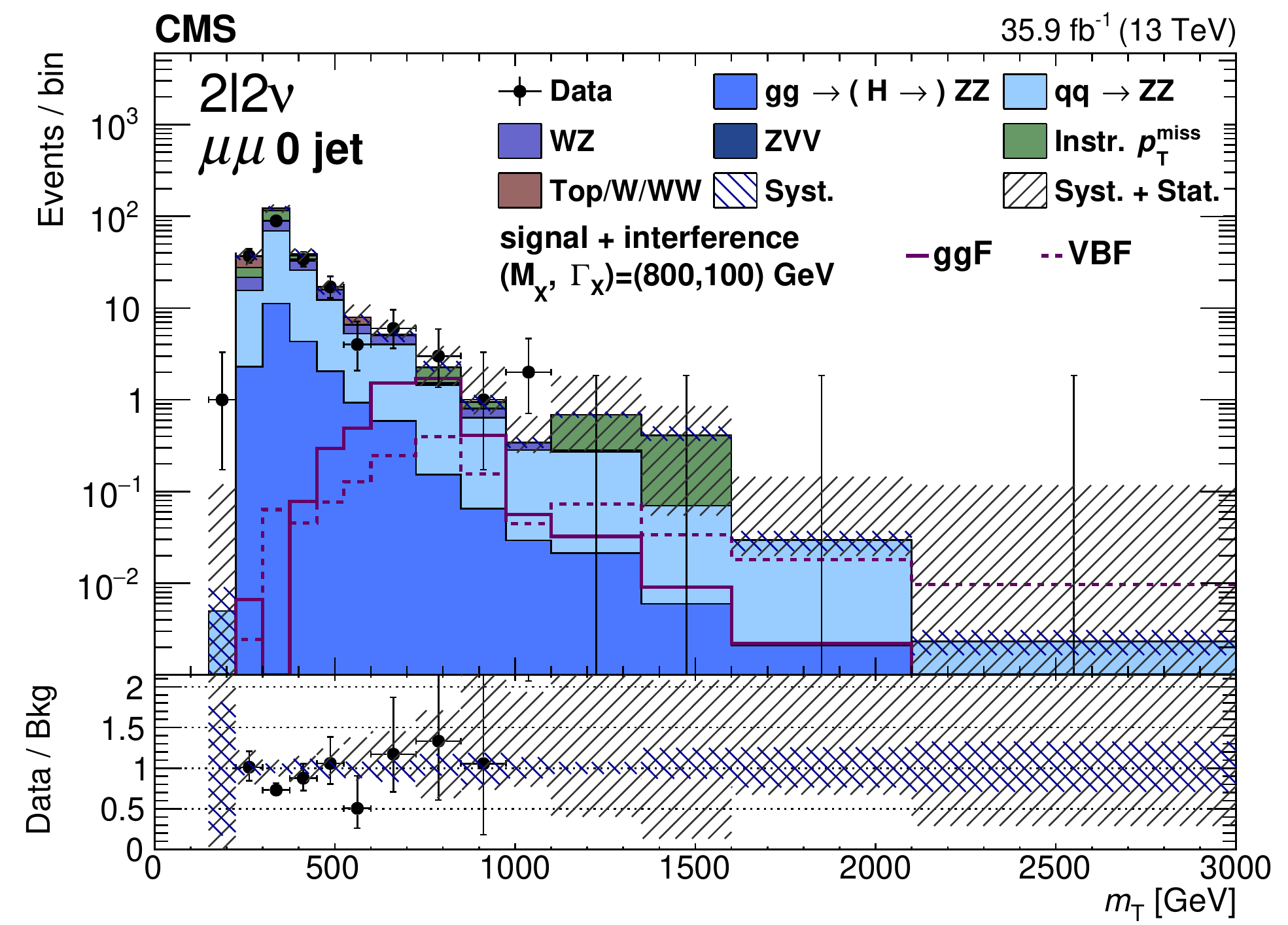}\\
\vspace{0.6cm}
   \includegraphics[width=0.45\textwidth]{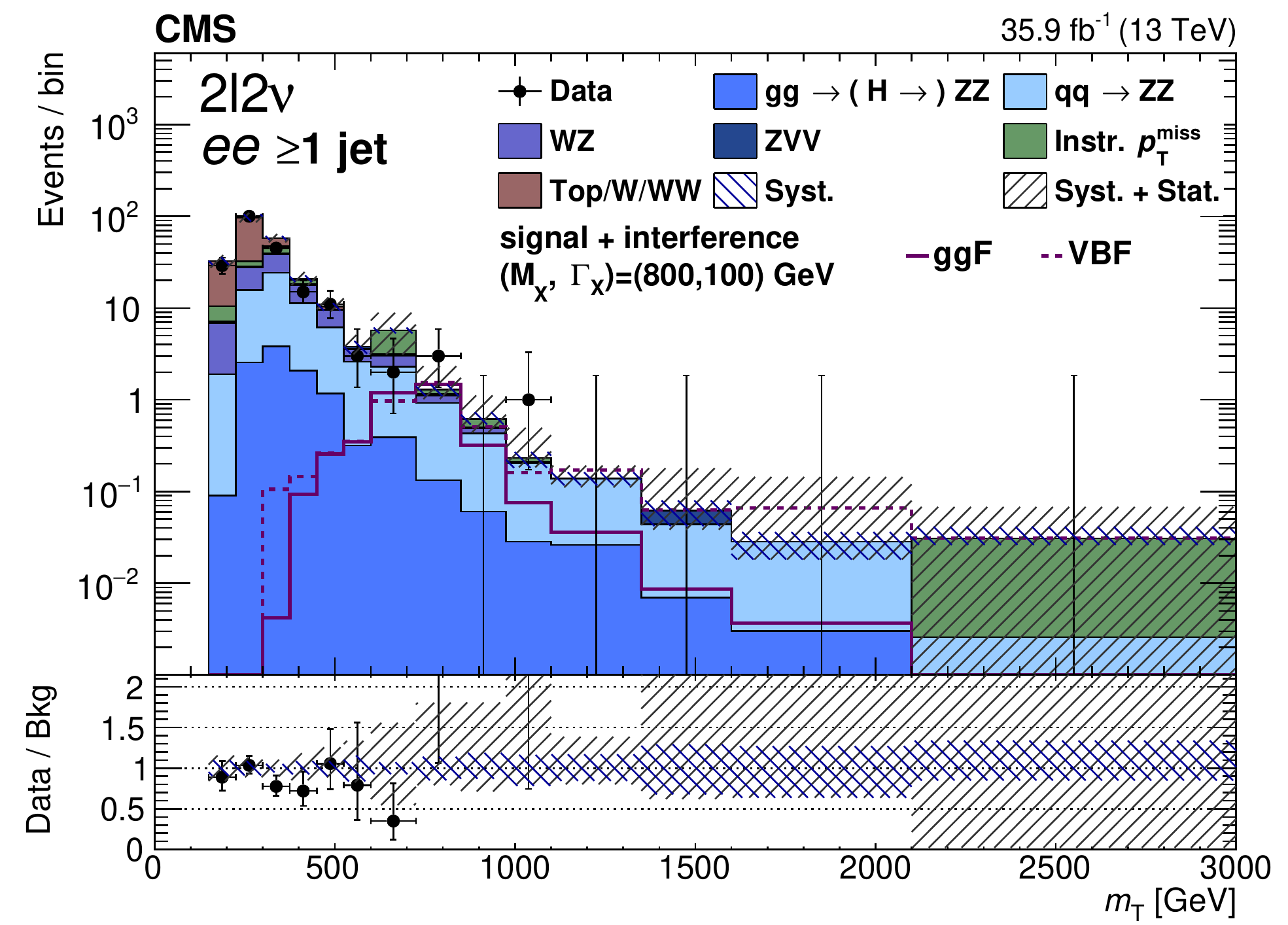}
   \includegraphics[width=0.45\textwidth]{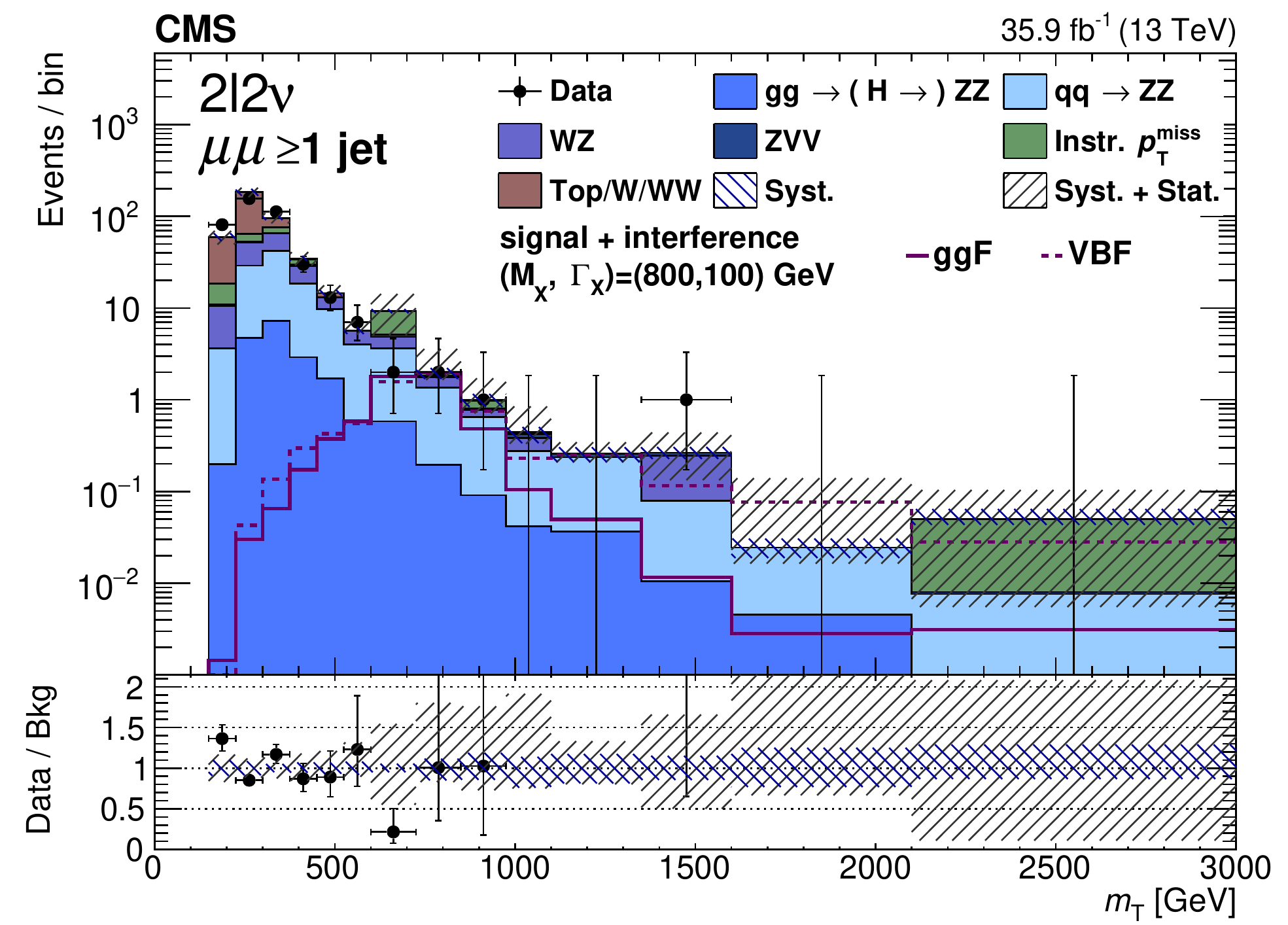} \\
\vspace{0.6cm}
   \includegraphics[width=0.45\textwidth]{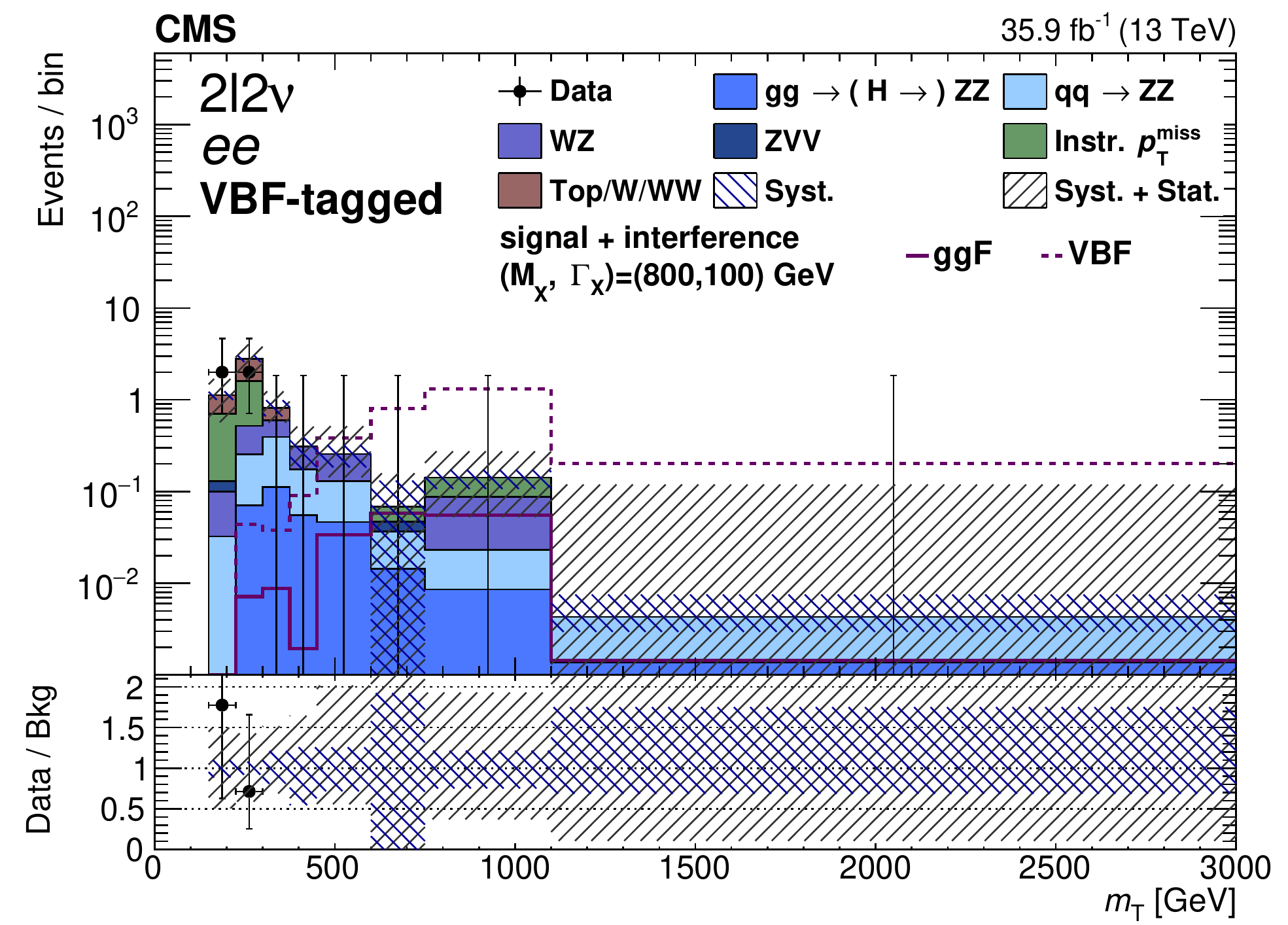}
   \includegraphics[width=0.45\textwidth]{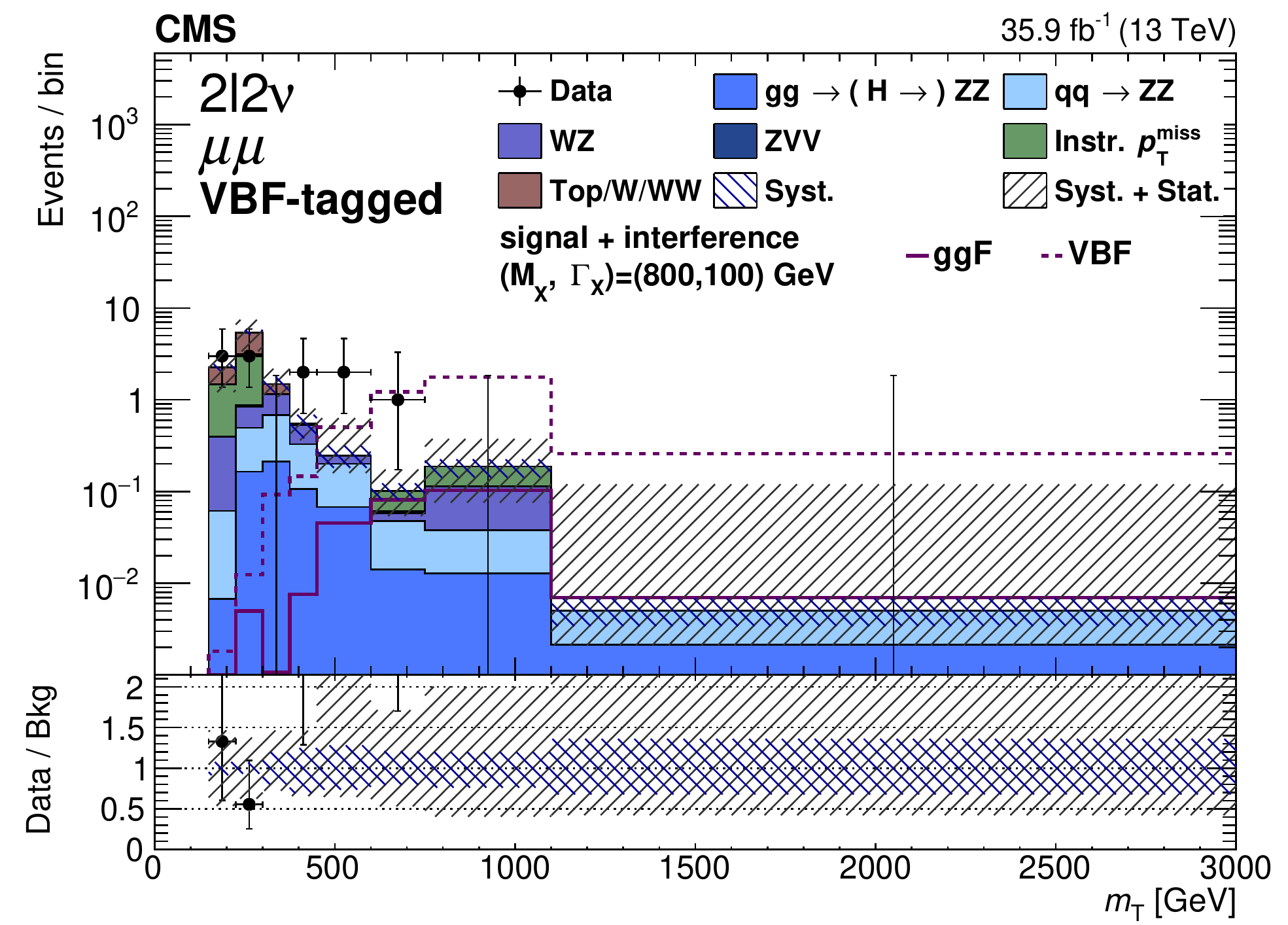}\\
\vspace{0.6cm}
\caption{ Distributions of the transverse mass \MT \ in the signal
region for the different analysis categories for the 2$\ell$2$\nu$ channel, in the \Pe\Pe (left)
and $\mu\mu$ final states (right). The points represent the data and the
stacked histograms the expected background. The open histograms show the
expected gluon fusion and VBF signals for the product of cross section and branching
fraction equal to $\sigma(\Pp\Pp\to\PH\to\cPZ\cPZ)=50\unit{fb}$. Lower panels show the ratio
of data to the expected background. The shaded areas show the systematic
and total combined statistical and systematic uncertainties in the background
estimation.}
   \label{fig:MT_cat}
\end{figure}

The last two categories are the most sensitive to the signal produced via \ggF but have different expected signal to background ratios. As a result of the above selection, events are split into six categories: \textbf{ $2\Pe2\nu$} or  \textbf{$2\mu2\nu$}, either 0-jet, $\geq 1$-jet or VBF-tagged. Fig.~\ref{fig:MT_cat} shows the \MT\ distributions for the signal and background processes superimposed, in the six event categories.

\section{Signal and background parameterization}
\label{Section_parameterization}

The goal of the analysis is to determine if a set of $\PX$ boson parameters
$\mX$, $\GX$, and $\sigma_i\mathcal{B}_{X\to\cPZ\cPZ}$ is consistent with the data, where $\sigma_i\mathcal{B}_{X\to\cPZ\cPZ}$ is the product of the signal production cross section and
the $\PX\to\cPZ\cPZ$ branching fraction in each production channel $i$ (gluon fusion or EW production). In practice, the $\sigma_i\mathcal{B}$ for $i=1,2$ are expressed in terms of
$\sigma_{\mathrm{tot}}\mathcal{B}_{X\to\cPZ\cPZ}$ and $f_{\mathrm{VBF}}$, where $\sigma_{\mathrm{tot}}$ is the sum of the cross sections in the two production channels. The confidence intervals on $\sigma_{\mathrm{tot}}\mathcal{B}_{X\to\cPZ\cPZ}$
are determined from profile likelihood scans for a given set of parameters $(\mX, \GX, f_{\mathrm{VBF}})$.
The extended likelihood function is defined for candidate events as
\begin{eqnarray}
\mathcal{L} =  \exp\Big( - \sum_{i} n_{vv}^i -\sum_i n_\text{bkg}^i  \Big)
\prod_k \prod_j
\Big(
  \sum_i n_{vv}^{i} \mathcal{P}^{i,k}_{vv}(\vec{x}_{j}; \mX, \GX)
+\sum_i n_\text{bkg}^{i} \mathcal{P}^{i,k}_\text{bkg}(\vec{x}_{j})
\Big),
\label{eq:likelihood}
\end{eqnarray}
where $n_{vv}^i$ and $n_\text{bkg}^i$ are the numbers of signal and background events in channel $i$. The observables $\vec{x}_j$ are defined for each event $j$ in category $k$
as discussed in Sections~\ref{sec:XZZ4l}, \ref{sec:XZZ2l2q}, and~\ref{sec:XZZ2l2nu}.
There are several signal and background types~$i$, defined for each production mechanism.
The background processes that do not interfere with the signal are described by the probability density functions (pdfs)
 $\mathcal{P}^{i,k}_\text{bkg}(\vec{x}_{j})$. The $vv\to \ff$ process is described by the pdf
$\mathcal{P}^{i,k}_{vv}(\vec{x}_{j}; \mX, \GX)$ for $vv=\Pg\Pg$ (gluon fusion) and $vv= $ VV (EW production).
This pdf describes the production and decay of the $\PX$ boson signal, SM background, including $\PH(125)$, and interference between all these contributions
and is parameterized as follows:
\begin{eqnarray}
\mathcal{P}^{i,k}_{vv} (\vec{x}_{j}; \mX, \GX) =
\mu_i  \mathcal{P}^{i,k}_{vv\to \PX\to \ff} (\vec{x}_{j}; \mX, \GX)
+ \sqrt{\mu_i}  \mathcal{P}^{i,k}_{\mathrm{int}} (\vec{x}_{j}; \mX, \GX)
+ \mathcal{P}^{i,k}_{vv\to\ff} (\vec{x}_{j})
,
\label{eq:psig}
\end{eqnarray}
where $\mu_i$ is the relative signal strength for production type $i$ defined as the ratio of
$\sigma_i\mathcal{B}$ with respect to a reference value, for which normalization of the pdf is determined.
The interference contribution $\mathcal{P}^{i,k}_{\mathrm{int}}$ scales as $\sqrt{\smash[b]{\mu_i}}$ and the pure signal as ${\mu_i}$,
while both depend on the signal parameters $\mX$ and $\GX$.
The likelihood defined in Eq.~(\ref{eq:likelihood}) is maximized with respect to the nuisance parameters,
which include the constrained parameters describing the systematic uncertainties.

\subsection{Signal model}
\label{sec:Signalmodel}

The parameterization of $\mathcal{P}^{i,k}_{vv} (\vec{x}_{j}; \mX, \GX)$ is performed using the MC simulation
discussed in Section~\ref{sec:MC} with the ME method.
In the case of the $\PX\to \cPZ\cPZ\to 4\ell$ or $2\ell2\Pq$ channels, a full reconstruction of the final state is possible. Therefore, the ideal differential distribution prior to detector
effects $\mathcal{P}^{\mathrm{ideal}}_{vv}$, equivalent to Eq.~(\ref{eq:psig}),
is parameterized using ME techniques and is further corrected for detector acceptance and resolution effects.
In the case of $\PX\to \cPZ\cPZ\to 2\ell2\nu$, this approach is not possible because of missing neutrinos:
MC simulation
is reweighted for each hypothesis of $\mX$, $\GX$, and $\sigma_i\mathcal{B}_{\PX \to \cPZ\cPZ}$, leading to template parameterization
of $\mathcal{P}^{i,k}_{vv}$ for each set of signal parameters. While ultimately the two approaches are equivalent, the former approach is more flexible in implementation, and the latter avoids
the intermediate step of ideal pdf parameterization.

In the $\PX\to \cPZ\cPZ\to 4\ell$ or $2\ell2\Pq$ channels, we parameterize the signal mass shape
as follows.
A pdf after detector effects ${\cal M}^{\mathrm{reco}}_{vv}(m_{\cPZ\cPZ})$ is implemented with the multiplicative efficiency function  ${\cal E}(m_{\cPZ\cPZ})$ and convolved with a mass resolution function
${\cal R}(m_{\cPZ\cPZ}|m^{\mathrm{Gen}}_{\cPZ\cPZ})$, both extracted from simulation of the \ggF and VBF processes:

\begin{eqnarray}
{\cal M}^{\mathrm{reco}}_{vv}(m_{\cPZ\cPZ})=
\left({\cal E}(m_{\cPZ\cPZ}^{\mathrm{Gen}}) {\cal M}_{vv}(m^{\mathrm{Gen}}_{\cPZ\cPZ}|\mX,\GX)\right) \otimes {\cal R}(m_{\cPZ\cPZ}|m^{\mathrm{Gen}}_{\cPZ\cPZ}).
\label{eq:signalpdf1D}
\end{eqnarray}

\begin{figure}[htbp]
\centering
\includegraphics[width=0.45\textwidth]{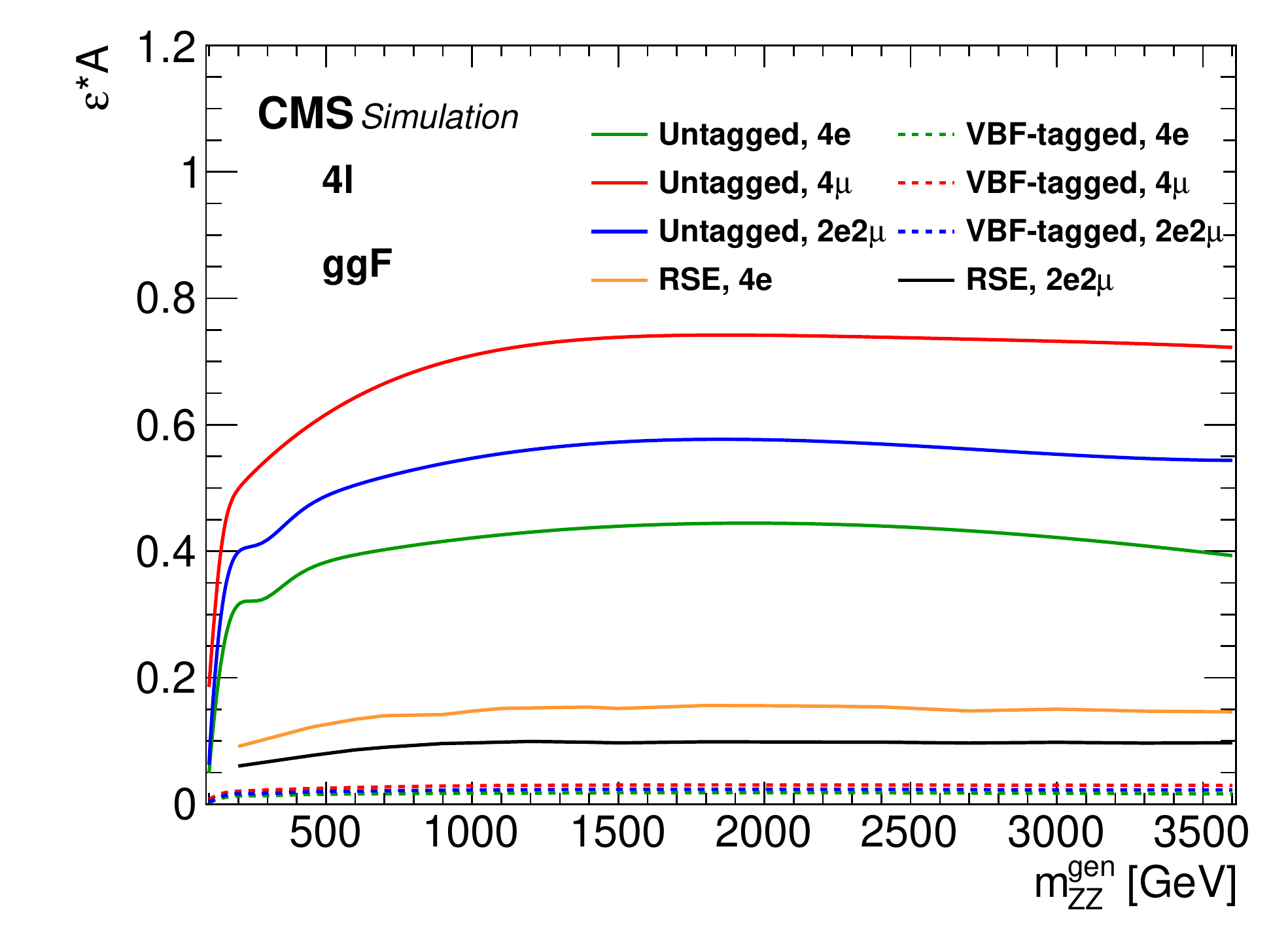}
\includegraphics[width=0.45\textwidth]{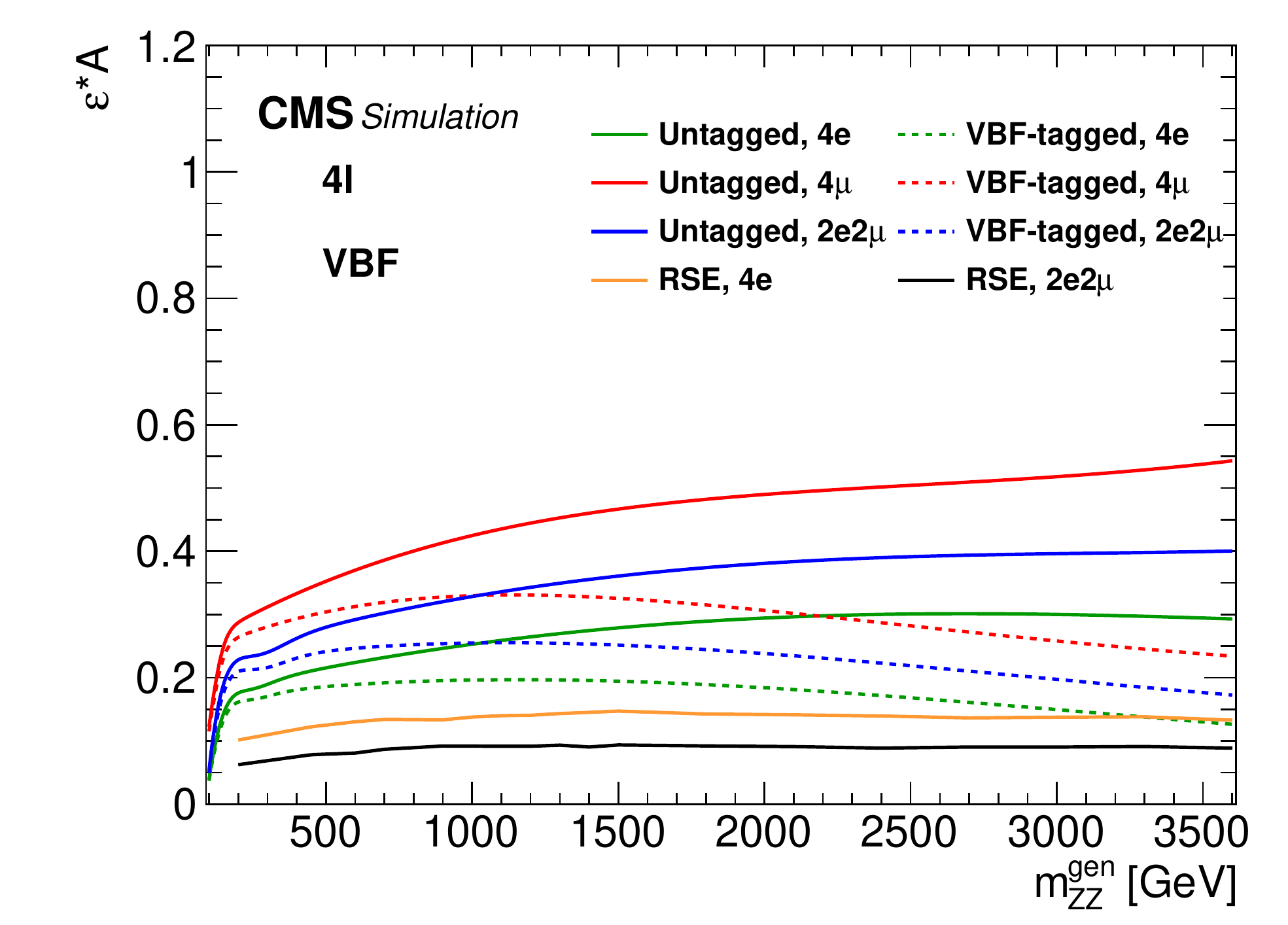}
\includegraphics[width=0.45\textwidth]{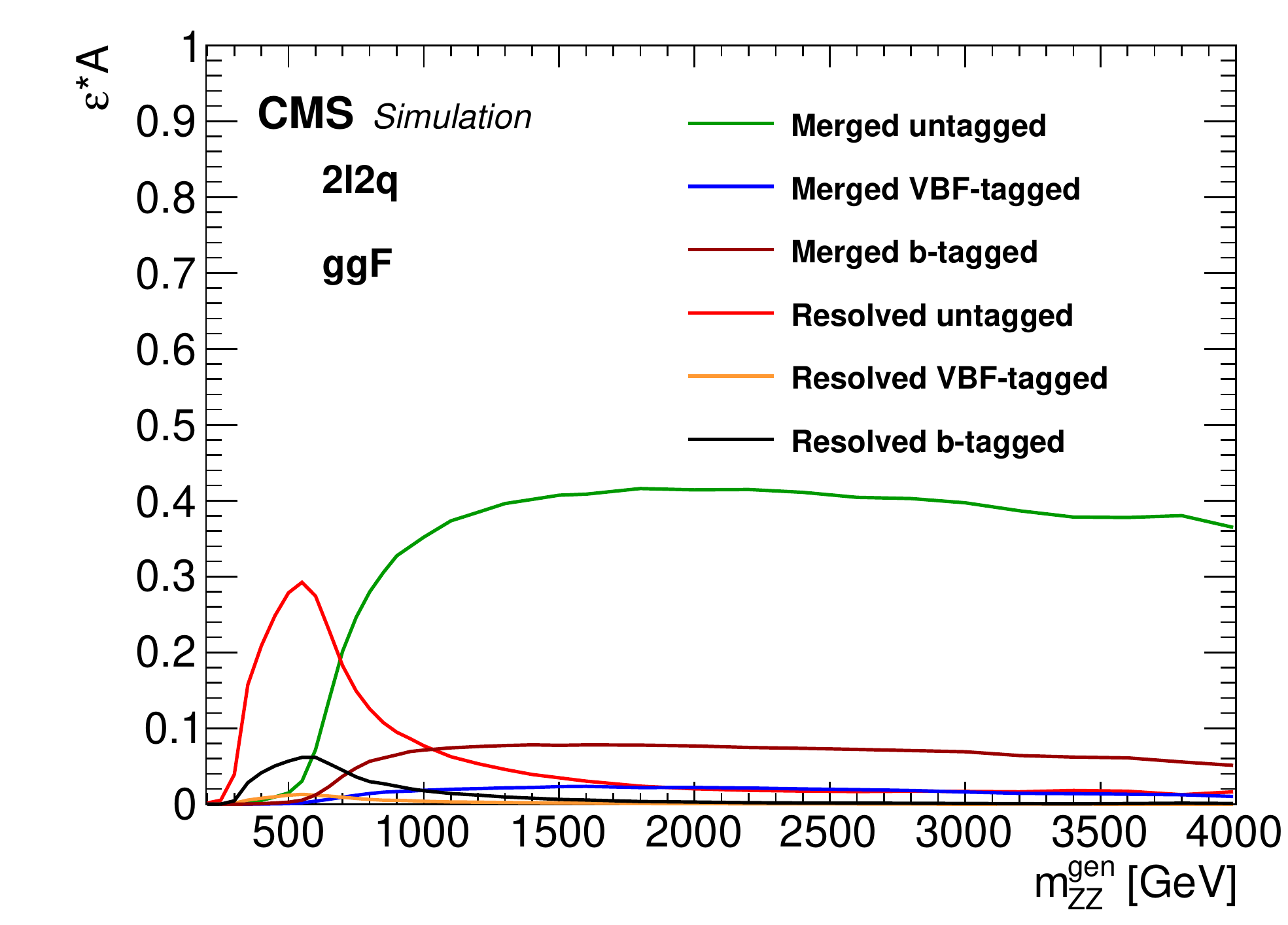}
\includegraphics[width=0.45\textwidth]{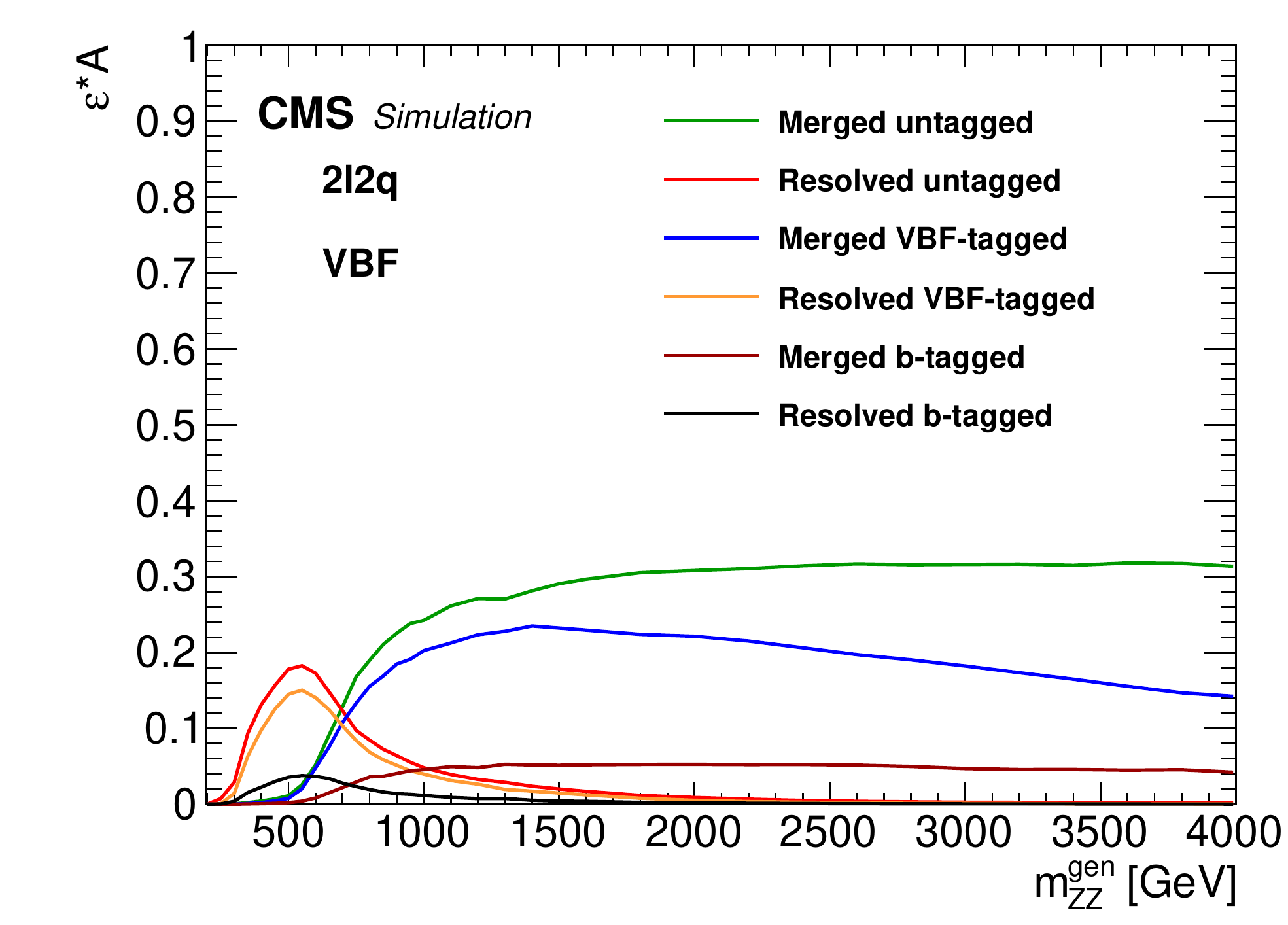}
\caption{
	The product of efficiency and acceptance for signal events to pass the $\PX \to \cPZ\cPZ \to 4\ell$ (upper plots) and $\PX \to \cPZ\cPZ \to 2\ell 2\Pq$ (lower plots) selection as a function of the generated mass $m_{\cPZ\cPZ}^{\mathrm{Gen}}$, from \ggF (left) and VBF (right) production modes.
\label{fig:eff}
}
\end{figure}

\begin{figure}[htbp]
\centering
\includegraphics[width=0.45\textwidth]{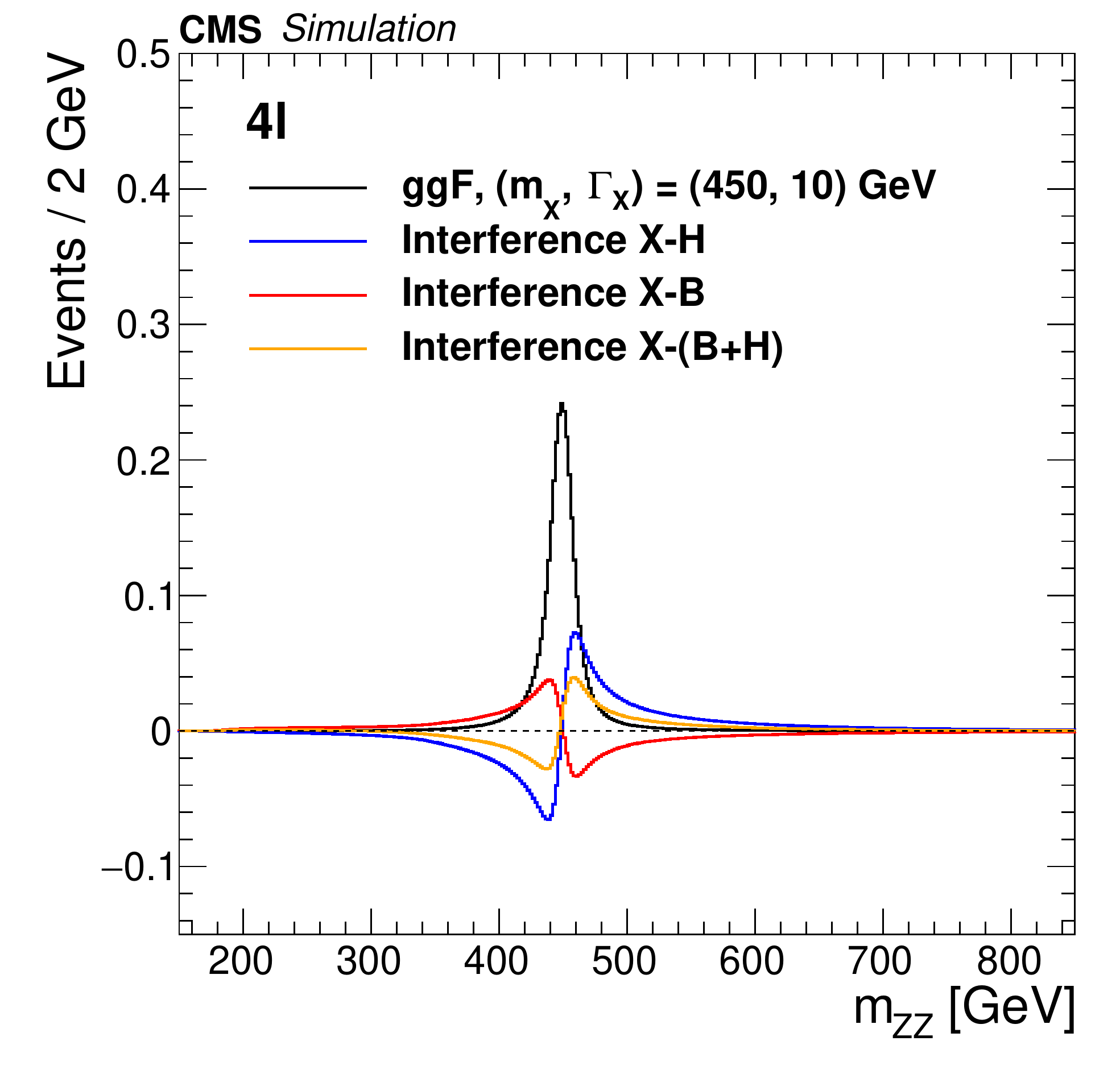}
\includegraphics[width=0.45\textwidth]{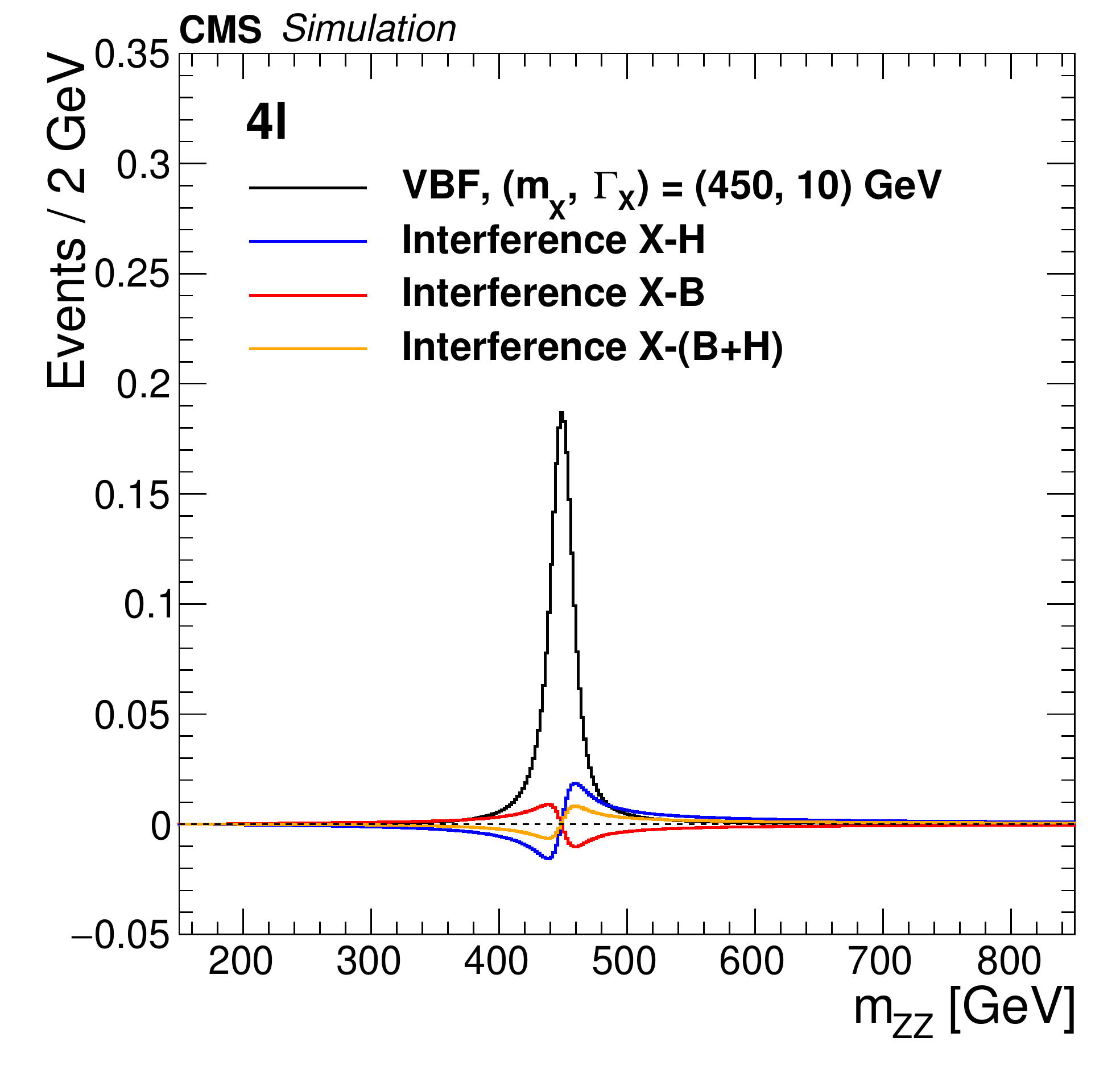}
\caption{
	Parameterizations of the four lepton invariant mass for \ggF (left) and VBF (right) production modes, for $\mX = 450\GeV$, $\GX= 10\GeV$. The interference contributions from $\PH(125)$ and $\Pg\Pg\to\cPZ\cPZ$ or $\PV\PV\to\cPZ\cPZ$ background are also shown. The signal cross section used corresponds to the limit obtained in the $4\ell$ final state.  
\label{fig:mzz_interference}
}
\end{figure}

The parameterizations of ${\cal R}(m_{\cPZ\cPZ}|m^{\mathrm{Gen}}_{\cPZ\cPZ})$ and ${\cal E}(m_{\cPZ\cPZ}^{\mathrm{Gen}})$ cover the mass
range from 100\GeV to 3.5\TeV. Figure~\ref{fig:eff} shows the efficiencies in the $\PX\to4\ell$ and $\PX\to2\ell2\Pq$ channels in the various categories. The resolution in the $4\ell$ final state is 1--2\% and 3--5\% in the $2\ell2\Pq$ final state. With the above ingredients, the $m_{\cPZ\cPZ}$ parameterization is shown in Fig.~\ref{fig:mzz_interference}, for a boson with $\mX = 450\GeV$, $\GX = 10\GeV$ decaying to four leptons. The interference contributions from $\PH(125)$ and $\Pg\Pg\to\cPZ\cPZ$ background are also shown.

The 2D signal distributions in the $4\ell$ and $2\ell2\Pq$ final states are built with the conditional template ${\cal T}(\Dbkg|m_{\cPZ\cPZ})$,
which describes the $\Dbkg$ discriminant distribution from Eq.~(\ref{eq:Dbkgmela}) or~(\ref{eq:Zjjmela})
for each value of $m_{\cPZ\cPZ}$:

\begin{eqnarray}
{\cal P}^{i, k}_{vv}(m_{\cPZ\cPZ},\Dbkg)={\cal M}^{\mathrm{reco}}_{vv}(m_{\cPZ\cPZ})  {\cal T}(\Dbkg|m_{\cPZ\cPZ}).
\label{eq:signalpdf2D}
\end{eqnarray}

The template ${\cal T}(\Dbkg|m_{\cPZ\cPZ})$ parameterization includes all detector effects
affecting the $\Dbkg$ distribution. A closure of the full model described by Eq.~(\ref{eq:signalpdf2D})
is achieved by comparing the model to the simulation for a number of
signal parameters.

\subsection{Background model}
\label{sec:Background}
Common backgrounds among the three final states include the $\Pg\Pg(\PV\PV)\to \cPZ\cPZ$ process, $\cPZ\cPZ$ produced via $\qqbar$ annihilation, as well as the
$\PW\cPZ$ production process.
The \ggF and EW production of the $\Pg\Pg(\PV\PV)\to \cPZ\cPZ$ background are treated together with the $\PX$ boson
signal and background, including interference between the corresponding amplitudes, as discussed in detail in Section~\ref{sec:Signalmodel}. Higher order corrections are applied to these processes as discussed in Section~\ref{sec:MC}.

The production of $\cPZ\cPZ$ via $\Pq\Paq$ annihilation is estimated using simulation. The fully differential cross section for the \qqZZ\ process is computed at NNLO~\cite{Grazzini2015407},
and the NNLO/NLO K factor as a function of $m_{\cPZ\cPZ}$ is applied to the {\sc POWHEG} sample.
This K factor varies from 1.0 to 1.2 and is 1.1 at $m_{\cPZ\cPZ}=125\GeV$.
Additional NLO EW corrections, which depend on the flavor of the initial state quarks and on kinematic properties, are also applied in the region $m_{\cPZ \cPZ}>2m_{\cPZ}$, where the corrections are computed~\cite{Gieseke:2014gka,Manohar:2016nzj,Baglio:2013toa}. The $\PW\cPZ$ production is estimated using simulation, where photon induced EW corrections are applied~\cite{Frixione:2015zaa,Frixione:2014qaa}.

The analysis specific background processes, or the ones whose contribution is derived from control samples in data, are discussed in the following sections.

\subsubsection{$\PX\to \cPZ\cPZ\to 4\ell$}

The most important background to the X signal in the $4\ell$ channel, in addition to the irreducible \cPZ\cPZ\,
arises from processes in which decays of heavy flavor hadrons, in flight decays of
light mesons within jets, or photon conversion or decay of charged
hadrons overlapping with $\pi^0$ decays are misidentified as leptons.
The main processes producing these backgrounds are
$\cPZ+{\mathrm{jets}}$,  $\ttbar+{\mathrm{jets}}$,
$\cPZ\gamma+{\mathrm{jets}}$, $\PW\PW+{\mathrm{jets}}$, and $\PW\cPZ+{\mathrm{jets}}$ production.
Collectively, we denote these as ``reducible'' backgrounds.
The contribution from the reducible background is estimated
using two independent methods  based on data from dedicated control regions.
The control regions are defined by a dilepton pair satisfying all the requirements of a $\cPZ_1$ candidate and two additional leptons, opposite sign (OS) or same sign (SS),
satisfying more relaxed identification criteria than the ones used for the selection and categorization for the signal events.
These four leptons are then required to pass the analysis
$\cPZ\cPZ$ candidate selection.
The event yield in the signal region is obtained by weighting the control region events
by the lepton misidentification probability, defined as the fraction of
non signal leptons that are identified by the analysis selection criteria.

The lepton misidentification probabilities are measured separately for electrons and muons from a control sample that requires a $\cPZ_1$ candidate
consisting of a pair of leptons, both passing the selection requirements used in the analysis, and exactly one additional lepton passing the relaxed selection.

The predicted yield in the signal region of the reducible background
is the result of a combination of the two methods described above.
The shape of the $m_{4\ell}$ distribution for the reducible background
is obtained by combining the prediction from the OS and SS methods and fitting the distributions
with empirical functional forms built from Landau~\cite{Landau:1944if} and
exponential distributions.

\subsubsection{$\PX\to \cPZ\cPZ\to 2\ell2\Pq$}

The majority of the background ($>$90\%) is composed of events from
$\cPZ + \text{jets}$ production,
where jets associated to the Drell--Yan production are misidentified as
coming from a hadronic \cPZ\ decay. Subdominant backgrounds comprise
events from \ttbar{} production and from diboson EW production.

The $\ttbar$ background is an important source of contamination
in the \cPqb\ tagged category. It is estimated from data using $\EMJWG$
events passing the same selection as for the signal.
This method accounts for other small backgrounds
(such as $\PW\PW + \text{jets}$, $\cPZ\to\TT + \text{jets}$, and
single top quark production) where the lepton flavor symmetry can be used as well.
Because of the limited number of events in the $\EMJWG$ control region, the \mZZ{} shapes are
taken from \ttbar\ simulation, and the statistical uncertainty in the control region is considered as the uncertainty in the background estimation.

In the $\cPZ +$ jets background, the misidentified hadronic \cPZ\ comes either from
the combinatoric background of $\cPZ + 2$ jets events where the dijet system happens to have an invariant mass
in the range compatible with that of the $\cPZ$ boson (resolved category)
or from an unusual parton shower and hadronization development for a single jet, leading to a configuration similar to that of the boosted $\PZ\to\qqbar$  decay (merged category). In both cases, and in each analysis category, a sideband region with a misidentified hadronic \cPZ\ mass close to that
of the signal region can be used to estimate the contribution of this background. To address the correlation
between the hadronic \cPZ\ mass and \mZZ{} in these configurations, a correction factor is estimated from simulation.

\par The alpha transfer factor $\alpha(\mZZ{})$, defined as
\begin{equation}
	\alpha(\mZZ{}) = \frac{N_\textrm{SIG}^\textrm{MC}(\mZZ{})}
	{N_\textrm{SB}^\textrm{MC}(\mZZ{})},
\end{equation}

is calculated as the ratio of the \mZZ{} distributions in the signal and sideband regions for $\cPZ + \text{jets}$ simulated events.
The alpha function is multiplied by the sideband \mZZ{} distribution to derive the $\cPZ + \text{jets}$ contribution in the signal region.
The $\cPZ + \text{jets}$ distribution from the sideband is obtained by subtracting the subdominant backgrounds from MC prediction. Both the shape and the yield for the $\cPZ + \text{jets}$ background are estimated using this method.

While a binned evaluation of the product of the alpha factor and the sideband yields would
be a complete estimate of the background, low event yields from data or simulation in specific
bins or event categories could induce large statistical fluctuations
in the bins with smaller event yields, occurring at large values of \mZZ.
We define a ``transition'' mass value \mZZtilde. For $\mZZ < \mZZtilde$, the
binned evaluation is used as mentioned above. For $\mZZ > \mZZtilde$,
in order to smooth the background estimation, the $\cPZ + \text{jets}$ shape is then fit using a sum of two exponential functions (a single exponential function) for the resolved jet untagged category (the remaining categories). A binned estimation
for $\mZZ > \mZZtilde$ is then obtained by integrating the smoothed
estimation in the corresponding intervals. The statistical uncertainty derived from the fit is propagated to the final result using the full covariance matrix.

\subsubsection{$\PX\to \cPZ\cPZ\to 2\ell2\nu$}
\label{sec:2l2nubkgest}

The $\cPZ + \text{jets}$ background is modeled from a control sample of events
with a single photon produced in association with jets ($\Pgg + \text{jets}$).
This choice has the advantage of making use of a large sample,
which captures the source of instrumental \ptmiss from the $\cPZ$ production in all important aspects, \ie\ production mechanism,
underlying event conditions, pileup scenario, and hadronic recoil.
By using the $\Pgg + \text{jets}$ expectation we avoid the need to use the prediction from simulation
for the instrumental background arising from the mismeasurement of jets.
Each $\Pgg + \text{jets}$ event must fulfill similar requirements as the dilepton events: no \cPqb\ tagged jets, no additional identified leptons, and a significant transverse momentum ($\PT\geq55\GeV$).

\begin{figure}[htp]
\centering
\includegraphics[width=0.45\textwidth]{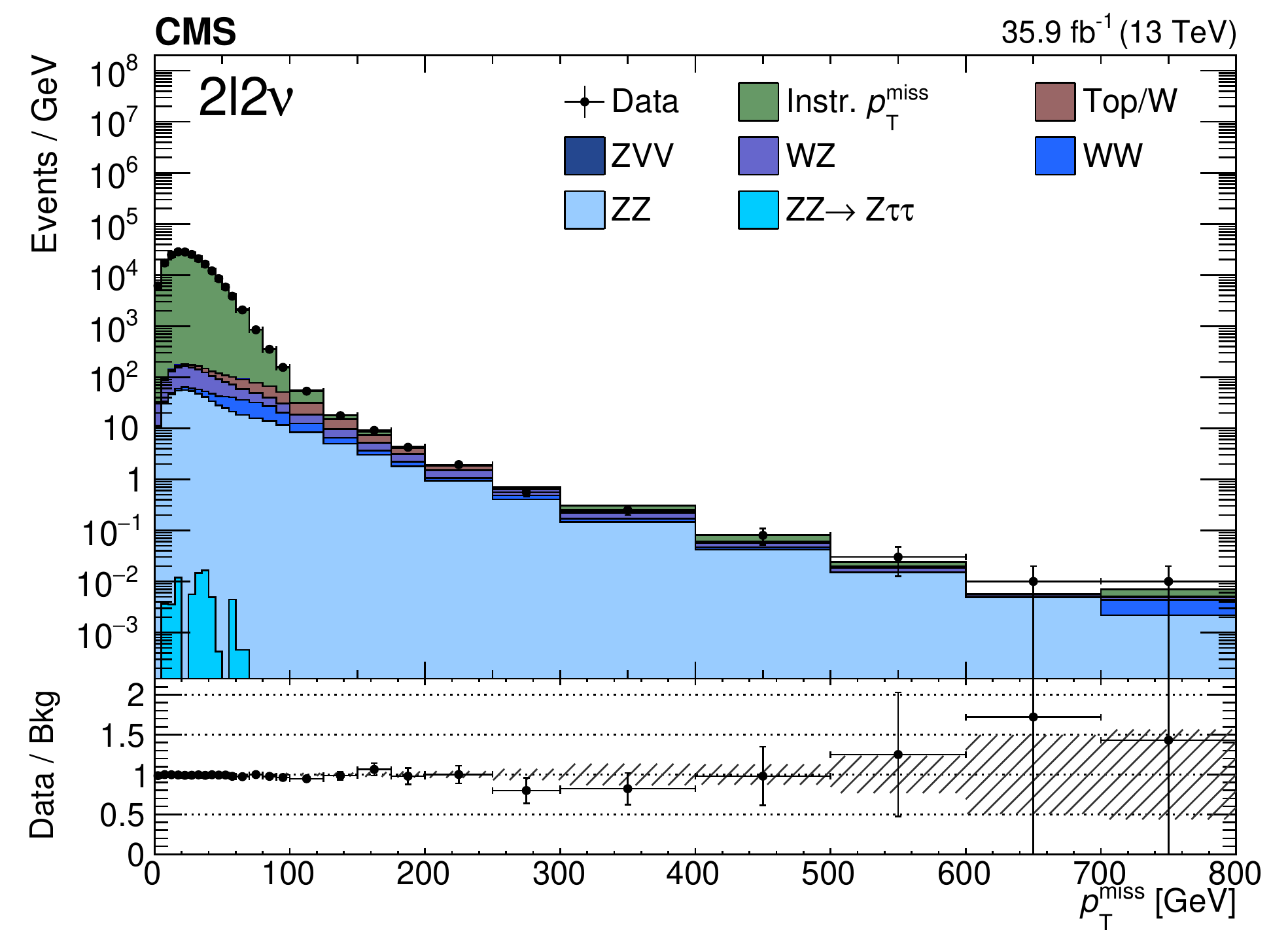}
\caption{Distribution of the missing transverse energy \ptmiss in the dilepton signal region. The points represent the data and the stacked histograms the expected backgrounds. The lower panel shows the ratio between data and background estimation.
}
\label{fig:2l2nu_met_mt}
\end{figure}

The kinematic properties and overall normalization of $\Pgg + \text{jets}$ events
are matched to $\cPZ + \text{jets}$ in data through an event by event
reweighting as a function of the boson $\PT$ in each of the event categories separately,
to account for the dependence of the \ptmiss on the associated hadronic activity.
Contamination of the photon data by processes that lead to a photon produced in association with genuine \ptmiss,
such as $\PW(\ell\cPgn)+\Pgg$ and $\PW(\ell\cPgn) + \text{jets}$ where the jet is mismeasured as a photon, and $\cPZ(\cPgn\cPgn)+\Pgg$ events, are subtracted using simulation.  The simulation of the \ptmiss in such events is more reliable than in  $\cPZ + \text{jets}$ as the \ptmiss is induced by a neutrino and not by detector features. After the $\PT$ reweighting and the \ptmiss requirement, these events represent less than 25\% of the photon sample.
This procedure yields a good description of the \ptmiss distribution in $\cPZ + \text{jets}$ events,
as shown in Fig.~\ref{fig:2l2nu_met_mt}, which compares the
\ptmiss distribution of the reweighted $\Pgg + \text{jets}$ events along with other backgrounds to the \ptmiss
distribution of the dilepton events in data.

To compute $\MT$ for each
$\Pgg + \text{jets}$ event, $\ptvecmiss(\ell\ell)$ is defined as the photon $\ptvecmiss$
and the value of $m(\ell\ell)$ is chosen according to a probability density function
constructed from the measured dilepton invariant mass distribution in data (dominated by $\cPZ + \text{jets}$ events).
The uncertainty in this background estimate includes a statistical contribution from the photon control sample and a contribution from the simulations used to subtract processes with photon and genuine \ptmiss, and is found to be equal to 100\% in the signal region. Another 10\% contribution comes from the degree of agreement between the $\Pgg + \text{jets}$ prediction and the \ptmiss distributions in a simulated dilepton sample. Uncertainties in the production cross section of the subtracted processes with genuine \ptmiss are also accounted for and are on the order of 25\%.

The background processes that do not involve a $\cPZ$ resonance (nonresonant background) are estimated using a control sample of events with dileptons of different flavor
($\Pe^{\pm}\Pgm^{\mp}$) that pass the analysis selection.
This background consists mainly of leptonic \PW\ decays from $\ttbar$, $\cPqt\PW$, and
$\PW\PW$ events. Small contributions from single top quark events produced in
$s$- and $t$-channels, $\PW + \text{jets}$ events in which the $\PW$ boson
decays leptonically and a jet is mismeasured as a lepton, and $\cPZ\cPZ$ or $\cPZ$ events where a $\cPZ$ decays into $\Pgt$ leptons, which produce light leptons and \ptmiss, are also included in this estimate.
This method cannot distinguish between the nonresonant background and
the contribution from $\PH \to \PW\PW \to 2\ell 2\cPgn$ events, which is treated
as a part of the nonresonant background estimate.
The numbers of nonresonant background events $N_{\Pgm\Pgm}$ and $N_{\Pe\Pe}$ in the $\Pep\Pem$ and $\Pgmp\Pgmm$ final states
are estimated by correcting the number of selected events $N_{\Pe\Pgm}$ in the $\Pe^{\pm}\Pgm^{\mp}$ final state.
The correction factor accounts for the difference in branching fractions, acceptance and efficiency between unlike flavor and same flavor dilepton events,
and is computed as:

\begin{equation}
N_{\Pgm\Pgm} = \frac{N_{\Pgm\Pgm}^\mathrm{SB}}{N_{\Pe\Pgm}^\mathrm{SB}}  N_{\Pe\Pgm}, \qquad
N_{\Pe\Pe} = \frac{N_{\Pe\Pe}^\mathrm{SB}}{N_{\Pe\Pgm}^{SB}}  N_{\Pe\Pgm},
\end{equation}

where $N_{\Pe\Pe}^\mathrm{SB}$, $N_{\Pgm\Pgm}^\mathrm{SB}$, and $N_{\Pe\Pgm}^\mathrm{SB}$ are the numbers of events
in a sideband control sample of $\Pep\Pem$, $\Pgmp\Pgmm$, and $\Pe^{\pm}\Pgm^{\mp}$ final states, respectively.
The sideband selection is defined by $40 < m(\ell\ell) < 70\GeV$ or $110 < m(\ell\ell) <200\GeV$,  $\ptmiss>70\GeV$,
and at least one \cPqb\ tagged jet.
The requirement of a \cPqb\ tagged jet is used to provide a sample enriched in top quark events and to suppress possible contamination from $\cPZ + \text{jet}$
events where a jet is misidentified as a lepton.
The correction factor measured in the sideband is $0.37\pm0.01 \stat$ and $0.68\pm 0.01 \stat$ for the $\Pe\Pe$ and $\Pgm\Pgm$ channels, respectively.
The uncertainty in the estimate of the nonresonant background is determined via MC closure tests using simulated events as well as by
comparing results calculated from sideband regions. The total error is within 13\%, which is assigned as the
systematic uncertainty in this method.

\section{Systematic uncertainties}
\label{sec:Systematics}

The three final states share common systematic uncertainties arising from the theoretical prediction,
reconstructed objects, and common backgrounds. Theoretical uncertainties that affect both the signal and background estimation
include uncertainties from the renormalization and factorization scales and the choice of the PDF set.
The uncertainties from the renormalization and factorization scale are determined by varying these scales independently by factors of
0.5 and 2 with respect to their nominal values, while keeping their ratio between 0.5 and 2. The uncertainties from the PDFs are obtained from the root mean squares of the variations, using different replicas of the default NNPDF set. An
uncertainty of 10\% in the K factor used for the $\ggZZ$ prediction is applied, which is derived from renormalization and factorization scale variations.
The uncertainty in the NNLO-to-NLO K factor for the $\cPZ\cPZ$ and $\PW\cPZ$ cross sections is about 10\%.
The renormalization and factorization scale and PDF uncertainties are evaluated from simulation, and are applied to the event categorization
and overall signal and background yields. A systematic uncertainty of 2\% in the $\cPZ$ boson branching fraction value is taken into account for the signal yields~\cite{deFlorian:2016spz}.

The uncertainty in the knowledge of the integrated luminosity of the data samples (2.5\%) introduces an uncertainty in the numbers of signal and background events passing the final selection. Uncertainties in the lepton identification and reconstruction efficiencies lead to 2.5\% uncertainties in the 4\PGm and 9\% in the 4\Pe\ final states for the $4\ell$ selection, 4--8\% (2\Pe\ and 2\PGm) for $2\ell2\Pq$ and 6--8\% for $2\ell2\nu$ in the normalizations of both signal and background. The uncertainties in the lepton energy scales are 0.01--0.1\% for muons and 0.3\% for electrons. A 20\% relative uncertainty in the signal resolution is assigned due to per lepton energy resolution in the $4\ell$ and $2\ell2\Pq$ final states. The jet energy scale (JES), jet energy resolution (JER) and jet reconstruction efficiency uncertainties affect both signal and background yields and represent the most important uncertainties for the $2\ell2\Pq$ signal shapes.
The systematic uncertainties that are common among the three final states are summarized in Table~\ref{tab:syst}.

 \begin{table}[h!t]
   \topcaption{Sources of uncertainties considered in each of the channels included in this
   analysis. Uncertainties are given in percent. The numbers shown as ranges represent the uncertainties in different final states or categories. Most uncertainties affect the normalizations of the background estimations or simulated event yields, and those that affect the shape of kinematic
   distributions as well are labeled with (*). }
   \label{tab:syst}
   \centering
   \begin{tabular}{>{\quad}lccc}
  \hline
  Source of uncertainty [\%] & $\PX\to\cPZ\cPZ$ & $\PX\to\cPZ\cPZ$ & $\PX\to\cPZ\cPZ$ \\
	  & $\to 4\ell$ & $\to 2\ell$2\Pq & $\to 2\ell 2\nu$ \\
  \hline
  \rowgroup{Experimental sources} \\
   Integrated luminosity &  2.5 & 2.5 &  2.5\\
  $\ell$ trigger and selection efficiency & 2.5--9 & 4--8 & 6--8\\
  $\ell$ momentum/energy scale (*)&  0.04--0.3& 0.1--0.3 &  0.01--0.3\\
  $\ell$ resolution (*)&  20 & 20 &  \NA \\
  JES, JER, $\ptmiss$ (*)& 1--30& 1--10  & 1--30\\
   \cPqb\ tagging/mistag &  \NA & 5--7 & 2--4 \\ [\cmsTabSkip]
    \rowgroup{Background estimates} \\
   $\cPZ + \text{jets}$ & 36--43 & 10--50 & 20--50\\
   top quark, $\PW\PW$ 	& \NA & 15 & 10\\
   $\PW\gamma^*$, $\PW\cPZ$ & \NA & 3--10 & 15\\[\cmsTabSkip]
   \rowgroup{Theoretical sources} \\
   Renorm./factor. scales  & 3--10 & 3--10 & 5--10\\
   PDF set & 3--4& 3--5 & 1--4 \\
  EW corrections ($\qqZZ$) (*)& 1 & 1 &2 \\
  NNLO (\ggZZ) K factor & 10 & 10 &10 \\
  \hline
   \end{tabular}

 \end{table}

In addition, each final state has channel specific uncertainties, mainly from the background estimations based on control samples in data, as well as from merged jet reconstruction.

\subsection{\texorpdfstring{$\PX\to \cPZ\cPZ\to 4\ell$}{to4l}}
Experimental uncertainties for this channel arise mainly from the reducible background estimation.
Impacts from the limited numbers of events
in the control regions as well as in the region region where the misidentification rates evaluated are taken into account. Additional sources of systematic uncertainty arise from the difference
in the composition of the sample from which the misidentification rate is computed and
the control regions of the two methods where the lepton misidentification probability is applied.
The systematic uncertainty in the $m_{4\ell}$ shape is determined by taking the envelope of
differences among the shapes from the OS and SS methods in the three different final states.
The combined systematic uncertainties are estimated to be about 36\% (4\PGm) to 43\% (4\Pe).

\subsection{\texorpdfstring{$\PX\to \cPZ\cPZ\to 2\ell 2\Pq$}{to2l2q}}
The dominant uncertainties in the signal selection efficiency for this channel arise from uncertainties in the efficiencies to tag the hadronic
jet as a \cPZ\ in the high mass boosted categories, and from uncertainties in the \cPqb\ tagging efficiency. The efficiency of the boosted boson tagging selection and its corresponding systematic uncertainty are measured from data using a sample enriched in $\ttbar$ events. Uncertainties in the signal efficiencies from the jet mass scale and resolution are 1--9\% and 7--13\% depending on the mass. $\tau_{21}$ selection scale factor and extrapolation lead to 8\% and 2--8\% uncertainties. The \cPqb\ tagging efficiencies and their corresponding systematic uncertainties are measured from data enriched in $\ttbar$ events. They account for 5--7\% uncertainties in the total signal efficiencies.

For the background estimated from data, the statistical uncertainty of the \Pepm\PGmmp control sample is propagated to an uncertainty in the $\ttbar$+$\PW\PW$ estimation. The alpha
method for the $\cPZ + \text{jets}$ background estimation depends on the uncertainty in the extrapolation factor and on the amount of data of the dijet mass or pruned jet mass sideband region. Jet energy scale and resolution affect the extrapolation factor $\alpha(\mZZ)$ by 3--10\% depending on the mass. In the low mass region, the statistical uncertainties in the simulated samples and mass sidebands in data are propagated to the binned alpha factor estimation. In the high mass region, they are obtained by the covariance matrix of the fit parameters of the sideband data \mZZ{} distributions.
Additional systematic uncertainties are derived from comparisons between the nominal $\cPZ + \text{jets}$ MC descriptions (exclusive LO samples with different associated parton multiplicities, and enriched in \cPqb\ quark production, all produced with \MGvATNLO) and the merged \MGvATNLO simulations at NLO. The same background estimation methods are used to derive an alternative
binned description of the $\cPZ + \text{jets}$ background, and appropriate nuisance parameters, symmetrized around zero, describe the variation between the nominal and alternative estimation.

For the two dimensional \ZJJMELA{} template shapes, two systematic uncertainties are considered for the signal samples: JES and JER variations, as well as comparison with identical MC samples where \HERWIGpp~\cite{Bahr:2008pv} with EE5C tune~\cite{CUETP8M1} is used for parton showering and hadronization instead of \PYTHIA.
For background templates, a conservative systematic uncertainty from the limited size of the MC samples and the consequent smoothing procedure is derived by using alternative
templates where the content of each two dimensional interval is replaced by the content
of the preceding or following interval in \mZZ{}. Background
systematic uncertainties are validated in an ``extended
sideband region'', which includes the sideband region used in the analysis,
as well as events failing the $\tau_{21}$ selection.
At masses above 1\TeV, 1$\sigma$ differences between data and simulation in this region are assigned as additional systematic uncertainties.

\subsection{\texorpdfstring{$\PX\to \cPZ\cPZ\to 2\ell 2\nu$}{to2l2nu}}

Various factors contribute to the experimental uncertainties that apply to
processes derived from MC simulation.
These include uncertainties in the trigger efficiency and lepton selection efficiencies.
The effects of lepton momentum scale and JES are also taken
into account and are propagated to the evaluation of \ptmiss.
The uncertainties in the \cPqb\ jet veto are estimated by measuring the \cPqb\ tagging
efficiency in data enriched in $\ttbar$ and are evaluated to be 2--4\% for processes estimated from simulation, namely signal and $\PW\PW$, $\PW\cPZ$ events. Uncertainties due to the modeling of pileup are evaluated by varying the total inelastic cross section by $\pm5\%$ around the nominal value.

Uncertainties in the background estimates  based on control regions in data are estimated as described in Section~\ref{sec:2l2nubkgest}.
For the Drell--Yan background a systematic uncertainty of $25$\% is combined with a statistical uncertainty from the size
of the photon + jet control sample of 10\% for the 0-jet and $\geq$1-jet categories, and of 50\% for the VBF-tagged category.
For the nonresonant background a 15\% uncertainty is applied.

\section{Results}
\label{sec:Results}

The search for a scalar resonance $\PX$ decaying to $\cPZ \cPZ$ is performed over the mass range
$130\GeV < m_{\PX} < 3$\TeV, where three final states are combined,
$\PX\to \cPZ\cPZ\to 4\ell$, $2\ell2\Pq$, and $2\ell2\nu$. Because of the different resolutions, efficiencies, and branching fractions, each final state contributes differently depending on the tested mass. The most sensitive final state between 130 and 500\GeV is $4\ell$ due to its best mass resolution, whereas in the intermediate region 500--700\GeV $2\ell2\nu$ is most sensitive, and for masses above 700\GeV $2\ell2\Pq$ is best.

In $\PX\to \cPZ\cPZ\to 4\ell$ and $2\ell2\Pq$, comparisons between the two dimensional (\mZZ{}, \ZJJMELA{}) distributions observed in data and expected from the sum of background predictions are made. We set upper limits on the production cross section of the resonance by combining all the event categories in each analysis.

In $\PX\to \cPZ\cPZ\to 2\ell2\nu$, using the resulting \MT \ distributions,
a shape based analysis is performed to extract the limits.
The shapes of the signal and $\PW\cPZ$, $\cPZ\cPZ$ backgrounds are taken from MC simulation, those of $\cPZ + \text{jets}$ are taken from data, and for nonresonant backgrounds, the \Pe\PGm control region is used to predict both shapes and normalizations of the \MT \ distributions in the signal region, as described in Section~\ref{sec:Background}.

We follow the modified frequentist prescription described in Refs.~\cite{Junk:1999kv,Read:2002hq,Cowan2011} (CL$_\mathrm{s}$ method), and an asymptotic approach with the profile likelihood ratio as the test statistic is used for upper limits. Systematic uncertainties are treated as nuisance parameters and profiled using lognormal priors.

\begin{figure}[htbp]
\centering
\includegraphics[width=0.48\textwidth]{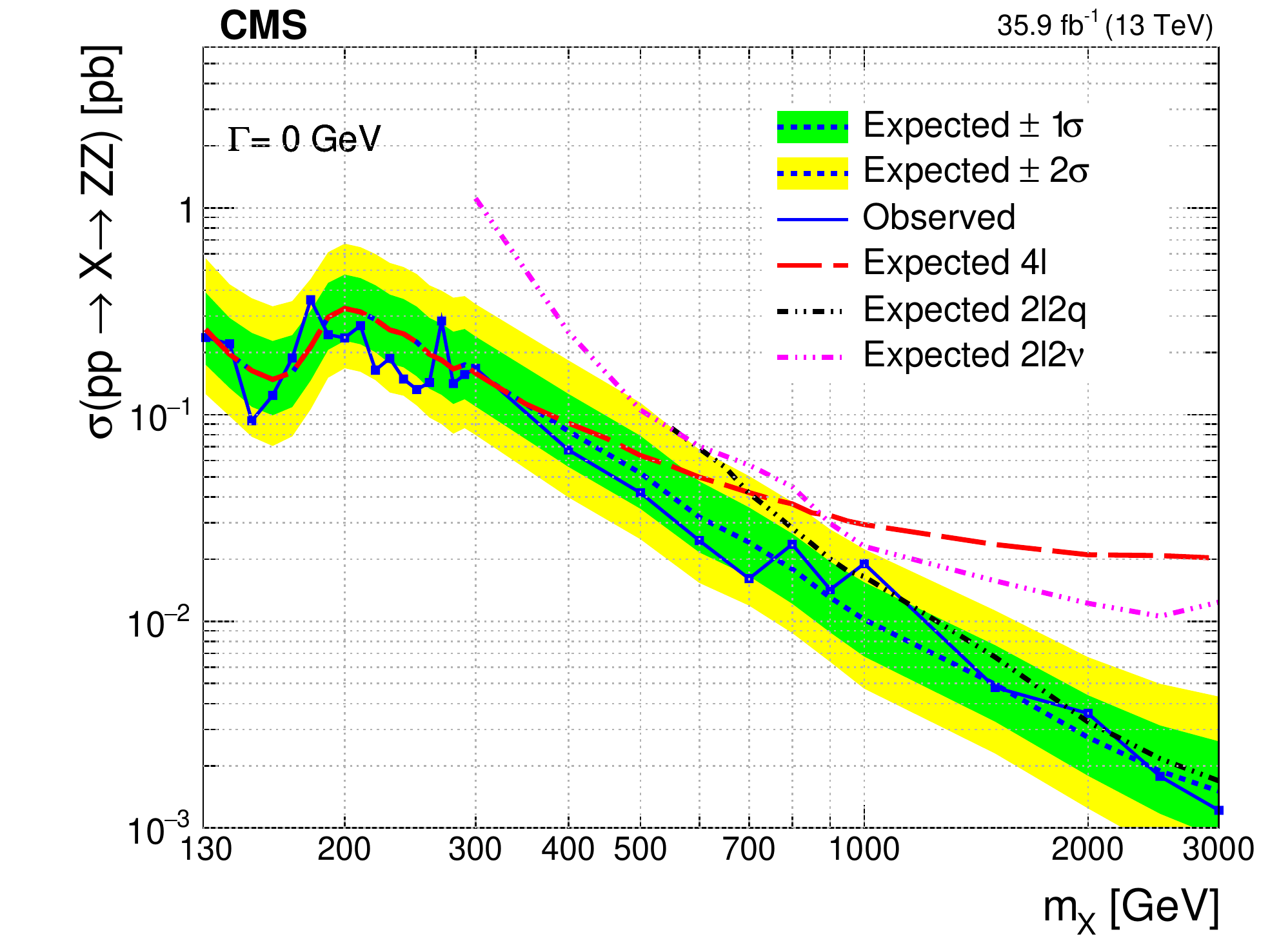}
\includegraphics[width=0.48\textwidth]{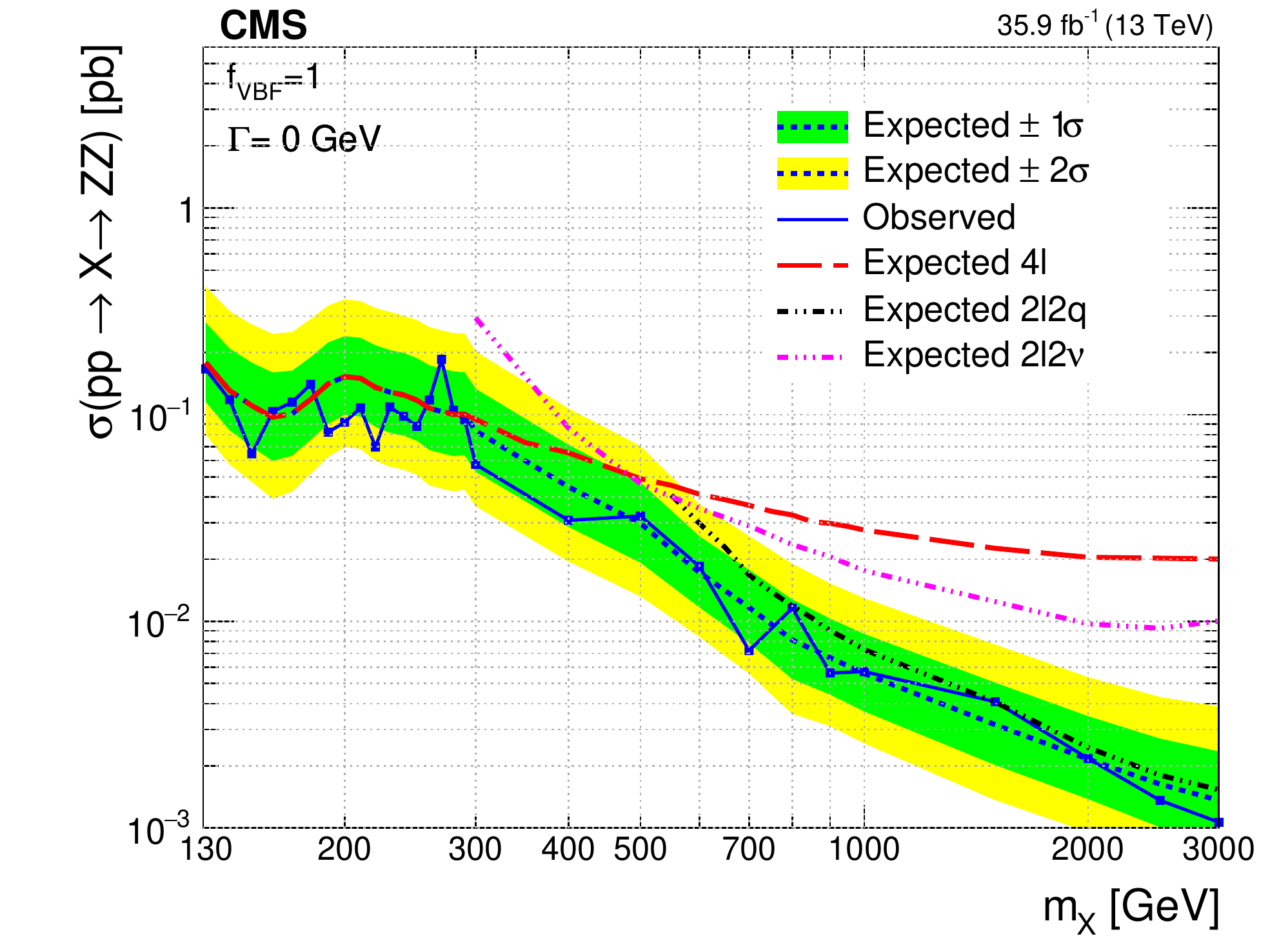}\\
\includegraphics[width=0.48\textwidth]{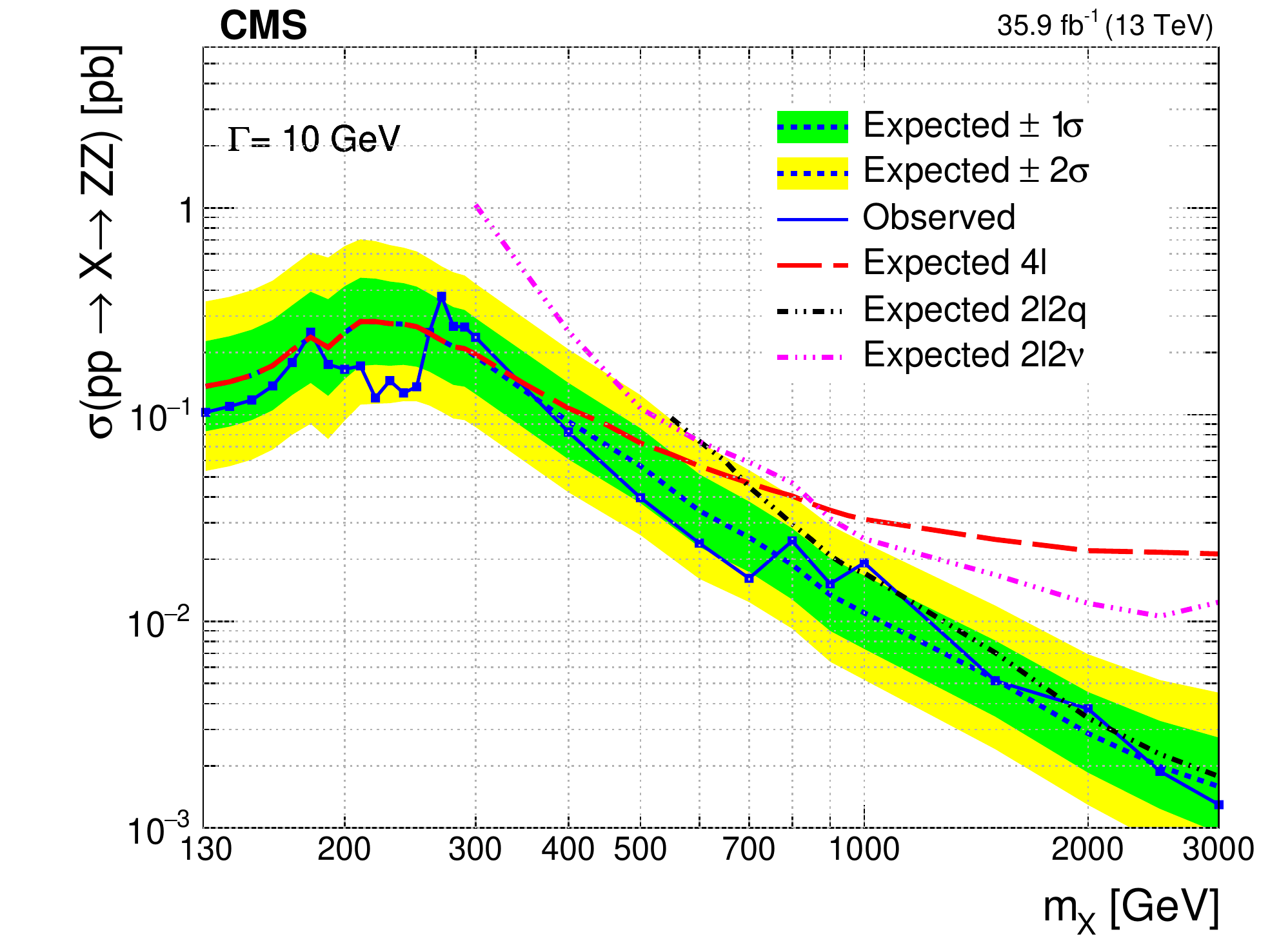}
\includegraphics[width=0.48\textwidth]{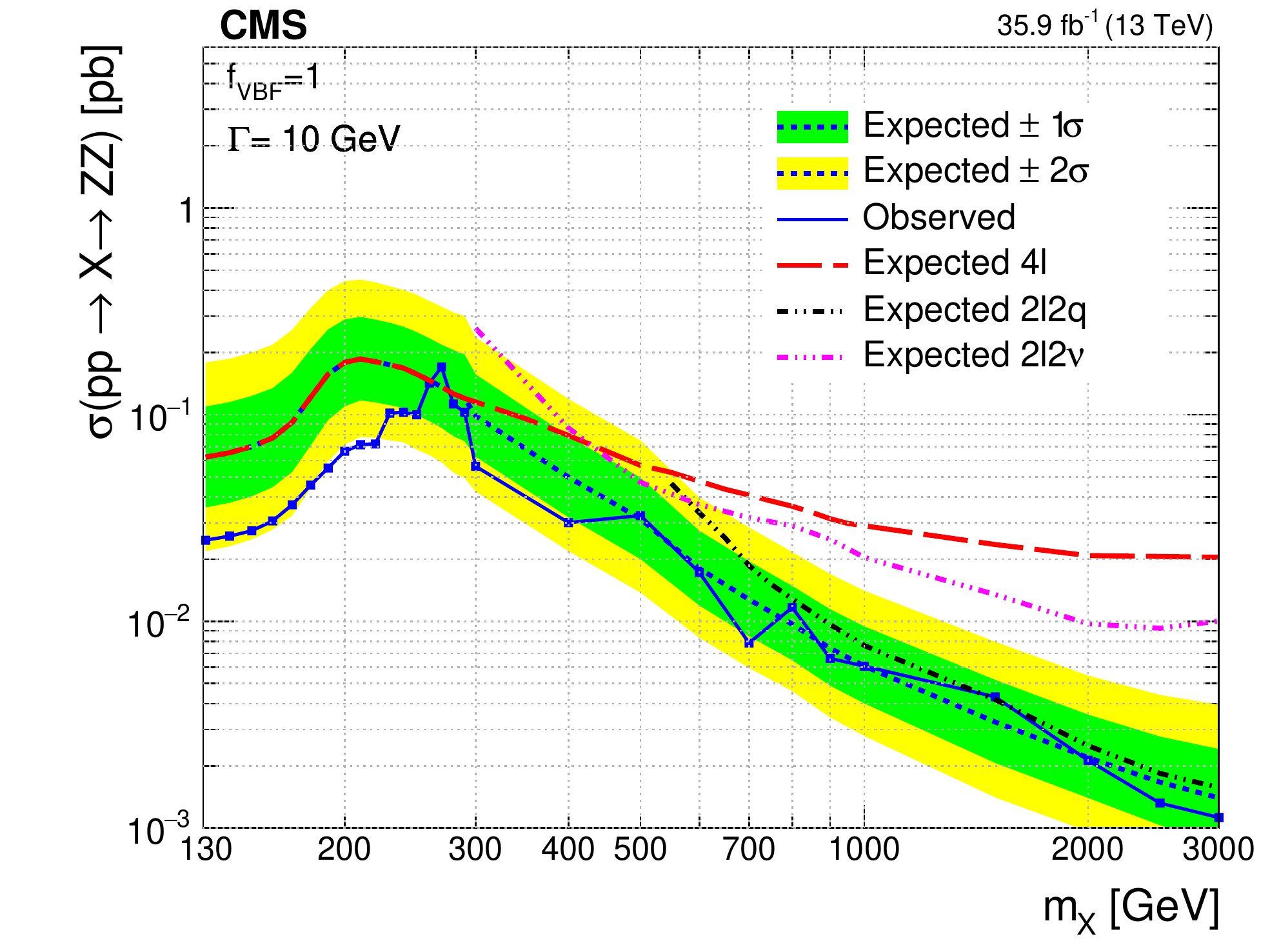}\\
\includegraphics[width=0.48\textwidth]{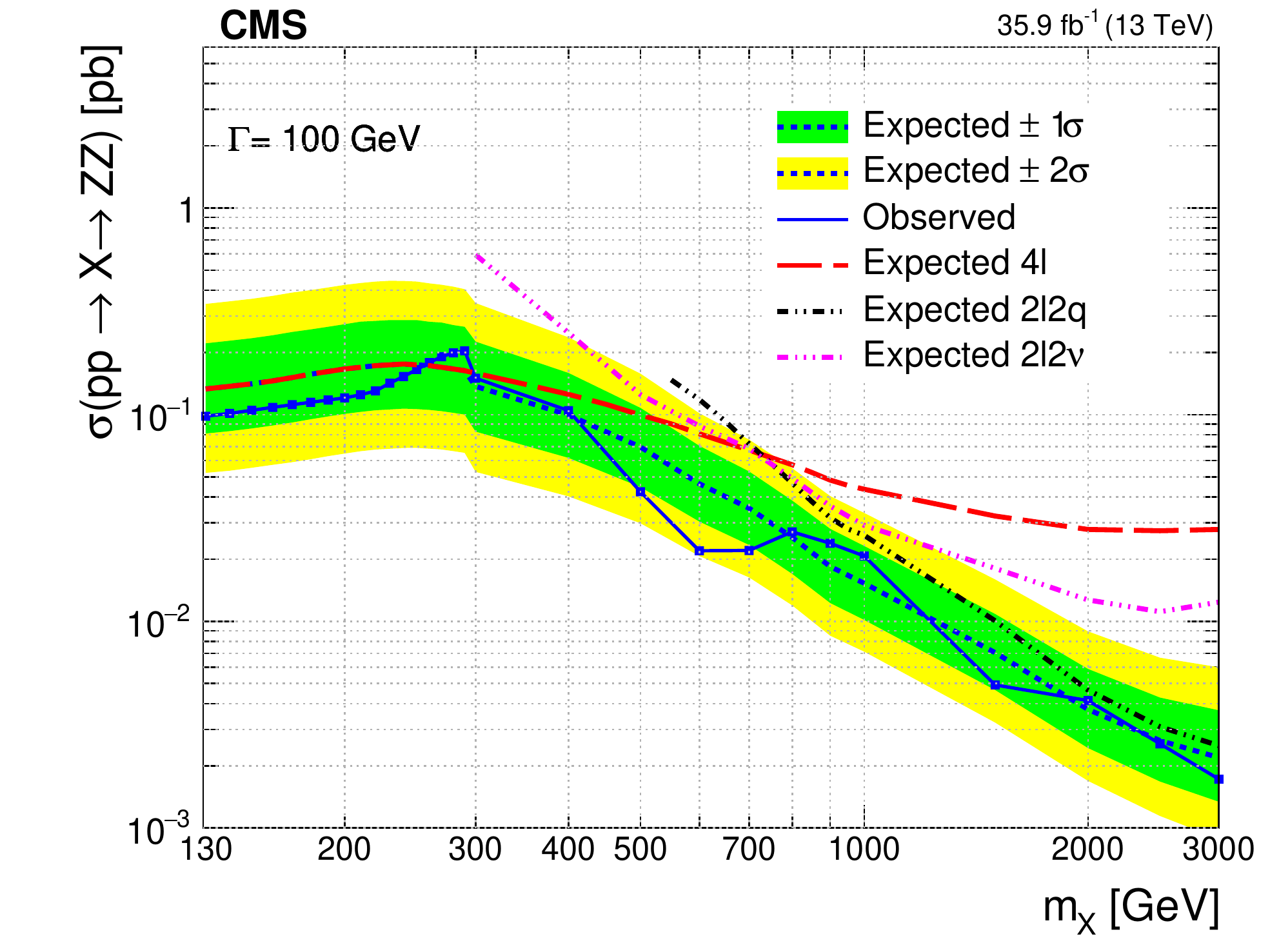}
\includegraphics[width=0.48\textwidth]{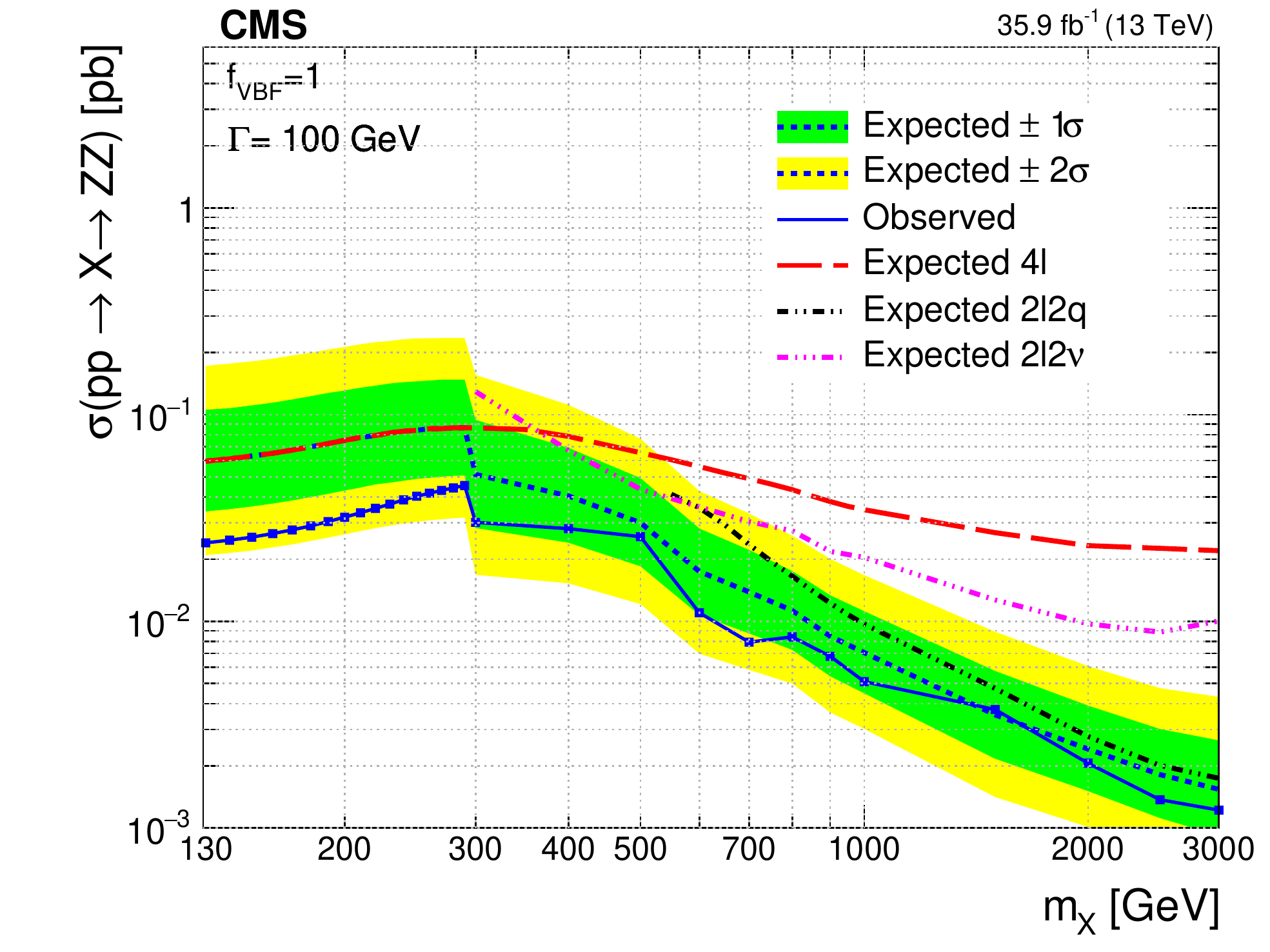}
\caption{
	Expected and observed upper limits at the 95\% \CL on the $\Pp\Pp\to\PX\to\cPZ\cPZ$ cross section as a function of $m_\PX$ and for several $\Gamma_\PX$ values with $f_{\mathrm{VBF}}$ as a free parameter (left) and fixed to 1 (right). The results are shown for $4\ell$, $2\ell2\Pq$, and $2\ell2\nu$ channels separately and combined. The reported cross section
corresponds to the signal only contribution in the absence of interference.
\label{fig:combinedresult}
}
\end{figure}

The width of the resonance $\GX$ is allowed
to vary, starting from the narrow width approximation (denoted as $\Gamma_{\PX}=0$) up to a
large width.
Production of the $\PX$ resonance is considered to be either in
\ggF or VBF, where ${\PV\PX}$ production is included according to the relative expectation of the ${\PV\PX}$ and VBF cross sections. No significant excess of events over the SM expectation is observed.
Figure~\ref{fig:combinedresult} shows upper limits at the 95\% confidence level (\CL) on the $\Pp\Pp\to\PX\to\cPZ\cPZ$ cross section
$\sigma_\PX \mathcal{B}_{{\PX}\to\cPZ\cPZ}$ as a function of $m_\PX$ for $\Gamma_\PX = $ 0, 10, and 100\GeV.

The expected and observed limits on the pure VBF production cross section are better than the inclusive ones, because the background is smaller in the dedicated VBF categories. In general, limits are better when assuming a narrow width signal, since the signal over background ratio is higher. However, in the mass region below 300\GeV, interference effects with background are more complicated and play a role in the evolution of the limit as a function of $\Gamma_{\PX}$.

For $m_{{\PX}} < 2 m_{\cPZ}$, while the signal events are produced on shell around $m_{{\PX}}$ for $\Gamma_{\PX}\sim0$, the majority of the events are produced off shell in the case of $\Gamma_{\PX}/m_{{\PX}}>1\%$. Thus the relevant background is quite different when $\Gamma_{\PX}$ varies. In the \ggF dominant category, for $130 < m_{{\PX}} < 140 $\GeV, the signal over background ratio is better in the relevant off shell region than in the on shell region, where signal events partly overlap with the $\PH(125)$ peak. This makes the sensitivity better for a wide resonance. For $150 < m_{\PX} < 180\GeV$, there is no overlap between the two on shell resonance peaks, so for a narrow resonance the signal over background ratio is larger and the limit is better. In the VBF category, the signal over background ratio is always smaller in the relevant off shell region compared to the on shell region, yielding a better sensitivity for a wide resonance. The downward fluctuation in the VBF limit for $\Gamma_{\PX}=10$ and 100\GeV, and $m_{\PX}<180\GeV$, reflects an overall deficit of events in the VBF category in the off shell region of $m_{4\ell}> 200\GeV$.

Above the 2$m_{\cPZ}$ threshold, for $180 < m_{\PX} < 250\GeV$, the net interference of the \ggF signal is positive around the peak, making the wide resonance sensitivity better. For the VBF signal, the enhancement from interference occurs at its right hand tail, where barely any background exists. This makes the limit for the wide VBF Higgs better in the range $m_{\PX}<300\GeV$. Above that, the background drops rapidly and the limits for narrow and wide resonances are compatible.

\begin{figure}[htbp]
\centering
\includegraphics[width=0.48\textwidth]{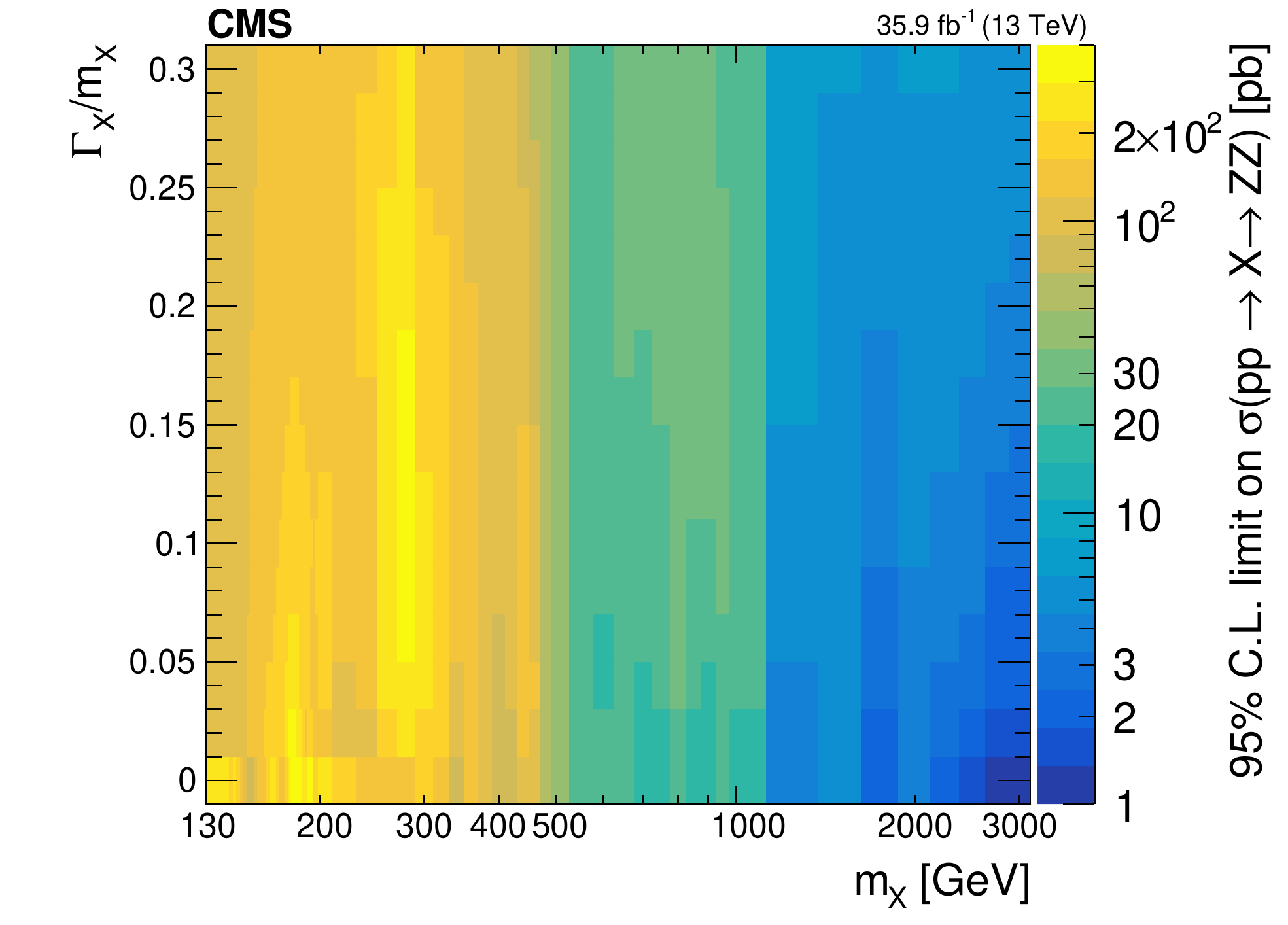}
\includegraphics[width=0.48\textwidth]{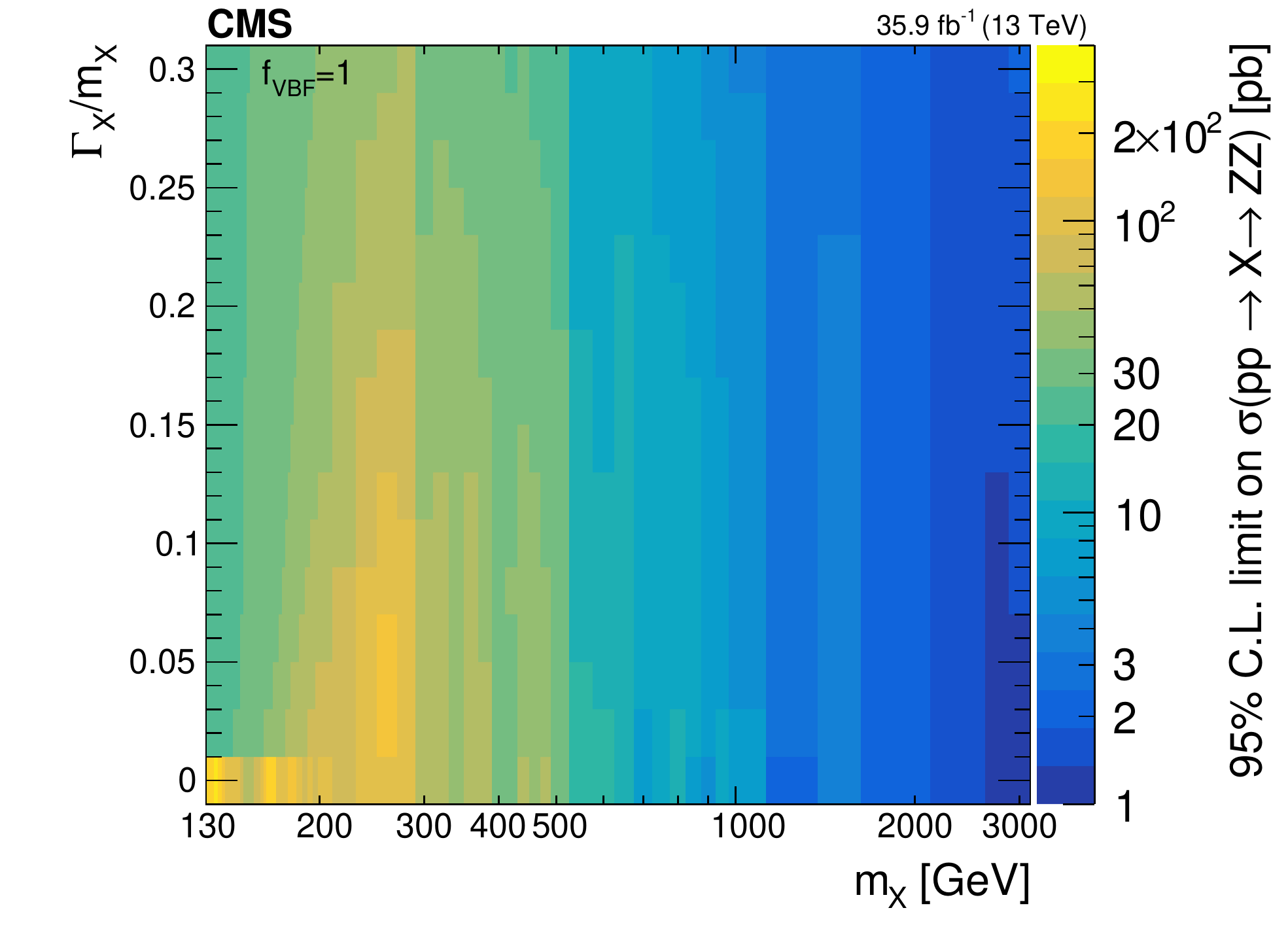}
\caption{
	Observed upper limits at the 95\% \CL on the $\Pp\Pp\to\PX\to\cPZ\cPZ$ cross section as a function of $m_\PX$ and $\Gamma_\PX/m_\PX$ values with $f_{\mathrm{VBF}}$ as a free parameter (left) and fixed to 1 (right). The results are shown for the $4\ell$, $2\ell2\Pq$, and $2\ell2\nu$ channels combined. The reported cross section
corresponds to the signal only contribution in the absence of interference.
\label{fig:2dscan}
}
\end{figure}

Figure~\ref{fig:2dscan} shows the scan of the observed upper limits at the 95\% \CL, as a function of $\mX$ and $\GX/\mX$. The mass is scanned from 130\GeV to 3\TeV and the relative width from 0 to 30\%. The results are provided with $f_{\mathrm{VBF}}$ profiled and fixed to unity. The excluded product of the cross section and branching fraction ranges from 1.2\unit{fb} at 3\TeV to 402.6\unit{fb} at 182\GeV in the case of $f_{\mathrm{VBF}}$ profiled, and from 1.0\unit{fb} at 3\TeV to 221.1\unit{fb} at 134\GeV in the VBF production mode.

\section{Summary}
\label{sec:Summary}

A search for a new scalar resonance decaying to a pair of $\cPZ$ bosons is performed for a range of masses
between 130\GeV and 3\TeV with the full data set recorded by the CMS experiment at 13\TeV during 2016
and corresponding to an integrated luminosity of \usedLumi.
Three final states $\cPZ\cPZ\to 4\ell$, $2\ell2\Pq$, and $2\ell2\nu$ are combined in the analysis, where $\ell = \Pe$ or \PGm.
Both gluon fusion and electroweak production of the scalar resonance are considered with a free parameter
describing their relative cross sections. A dedicated categorization of events based on the kinematic properties of
the associated jets is used to improve the sensitivity of the search. A description of the interference between signal and background amplitudes for a resonance of an arbitrary width is included. No significant excess of events over the SM expectation is observed and limits are set on the product of the cross section and the branching fraction for its decay to $\cPZ\cPZ$ for a wide range of masses and widths, and for different production mechanisms.

\begin{acknowledgments}
\hyphenation{Bundes-ministerium Forschungs-gemeinschaft Forschungs-zentren Rachada-pisek}

We thank Markus Schulze for optimizing the \textsc{MCFM} and \textsc{JHUGen} matrix element library for this analysis. We congratulate our colleagues in the CERN accelerator departments for the excellent performance of the LHC and thank the technical and administrative staffs at CERN and at other CMS institutes for their contributions to the success of the CMS effort. In addition, we gratefully acknowledge the computing centres and personnel of the Worldwide LHC Computing Grid for delivering so effectively the computing infrastructure essential to our analyses. Finally, we acknowledge the enduring support for the construction and operation of the LHC and the CMS detector provided by the following funding agencies: the Austrian Federal Ministry of Science, Research and Economy and the Austrian Science Fund; the Belgian Fonds de la Recherche Scientifique, and Fonds voor Wetenschappelijk Onderzoek; the Brazilian Funding Agencies (CNPq, CAPES, FAPERJ, and FAPESP); the Bulgarian Ministry of Education and Science; CERN; the Chinese Academy of Sciences, Ministry of Science and Technology, and National Natural Science Foundation of China; the Colombian Funding Agency (COLCIENCIAS); the Croatian Ministry of Science, Education and Sport, and the Croatian Science Foundation; the Research Promotion Foundation, Cyprus; the Secretariat for Higher Education, Science, Technology and Innovation, Ecuador; the Ministry of Education and Research, Estonian Research Council via IUT23-4 and IUT23-6 and European Regional Development Fund, Estonia; the Academy of Finland, Finnish Ministry of Education and Culture, and Helsinki Institute of Physics; the Institut National de Physique Nucl\'eaire et de Physique des Particules~/~CNRS, and Commissariat \`a l'\'Energie Atomique et aux \'Energies Alternatives~/~CEA, France; the Bundesministerium f\"ur Bildung und Forschung, Deutsche Forschungsgemeinschaft, and Helmholtz-Gemeinschaft Deutscher Forschungszentren, Germany; the General Secretariat for Research and Technology, Greece; the National Scientific Research Foundation, and National Innovation Office, Hungary; the Department of Atomic Energy and the Department of Science and Technology, India; the Institute for Studies in Theoretical Physics and Mathematics, Iran; the Science Foundation, Ireland; the Istituto Nazionale di Fisica Nucleare, Italy; the Ministry of Science, ICT and Future Planning, and National Research Foundation (NRF), Republic of Korea; the Lithuanian Academy of Sciences; the Ministry of Education, and University of Malaya (Malaysia); the Mexican Funding Agencies (BUAP, CINVESTAV, CONACYT, LNS, SEP, and UASLP-FAI); the Ministry of Business, Innovation and Employment, New Zealand; the Pakistan Atomic Energy Commission; the Ministry of Science and Higher Education and the National Science Centre, Poland; the Funda\c{c}\~ao para a Ci\^encia e a Tecnologia, Portugal; JINR, Dubna; the Ministry of Education and Science of the Russian Federation, the Federal Agency of Atomic Energy of the Russian Federation, Russian Academy of Sciences, the Russian Foundation for Basic Research and the Russian Competitiveness Program of NRNU ``MEPhI"; the Ministry of Education, Science and Technological Development of Serbia; the Secretar\'{\i}a de Estado de Investigaci\'on, Desarrollo e Innovaci\'on, Programa Consolider-Ingenio 2010, Plan de Ciencia, Tecnolog\'{i}a e Innovaci\'on 2013-2017 del Principado de Asturias and Fondo Europeo de Desarrollo Regional, Spain; the Swiss Funding Agencies (ETH Board, ETH Zurich, PSI, SNF, UniZH, Canton Zurich, and SER); the Ministry of Science and Technology, Taipei; the Thailand Center of Excellence in Physics, the Institute for the Promotion of Teaching Science and Technology of Thailand, Special Task Force for Activating Research and the National Science and Technology Development Agency of Thailand; the Scientific and Technical Research Council of Turkey, and Turkish Atomic Energy Authority; the National Academy of Sciences of Ukraine, and State Fund for Fundamental Researches, Ukraine; the Science and Technology Facilities Council, UK; the US Department of Energy, and the US National Science Foundation.

Individuals have received support from the Marie-Curie programme and the European Research Council and Horizon 2020 Grant, contract No. 675440 (European Union); the Leventis Foundation; the A. P. Sloan Foundation; the Alexander von Humboldt Foundation; the Belgian Federal Science Policy Office; the Fonds pour la Formation \`a la Recherche dans l'Industrie et dans l'Agriculture (FRIA-Belgium); the Agentschap voor Innovatie door Wetenschap en Technologie (IWT-Belgium); the F.R.S.-FNRS and FWO (Belgium) under the ``Excellence of Science - EOS" - be.h project n. 30820817; the Ministry of Education, Youth and Sports (MEYS) of the Czech Republic; the Lend\"ulet (``Momentum") Programme and the J\'anos Bolyai Research Scholarship of the Hungarian Academy of Sciences, the New National Excellence Program \'UNKP, the NKFIA research grants 123842, 123959, 124845, 124850 and 125105 (Hungary); the Council of Scientific and Industrial Research, India; the HOMING PLUS programme of the Foundation for Polish Science, cofinanced from European Union, Regional Development Fund, the Mobility Plus programme of the Ministry of Science and Higher Education, the National Science Center (Poland), contracts Harmonia 2014/14/M/ST2/00428, Opus 2014/13/B/ST2/02543, 2014/15/B/ST2/03998, and 2015/19/B/ST2/02861, Sonata-bis 2012/07/E/ST2/01406; the National Priorities Research Program by Qatar National Research Fund; the Programa de Excelencia Mar\'{i}a de Maeztu and the Programa Severo Ochoa del Principado de Asturias; the Thalis and Aristeia programmes cofinanced by EU-ESF and the Greek NSRF; the Rachadapisek Sompot Fund for Postdoctoral Fellowship, Chulalongkorn University and the Chulalongkorn Academic into Its 2nd Century Project Advancement Project (Thailand); the Welch Foundation, contract C-1845; and the Weston Havens Foundation (USA).

\end{acknowledgments}

\bibliography{auto_generated}

\cleardoublepage \appendix\section{The CMS Collaboration \label{app:collab}}\begin{sloppypar}\hyphenpenalty=5000\widowpenalty=500\clubpenalty=5000\input{HIG-17-012-authorlist.tex}\end{sloppypar}
\end{document}

%% file: HIG-17-012-authorlist.tex
\vskip\cmsinstskip
\textbf{Yerevan Physics Institute,  Yerevan,  Armenia}\\*[0pt]
A.M.~Sirunyan,  A.~Tumasyan
\vskip\cmsinstskip
\textbf{Institut f\"{u}r Hochenergiephysik,  Wien,  Austria}\\*[0pt]
W.~Adam,  F.~Ambrogi,  E.~Asilar,  T.~Bergauer,  J.~Brandstetter,  E.~Brondolin,  M.~Dragicevic,  J.~Er\"{o},  A.~Escalante Del Valle,  M.~Flechl,  M.~Friedl,  R.~Fr\"{u}hwirth\cmsAuthorMark{1},  V.M.~Ghete,  J.~Grossmann,  J.~Hrubec,  M.~Jeitler\cmsAuthorMark{1},  A.~K\"{o}nig,  N.~Krammer,  I.~Kr\"{a}tschmer,  D.~Liko,  T.~Madlener,  I.~Mikulec,  E.~Pree,  N.~Rad,  H.~Rohringer,  J.~Schieck\cmsAuthorMark{1},  R.~Sch\"{o}fbeck,  M.~Spanring,  D.~Spitzbart,  A.~Taurok,  W.~Waltenberger,  J.~Wittmann,  C.-E.~Wulz\cmsAuthorMark{1},  M.~Zarucki
% ...
\vskip\cmsinstskip
% ...
1:~Also at Vienna University of Technology,  Vienna,  Austria\\
% ...
